\pdfoutput=1
\documentclass[acmsmall,nonacm,screen,appendix]{acmart}

\definecolor{Violet}{HTML}{5836A5}

\newcommand{\ourlogicname}{Violet}
\newcommand{\ourlogic}{\texorpdfstring{\textsf{\textcolor{Violet}{Violet}}}{Violet}}
\let\thelogic\ourlogic

\newcommand{\Bluebell}{\textsc{Bluebell}\xspace}

\providecommand{\condition}[1][]{partition#1\xspace}
\providecommand{\Condition}[1][]{Partition#1\xspace}
\usepackage[capitalize]{cleveref}

\crefname{rule}{rule}{rules}
\Crefname{rule}{Rule}{Rules}
\crefname{step}{step}{steps}
\Crefname{step}{Step}{Steps}
\crefname{line}{Line}{Lines}
\crefrangelabelformat{line}{#3#1#4--#5#2#6}

\usepackage{thmtools}
\usepackage{thm-restate}
\AtEndPreamble{\theoremstyle{acmdefinition}\newtheorem{remark}[theorem]{Remark}\theoremstyle{acmplain}}

\usepackage{booktabs}
\usepackage{microtype}
\usepackage{wrapfig}
\usepackage{natbib}
\usepackage[inline,shortlabels]{enumitem}
\usepackage{mathpartir}

\usepackage{style/comments}
\usepackage{style/notation}
\usepackage{style/appendix}
\usepackage{style/tikztricks}

\additionalmaterial[The Appendix]{Appendix}

\acmDOI{10.1145/3871749}

\ifdraft

\else
\fi

\ifdraft
  \fi

\begin{document}

\title{\ourlogicname: A Logical Framework for Information Security via Possibilistic Conditional Independence}
\ifappendix
  \subtitle{(Extended Version)}
\fi

 \author{Flavio Ascari}
 \orcid{0000-0003-4624-9752}
 \email{flavio.ascari@uni-konstanz.de}
 \affiliation{\institution{University of Konstanz}
  \city{Konstanz}
  \country{Germany}
 }
\author{Emanuele D'Osualdo}
\orcid{0000-0002-9179-5827}
\email{emanuele.dosualdo@uni-konstanz.de}
 \affiliation{\institution{University of Konstanz}
    \city{Konstanz}
    \country{Germany}
}
\author{Deepak Garg}
\orcid{0000-0002-0888-3093}
\email{dg@mpi-sws.org}
\affiliation{\institution{MPI-SWS}
    \city{Saarbrücken}
    \country{Germany}
}
\author{Ana Oliveira da Costa}
\orcid{0000-0002-8741-5799}
\email{ana.costa@ist.ac.at}
\affiliation{\institution{IST Austria}
    \city{Vienna}
    \country{Austria}
}

\begin{abstract}
Language-based information flow control is varied by necessity: no single security notion fits every setting. Perfect security forbids all flow from secret to public, but real systems must release information in controlled ways, and the appropriate declassification policies depend on the application context. This diversity of definitions is matched by a proliferation of verification methods, from type systems and logical relations to bisimulations, relational program logics, and hyperlogics.

Our thesis is that underneath this variety sits a single notion: Conditional independence between secret and public information.
As a conditional version of Sutherland's non-deducibility, conditional independence captures perfect security, generalizes cleanly to declassification, and applies uniformly to nondeterministic programs.
We realize this potential by internalizing conditional independence in a modal separation logic, \thelogic. Reading the separating conjunction as independence and a partitioning modality as conditioning, \thelogic\ becomes at once a language for \emph{specifying} security policies and a proof system for \emph{verifying} programs against them. Because assertions are predicates over sets of states, reasoning stays unary in feel despite a relational target. From a small set of general-purpose rules over these connectives, we specify and verify perfect security and several declassification policies across deterministic and nondeterministic programs, and we recover a standard information flow type system as derived rules.
\end{abstract}



\maketitle

\section{Introduction}
\label{sec:intro}

Information flow control (IFC), or the problem of verifying that a
program releases sensitive information only in accordance with a
security policy, is of paramount importance and widely studied. In its
strictest form, information flow control stipulates that sensitive
information, often called ``high'' information, should not flow to, or
influence, attacker-visible outputs, also called ``low'' outputs. This
property is formalized as
\emph{noninterference}~\cite{DBLP:journals/jcs/VolpanoIS96,DBLP:conf/sp/GoguenM82a},
which states that two runs of a program that differ only in high
inputs should produce equal low outputs. There is nearly universal
agreement on noninterference as the strongest possible information
security property.

In practice, however, absolute segregation of high and low information
is rarely the goal. Instead, we want the high information to be
\emph{declassified} and for it to influence low outputs but only in
policy-specified ways. This is where the formal literature becomes
highly nuanced: Of necessity, there are several, vastly different
classes of declassification policies, roughly divided based on how the
policy specifies the declassification conditions (``what''-,
``where''-,
``when''-declassification~\cite{DBLP:journals/jcs/SabelfeldS09,DBLP:conf/isss2/SabelfeldM03,DBLP:conf/pldi/AskarovS07,DBLP:conf/sp/AskarovS07,DBLP:conf/popl/BrobergS10,DBLP:conf/esop/BrobergS06,DBLP:conf/ccs/ChongM04,DBLP:conf/csfw/MyersSZ04,DBLP:conf/ccs/CecchettiMA17,DBLP:journals/pacmpl/MenzHLG23,DBLP:conf/ifip1-7/Boudol08})
and, correspondingly, several very different definitions of
information flow security that rely on a variety of techniques
spanning relations on pairs of runs like
bisimulation~\cite{DBLP:conf/csfw/Mantel00,DBLP:conf/esorics/Mantel00,DBLP:conf/isss2/SabelfeldM03,DBLP:journals/jcs/VolpanoIS96,DBLP:conf/ccs/ChongM04,DBLP:conf/ccs/CecchettiMA17,DBLP:conf/popl/PottierS02,DBLP:conf/ifip1-7/Boudol08,DBLP:conf/icfp/BuirasVR15},
logical relations and
PERs~\cite{DBLP:conf/popl/HeintzeR98,DBLP:conf/csfw/Rajani018,DBLP:journals/pacmpl/MenzHLG23,DBLP:journals/toplas/NanevskiBG13,DBLP:journals/lisp/SabelfeldS01},
attacker knowledge
(epistemology)~\cite{DBLP:conf/sp/AskarovS07,DBLP:conf/popl/BrobergS10,DBLP:conf/pldi/AskarovS07},
and denotational models~\cite{DBLP:conf/popl/AbadiBHR99}.  The
literature on methods for the verification of information flow
properties is similarly diverse and spans static
analysis~\cite{DBLP:journals/pacmpl/LambaTBBBL24,hammer09ijis,DBLP:conf/sac/MastroeniP19},
type
systems~\cite{DBLP:journals/jcs/VolpanoIS96,DBLP:conf/isss2/SabelfeldM03,DBLP:conf/pldi/AskarovS07,DBLP:conf/csfw/Rajani018,DBLP:conf/popl/Myers99,DBLP:conf/ccs/ChongM04,DBLP:conf/ccs/CecchettiMA17,DBLP:conf/popl/AbadiBHR99,DBLP:conf/popl/PottierS02,DBLP:journals/pacmpl/MenzHLG23,DBLP:conf/ifip1-7/Boudol08,DBLP:conf/icfp/BuirasVR15,DBLP:conf/popl/HuntS06},
unary program logics~\cite{DBLP:journals/mscs/BartheDR11}, relational
program
logics~\cite{DBLP:journals/scp/AmtoftB07,GregersenBTB21,DBLP:journals/toplas/NanevskiBG13,DBLP:conf/sas/AmtoftB04,DBLP:conf/sp/BanerjeeNR08},
and
hyperlogics~\cite{DBLP:conf/popl/AssafNSTT17,DBLP:journals/sLogica/GradelV13,DBLP:conf/mfcs/KrebsMV018,DBLP:conf/fsttcs/VirtemaHFK021,DBLP:journals/iandc/KontinenSV25}.

\paragraph{Problem and Main Ideas}
We ask whether the different notions of information flow security
share a common structure and, if so, whether that structure would lead
to a logical framework that can express the different definitions and
derived inference rules to verify those properties. We answer both
questions in the affirmative: We argue that there is indeed such a
common structure and, in fact, one that has been proposed in prior
work, but never really exploited for our objectives.

Our starting point is a generalization of noninterference by
Sutherland~\cite{sutherland1986}, now called \emph{non-deducibility}. Working in
a nondeterministic setting, Sutherland formalized (perfect)
information flow security as the non-deducibility of anything about
high information from the low outputs. Formally, if $S$ is the set of
states of a program (containing both high information and low outputs)
and $\pi_H(S)$ and $\pi_L(S)$ are, respectively, the projections of
$S$ to the high state and low outputs, then non-deducibility requires
$S = \pi_H(S) \times \pi_L(S)$.  Stated differently, restricting one's
view to a subset of the low outputs does not change the possible
values of high state, so the high state and low outputs must be
\emph{independent}. Remarkably, this definition applies to both
deterministic and nondeterministic systems and, on the former, it
coincides with noninterference.

Our first observation is that a further generalization of
non-deducibility, namely, \emph{conditional independence}, is an
expressive basis for representing a variety of notions of
declassification security.
Take for example the set of states
$ S = \set{(z,h,l) | h = z \lor h = z+1, l = z \lor l = z+1, z\in \Int } $
over three variables \p{h} (high), \p{l} (low) and \p{z}.
The possible values of \p{h} include all integers,
yet knowing the value of \p{l} would pin down the value of \p{h} up to an error
of at most~$\pm 1$, which makes \p{l} not independent of \p{h}.
However, if we fix \p{z} to some value~$v$, the possible values
of $(\p{h}, \p{l})$ are $\set{v,v+1}\times\set{v,v+1}$:
\p{h} and \p{l} are independent conditionally on \p{z}.
A policy where \p{z} is declassified stipulates that after the release
of \p{z}'s value, no \emph{additional} information is leaked about high
through low;
the act of declassifying \p{z} can be modelled as pinning its value,
and the policy can then be expressed as the independence of high and low
conditionally on the leaked value of \p{z}.

Our second insight,
inspired by recent advances in logics for probabilistic programs
\cite{BartheHL20,BaoDH021,Li0H23,DBLP:journals/pacmpl/BaoDF25},
is that this notion of \emph{independence can be
internalized as separating conjunction} in a separation logic over a
suitable domain based on Boolean algebras over sets, which we
call \emph{possibility spaces}.
The separating conjunction $P_1 * P_2$
over this domain models independence as we want it:
Roughly, in the special case where $P_1$ and $P_2$ predicate over disjoint
regions of state,
$P_1*P_2$ holds on a set of states~$S$ if it can be decomposed into
a product $S_1 \times S_2$ such that $P_i$ holds on $S_i$, for $i=1,2$.
Our definition is more general in that our separation is not tied to heap
partitioning: If \p{r} can take both values in
$\{\p{true},\p{false}\}$, then $\p{x}=1$ and $\p{x} \oplus \p{r} = 1$
are independent in our logic, even though they share the variable \p{x}.

Next, to model conditional independence, we introduce a
\emph{\condition[ing] modality}, $\cond{x}{V}{P(x)}$, which holds on a
set of states~$S$ if~$S$ can be partitioned into sets~$\set{S_{v}}_{v
  \in V}$ such that for every $v\in V$, $P(v)$ holds on $S_{v}$.
We can then state independence of \p{y} and \p{z} conditional on \p{x} by combining independence-as-separation and the \condition[ing] modality:
for instance,
independence of \p{h} and \p{l} conditional on \p{z} can be expressed as
$
  \Exists V.
    \cond{v}{V}{(\sure{\p{z}=v} * \own{\p{h}} * \own{\p{l}})}
$;
the assertion states that it is possible to see the
possible states~$S \subs \Int^3$
as a partition $\Dunion_{v\in V} \set{v} \times S^v_{\p{h}} \times S^v_{\p{l}}$
where $v \in V$ pins the value of \p{z}, and $S^v_{\p{h}}$ and $S^v_{\p{l}}$ are
the possible values of \p{h} and \p{l} respectively, fixing $\p{z}=v$.
By generalizing to a \p{z} representing a
declassification condition, we can express various
declassification-based security properties in the logic.

We call the resulting logic {\ourlogic}. Unlike many prior logics for
establishing noninterference, which are relational, reasoning in
{\ourlogic} is closer to unary logics, thanks to our assertions being
predicates over sets of states, not individual states.

Finally, {\ourlogic} has a small set of intuitive proof rules for the
separating conjunction and partitioning modality,
enabling not just the \emph{expression} of security policies,
but their \emph{verification} within the framework.
We show the rules to be
sufficient for verifying programs that use different notions of
declassification, and also simulate entire proof systems for
IFC from prior work as derived rules within the
logic. Consequently, {\ourlogic} can also be used to combine proofs in
existing verification systems with proofs carried out natively in
{\ourlogic}, in the spirit of semantic typing.

\paragraph{Summary of Contributions}
Overall, we make the following contributions. 1) The observation that
conditional independence is a common foundation for representing many
information flow properties, with or without declassification,
nondeterminism and locally encapsulated high state (which is hard to
model relationally); 2) The design of {\ourlogic}, a separation logic
over possibility spaces, which internalizes independence as the
separating conjunction and conditioning as the \condition[ing]
modality; 3) Several examples and the derivation of standard typing
rules for noninterference to show that {\ourlogic}'s inference rules
are sufficient to prove various information flow properties, in both
deterministic and nondeterministic programs, even with features that
are hard to model in prior work, e.g., secret locally encapsulated
state.
The whole development has been mechanized in Rocq~\cite{artifact}. \section{Key Ideas}
\label{sec:key-ideas}

The main thesis of \ourlogic\ is that,
by internalizing conditional independence in a separation logic,
one obtains a versatile and abstract logical system in which
security notions can be
\begin{enumerate*}[a)]
  \item \emph{expressed} in terms
        of a handful of simple connectives, and
  \item \emph{reasoned about} by deriving the principles applicable to specific
        notions from general-purpose reasoning principles that apply to the connectives.
\end{enumerate*}
We explain how \ourlogic\ obtains this in three steps:
we first focus on specifying security, then on verification, and finally
on how to build abstractions (\eg security types) on top of the base framework.

\subsection{Specifying Security Notions}
\label{sec:key-ideas:spec}

\begin{figure}
  \adjustfigure[\small]
  \lstset{gobble=2}
  \begin{codecols}
  \begin{sourcecode*}
  def ex1:
    v := h;
    l := (h+v) mod 2
  \end{sourcecode*}
\begin{sourcecode*}
  def ex2:
    v := 0;
    l := (h+v) mod 2
  \end{sourcecode*}
\begin{sourcecode*}
  def ex3:
    v := rnd({0,1});
    l := (h+v) mod 2
  \end{sourcecode*}
  \end{codecols}
  \Description{The three programs show that the same piece of code l := (h+v) mod 2 can be either secure or insecure, depending on the context, and we show how \thelogic{} deals with these cases.}
  \caption{Running examples}
  \label{fig:running-ex}
\end{figure}

\subsubsection{Security via Independence}

To explore our approach to specification,
we consider the standard problem of limiting flows between two security levels:
$\high$ for sensitive data, and~$\low$ for public data.
Consider the examples in \cref{fig:running-ex}:
\p{h} stores a high input, \p{l} is a low location,
and \p{v} is a local variable.
The security notion we want the programs to comply with
is standard \emph{noninterference}:
if we run the program from two states that may
differ only on the $\high$~input, we should observe the same $\low$
output.
For instance, \p{ex1} satisfies the policy, as \p{l} will always be assigned 0
regardless of the value of \p{h}.
By contrast, \p{ex2} violates the policy because \p{l} will contain the last
bit of \p{h}, and thus will vary with the high input.

The noninterference property can be rephrased using the concept of independence.
Two expressions $\expr_1$ and $\expr_2$ are \emph{independent}
if learning the value of $\expr_1$
does not restrict the possible values of $\expr_2$.
More formally, an expression can be seen as a function from states~$\State$ to values~$\Val$, $ \expr_1,\expr_2 \from \State \to \Val $;
let $S \subs \State$ be a set of possible states;
then $\expr_1$ is independent of $\expr_2$ if
$
  \angles{\expr_1,\expr_2}(S) \is
  \set{ (\expr_1(\state), \expr_2(\state)) | s \in S }
  =
  \expr_1(S) \times \expr_2(S).
$
Here $\expr_1(S)$ represents the set of possible values of $\expr_1$;
when the expressions are independent,
learning that we are in a state where~$\expr_1$ evaluates to a specific~$v$
only restricts the possible states to some~$S'$ where
$
  \angles{\expr_1,\expr_2}(S') = \set{v} \times \expr_2(S)
$, which does not restrict $\expr_2(S)$.
In our examples, the fact that we declared \p{h} as $\high$
can be represented as the fact that the set of possible initial states
contains a state for each possible high input $h \in H$.
Then, let $S_1$ be the set of output states after \p{ex1} has terminated.
The expression \p{l} evaluates to 0 in every $s\in S_1$, and thus
$\angles{\p{l}, \p{h}}(S_1) = \set{0} \times H$, which does not refine our
knowledge of what value \p{h} might store beyond what was known initially.
By contrast, for the set~$S_2$ of states produced by \p{ex2}
we have $\angles{\p{l}, \p{h}}(S_2) = (\set{0} \times 2\Int) \union (\set{1} \times (2\Int+1)) \ne \set{0,1} \times \Int$.

The connection between independence and noninterference is well known,
but \ourlogic\ elevates the concept to the primary tool of specification
and reasoning for security policies.
In \ourlogic\ we internalize independence as \emph{separating conjunction}
$P_1 * P_2$.
Since independence is not a property of single states,
our assertions should be interpreted as constraining the set of possible states.
For example, the \ourlogic\ assertion
$\In{\expr}{V}$
holds when the set of possible values of expression~$\expr$
is \emph{precisely}~$V$: only values in $V$ are possible,
and for each $v \in V$, there is some possible $\state$
where $\expr(\state) = v$.
As for separation, when $P_1$ and $P_2$ predicate over disjoint parts of memory,\footnote{We use this restriction for simplicity here,
  but the actual model presented in \cref{sec:model}
  is more sophisticated and
  allows $P_1$ and $P_2$
  to predicate over overlapping regions of memory.}
the assertion $P_1 * P_2$ holds if the possible states can be seen
as a set $S_1 \times S_2$ where
$S_i$ is the portion of state that is relevant for $P_i$ for $i=1,2$,
and $P_1$ holds on $S_1$ and $P_2$ holds on $S_2$.
With this, we can express independence of \p{h} and \p{l} by
$ \In{\p{l}}{L} * \In{\p{h}}{H} $ for some sets $H,L$;
since the precise sets are irrelevant for stating independence,
we use the abbreviation $\own{\expr} \is \Exists V.\In{\expr}{V} $
and write $\own{\p{l}} * \own{\p{h}}$.
Noninterference of \p{ex1} can then be expressed as a triple:
\begin{equation}
  \forall H \subs \Int\st
  \T
  {\In{\p{h}}{H}}
    {\p{ex1}}
  { \own{\p{l}} * \own{\p{h}} }
\label{triple:vhout-secure}
\end{equation}
The precondition asserts that \p{h} has initial possible values~$H$, and after running $\p{ex1}$ the possible states see \p{l} and \p{h}
as independent.

\subsubsection{Nondeterminism and Secret Generation}
Having independence as separation in assertions is not just useful to
provide guarantees on the output states, but to state assumptions too.
For example, \code{l := (h+v) mod 2} would be secure if
\p{v} were an unknown value in
$\set{0,1}$ and its value were independent of \p{h};
as a \ourlogic\ assertion: $ \In{\p{h}}{H} * \In{\p{v}}{\set{0,1}} $.
In that case, for every possible value~$l$ of \p{l}
and~$h$ of \p{h}, there is a possible value of \p{v} that would make
the pair $(l,h)$ possible in the output states.
In \ourlogic, this fact can be represented as a valid triple:
\begin{equation}
  \forall H \subs \Int\st
  \T
  {\In{\p{h}}{H} * \In{\p{v}}{\set{0,1}}}
    {\code{l := (h+v) mod 2}}
  { \own{\p{l}} * \own{\p{h}} }
\label{triple:vhindep-secure}
\end{equation}

A notable advantage of the formulation of noninterference via
independence is that it supports nondeterminism out of the box.
For example, we allow our programming language to call \code{rnd($V$)},
which returns an arbitrary element of the set of values~$V$.
In the \p{ex3} program in \cref{fig:running-ex} we use \code{rnd}
to model the generation of a new boolean secret in \p{v}.
The axiom associated with \code{rnd} is the triple
$
  \T{\True}{\code{x:=rnd($V$)}}{\In{\p{x}}{V}}.
$
With a single application of the \ref{rule:frame} rule
(which we discuss in \cref{sec:key-ideas:verif})
we can establish the independence of \p{h} and \p{v}
after the first line of example~\p{ex3}:
\begin{equation}
  \T
  { \In{\p{h}}{H} }
    {\code{v:=rnd(\{0,1\})}}
  { \In{\p{h}}{H} * \In{\p{v}}{\set{0,1}} }
\label{triple:vrnd}
\end{equation}
Together, \eqref{triple:vhindep-secure} and \eqref{triple:vrnd}
prove that \p{ex3} is secure, \ie
$
  \T
  { \In{\p{h}}{H} }
    {\p{ex3}}
  { \own{\p{l}} * \own{\p{h}} }.
$

\subsubsection{Noninterference}
Using the tools outlined above, it is actually possible to faithfully encode
\emph{nondeterministic termination insensitive noninterference}~(TINI)
as a triple in \ourlogic.
To achieve generality, the programming language used in \ourlogic\ is
purposely generic and has no inherent notion of security labels built in.
The language instead is equipped with an \emph{append-only log},
which can be used to record security-relevant events
using the \code{log!(v)} instruction.
The intended meaning of such events is then determined by how the assertions in the specification use them to encode the desired security notion.
This gives us a clean \emph{intermediate language} which can serve as the target of a compiler from a security-oriented source language with suitable security-label annotations.
We show how this works in \cref{sec:abstractions} where we compile the language
and type system of Volpano-Irvine-Smith into \ourlogic.

To model TINI, we follow a simple strategy.
To generically model $\high$ \emph{input} we reserve the first item of the log
to be an entry $(\high, \vec{h})$ where $\vec{h}$ is a sequence of values.
We write~$\highIn$ for the expression that returns such $\vec{h}$ by reading it from the log.
Since the log is append-only, referring to $\highIn$ unambiguously refers to
the input, thus avoiding laundering.
Similarly,
we log low output by logging \code{(low, $\vec{v}$)}, and
the expression~$\lowOf$ extracts the sequence of \pre\low-tagged
values in the log.
All the program variables~$\pv{x}$ are assumed to \emph{initially} only store $\low$ values.

With these conventions we can characterize TINI as the triple:
\begin{equation}
  \forall \tau, \vec{v}\st
  \T
  { \sure{\lowOf = \obs} * \sure{\pv{x} = \vec{v}} * \own{\p{log}} }
    {\prog}
  { \own{\lowOf} * \own{\highIn} }
\tag{$\TINI$}
\label{triple:tini}
\end{equation}
Here $\sure{\varphi}$ states that the predicate over states $\varphi$
holds on every possible state.
The idea of the triple is that the precondition selects an arbitrary
set of initial states, asking that it only contains states
where $\lowOf$ is some arbitrary but deterministic $\obs$
and $\pv{x}$ is some $\vec{v}$:
this means all the input states agree on the low inputs.
The rest of the \p{log} (including $\highIn$)
is allowed to be arbitrarily indeterminate.
Note that since the low values are deterministic, they are trivially independent
of \p{log}, hence the use of separating conjunction.
The postcondition then requires the \pre\low-tagged contents of the log
(which are added every time $\prog$ makes a low output)
to be independent of the high input.

As an example, the program
\p{ex3} from \cref{fig:running-ex}
would be instrumented by initializing \p{h} to $\highIn$ and
emitting \p{l} to the low log:
\code{h:=$\highIn$;ex3;log!(low,l)}.
The resulting program can be easily seen to satisfy \eqref{triple:tini}
by combining the effect of the instrumentation and the triple
$
  \T
  { \In{\p{h}}{H} }
    {\p{ex3}}
  { \own{\p{l}} * \own{\p{h}} }.
$
As we prove in \cref{sec:modelling-security}, the semantics of the \eqref{triple:tini}
triple exactly characterizes TINI as usually expressed in terms of comparing
pairs of executions~\cite{DBLP:journals/jsac/SabelfeldM03}.

\subsubsection{Release via Conditioning}

On its own, independence is a very blunt instrument:
only a few programs can achieve TINI.
For example we might be in a situation where it would be fine to declassify the
last two bits of \p{h}, making the \p{ex2} program secure.
The good news is that \ourlogic\ is perfectly capable of representing such
declassification scenarios.
First, we model declassification events by appending to the log
the declassified value with a $\decl$ tag; applied to our \p{ex2} example:
\[
  \p{declex} \is \bigl(\code{h:=$\highIn$;v:=0;log!(decl, $\highIn$ mod 4);log!(low,(h+v) mod 2)}\bigr)
\]
Second, we still require, in the postcondition, independence of
$\lowOf$ and $\highIn$, but only when looking at a subset of states
that agree on what value has been disclosed. In line with
the literature, we call this \emph{conditional independence} of
$\lowOf$ and $\highIn$---conditional on the declassification events.
Intuitively, if we restrict our focus to the subset of output states
which saw the same value~$d$ having been declassified,
we are selecting the states
that cannot be distinguished up to information that was allowed to be
released.  By requiring $\lowOf$ and $\highIn$ to be independent
conditionally on~$d$, we allow the release of~$d$ to refine
the possible values $L_d$ of $\lowOf$ and~$H_d$ of $\highIn$, but the
set of possible values (given~$d$) is still $L_d\times H_d$, \ie no
further information was leaked on top of~$d$.

In \ourlogic, we formalize conditional independence by combining
separation with a modality~$\condmod$ that we call the \emph{\condition[ing] modality}.
Semantically, $\cond{x}{V}{P(x)}$ holds on a set of states~$S$
if~$S$ can be partitioned into sets~$\set{S_{v}}_{v \in V}$ such that
for every $v\in V$, $P(v)$ holds on $S_{v}$.
We can then state independence of \p{y} and \p{z} conditional on \p{x}
by the assertion
$
  \Exists V. \cond{v}{V}{(\sure{\p{x}=v} * \own{\p{y}} * \own{\p{z}})}
$
which states:
for some appropriate set of values~$V$,
we can partition the current set of states into a set~$S_v$ for each $v\in V$
so that~\p{x} is~$v$ in every state of~$S_v$
(\ie $S_v$ selects the states which cannot be distinguished by the value of \p{x});
then in each~$S_v$, \p{y} and \p{z} are independent.
The pattern
$ \Exists V. \cond{v}{V}{\sure{\expr=v} * P(v)} $,
where~$\expr$ is an expression,
is so common that we abbreviate it with
$\vcond[v]{\expr}{P(v)}$
and with
$\vcond{\expr}{P}$ when~$P$ does not depend on~$v$.
In our example,
$\vcond{\p{x}}{(\own{\p{y}} * \own{\p{z}})}$.

Equipped with this new connective, we can then express the declassification
policy of our example program
as a triple:
\begin{equation}
  \T
  { \sure{\lowOf = \obsL} * \sure{\pv{x} = \vec{v}} * \own{\p{log}} }
    {\p{declex}}
  { \vcond{\declOf}{\bigl(\own{\lowOf} * \own{\highIn}\bigr)} }
  \label{triple:declex-secure}
\end{equation}
where $\declOf$ is an expression that extracts the \pre\decl-tagged values
from the log.
The triple is valid for this program:
since fixing the declassified last two bits of \p{h}
would restrict the possible value of the last bit of \p{h}
(released in the $\lowOf$ log)
to a singleton, the conditional independence between $\lowOf$ and $\highIn$ holds
trivially.

\subsubsection{Beyond `What' Declassification}
\label{sec:key-ideas:spec-beyond}

Conditional independence in \ourlogic\ is a very versatile tool
to specify security policies, and allows one to flexibly add or tweak
aspects of the policies, enabling the exploration of the policy design space
in a uniform framework.
To give an example, we can tweak the ``what'' declassification
policy expressed in \eqref{triple:declex-secure} so that it expresses
the ``when'' declassification constraint that high values should stay
secret until declassified, thus ruling out the version of \p{declex}
with the last two instructions swapped
(which would still satisfy \eqref{triple:declex-secure}).
To express this we simply require that the
policy should hold at each \emph{prefix} of the log:
\begin{equation}
  \All n \geq n_0.
  \T
  { \sure{\lowOf = \obsL} * \sure{\pv{x} = \vec{v}} * \sure{\obsfun{len}=n_0} * \own{\p{log}} }
    {\prog}
  {
      \vcond{\restr{\declOf}{n}}{\bigl(
        \own{\restr{\lowOf}{n}} * \own{\highIn}
      \bigr)}
  }
  \label{ideas:triple:prefix-close}
\end{equation}
The precondition is tweaked so the initial log (containing just low/high inputs)
has a fixed length $n_0$;
the postcondition uses $\restr{\obsfun{expr}}{n}$ which represents the expression $\obsfun{expr}$ evaluated on the log truncated at the \pre n-th position;
we only consider $n \geq n_0$ because those are the points during program execution.
While \p{declex} still satisfies the specification,
the version with the low output and the declassification swapped would not:
at $n_0+1$ we would see the low output but no declassification,
making $\restr{\declOf}{n_0+1}$ trivial
and $\restr{\lowOf}{n_0+1}$ dependent on $\highIn$.

In \cref{sec:modelling-security}
we show that the approach scales to more complex notions:
TINI with declassification over an arbitrary lattice,
``when'' declassification,
timing side-channels, and
Paralocks~\cite{DBLP:conf/popl/BrobergS10},
which expresses a particularly dynamic notion of declassification.
In \cref{sec:verification} we also exploit the support for nondeterminism
to model dynamic creation of secrets and high/low input.

\subsection{Verifying Security Notions}
\label{sec:key-ideas:verif}

A key strength of \ourlogic\ is the support for reasoning about
security policies in a \emph{program logic} that natively manipulates
the conditional independence assertions we used in the specifications.
In particular, the key connectives $E \pmapsto V$, $P_1 * P_2$, and
$\cond{v}{V}{P(v)}$ enjoy a rich set of natural rules with which it is
possible to carry out proofs without breaking the nice abstraction
offered by the assertion language.
In this section we illustrate the main reasoning patterns that enable
verification of programs against security specifications.

\begin{figure}
  \adjustfigure[\small]
  \begin{proofrules}
    \infer*[lab={assign}]{}{
	\T
	{\sure{\varphi\upd{\p{x}->\aexpr}}}
	{\code{x:=$\aexpr$}}
	{\sure{\varphi}}
}
     \label{rule:assign}

    \infer*[lab=rnd]{}{
      \T{\True}{\code{v:=rnd($V$)}}{\p{v} \pmapsto V}
    }
    \label{rule:rnd}

    \infer*[lab=frame]{
      \T{P}{\prog}{Q}
    }{
      \T{P*R}{\prog}{Q*R}
    }
    \label{rule:frame}

    \infer*[lab=if-then]{
      \T
      {P*\sure{\bexpr\ne 0}}
      {\prog_1}
      {Q}
    }{
      \T
      {P*\sure{\bexpr \ne 0}}
      {\code{if}\ \bexpr\ \code{then}\ \prog_1\ \code{else}\ \prog_2}
      {Q}
    }
    \label{rule:if-then}

    \infer*[lab=if-else]{
      \T
      {P*\sure{\bexpr = 0}}
      {\prog_2}
      {Q}
    }{
      \T
      {P*\sure{\bexpr=0}}
      {\code{if}\ \bexpr\ \code{then}\ \prog_1\ \code{else}\ \prog_2}
      {Q}
    }
    \label{rule:if-else}

    \infer*[lab=$\oplus$-triple]{
      \forall v\in V\st
      \T
      {\sure{\expr=v}*P(v)}
      {\prog}
      {\sure{\expr=v}*Q(v)}
      \\
      \mayterm{\prog}
    }{
      \T
      {\cond{v}{V}{\bigl(\sure{\expr=v}*P(v)\bigr)}}
      {\prog}
      {\cond{v}{V}{\bigl(\sure{\expr=v}*Q(v)\bigr)}}
    }
    \label{rule:oplus-triple}
  \end{proofrules}
  \caption{Selected rules of \ourlogic.}
  \label{fig:violet-rules}
\end{figure}

\subsubsection{Reasoning with Conditional Independence}

To illustrate the core patterns of reasoning in \ourlogic,
we sketch the proof of noninterference for \p{ex3}
from the previous section:
\begin{equation}
  \T
  { \In{\p{h}}{H} }
    {\p{ex3}}
  { \own{\p{l}} * \own{\p{h}} }
  \label{ideas:walkthrough:triple}
\end{equation}

\noindent
The idea of the proof is to break down the program into the two instructions,
and prove:
\begin{enumerate}[(a)]
  \item the assignment to \p{v} makes \p{v} and \p{h} independent;
        \label[step]{step:ideas-proof:v}
  \item the independence of \p{v} and \p{h} can be transformed
        into independence of \code{(h+v) mod 2} and \p{h};
        \label[step]{step:ideas-proof:indep}
  \item the assignment of \code{(h+v) mod 2} to \p{l} makes \p{l} and \p{h} independent.
        \label[step]{step:ideas-proof:log}
\end{enumerate}
The proof of \cref{step:ideas-proof:v} employs the \ref{rule:rnd} rule
and the \ref{rule:frame} rule from Separation Logic,
shown in \cref{fig:violet-rules}.
The \ref{rule:rnd} rule simply introduces the fact that
the assignment leads to a set of states such that
in every state \p{v} takes some value in~$V$,
but also that for every value $v\in V$ there is a state where \p{v}
is assigned~$v$.
The \ref{rule:frame} rule allows adding independent resources~$R$
in the precondition and postcondition of a triple.
For the purposes of this overview, \ref{rule:frame} should be considered
only applicable if $\pvar(R)\inters\mods(\prog) = \emptyset$;
thanks to permissions, a technical device we explain in \cref{sec:model} but omit here for simplicity,
\ourlogic's \ref{rule:frame} does not in fact need a side condition.

By applying \ref{rule:frame} to \ref{rule:rnd} with frame
$ R = (\In{\p{h}}{H}) $,
we obtain a triple representing our \cref{step:ideas-proof:v}:
$
    \T
    { \In{\p{h}}{H} }
      {\code{v:=rnd(\{0,1\})}}
    { \In{\p{h}}{H} * \In{\p{v}}{\set{0,1}} }.
$
Notice how the independence between \p{h} and \p{v} arises purely by framing.

\begin{figure}
  \adjustfigure[\small]
  \begin{proofrules}
    \infer*[lab=$\oplus$-unit-r]{}{
  \In{\expr}{X}
  \lequiv
  \cond{\val}{X} \sure{\expr=\val}
}
     \label{rule:unit-r}

    \infer*[lab=$\oplus$-unit-l]{}{
    P(\val_0)
    \lequiv
    \cond{\val}{\set{ \val_0 }} P(\val)
}
     \label{rule:unit-l}

    \infer*[lab={$\oplus$-frame}]{}{
  F * \cond{\val}{X} P(\val)
  \entails
  \cond{\val}{X} (F * P(\val))
}
     \label{rule:oplus-frame}

    \infer*[lab={$\oplus$-cons}]{
    \All v \in X. P(\val) \entails Q(\val)
}{
    \cond{\val}{X}{P(\val)}
    \entails
    \cond{\val}{X}{Q(\val)}
}
     \label{rule:oplus-cons}

    \infer*[lab={$\oplus$-unframe}]{
    F : \precise \\
    \partitions[\expr]{\val}{X}{P}
}{
    \cond{v}{X} (F * P(v))
    \entails
    F * \cond{v}{X} P(v)
}
     \label{rule:oplus-unframe}

    \infer*[lab={$\oplus$-transf}]{
    f \from Y \to X \text{ bijective}
}{
    \cond{v}{X}{ P(v) }
    \entails
    \cond{w}{Y}{ P(f(w)) }
}
     \label{rule:oplus-transf}
  \end{proofrules}
  \caption{Selected entailment rules of \ourlogic.}
  \label{fig:violet-ent-rules}
\end{figure}

The crucial step of the proof is \ref{step:ideas-proof:indep},
which in \ourlogic\ can be formalized as an entailment:
\begin{equation*}
  \In{\p{h}}{H} * \In{\p{v}}{\set{0,1}}
  \entails
  \In{\p{h}}{H} * \In{(\code{(h+v) mod 2})}{\set{0,1}}
\end{equation*}
Remarkably, the proof of this entailment can be carried out entirely
within \thelogic's proof system thanks to a handful of fundamental rules
of the connectives.
The rough proof strategy is as follows:
\begin{enumerate}[i)]
  \item we want to consider two cases:
  \begin{itemize}
    \item $\code{h mod 2} = 0$:
      then $\code{(h+v) mod 2} = \p{v}$
    \item $\code{h mod 2} = 1$:
      then $\code{(h+v) mod 2} = 1-\p{v}$
  \end{itemize}
  \item in each case we have $\In{(\code{(h+v) mod 2})}{\set{0,1}}$;
  \item since that fact does not depend on \code{h mod 2},
  we want to conclude it is independent of \p{h}.
\end{enumerate}
These manipulations require some entailment rules
shown in \cref{fig:violet-ent-rules}.
\Cref{rule:unit-r} says that stating $\expr$ can take the values in~$X$
is equivalent to stating that the set of states can be partitioned into
one subset $S_v$ for every $v\in X$ where~$E$ evaluates to~$v$ in each state in $S_v$.
The \ref{rule:oplus-frame} rule says that if $F$ is independent of a partitioning, then it is also independent of all the subsets in the partition
individually.
Combined, these rules allow us to reason by cases on the value of an expression,
for example \p{h}:
\begin{eqexplain}
  \In{\p{h}}{H}
  *
  \In{\p{v}}{\set{0,1}}
\whichentails*
  \bigl(\cond{h}{\set{0,1}}{\sure{\p{h}=h}}\bigr)
  *
  \In{\p{v}}{\set{0,1}}
\byrule{unit-r}
\whichentails
  \cond{h}{H}{
  \bigl(
    \sure{\p{h}=h}
    *
    \In{\p{v}}{\set{0,1}}
  \bigr)}
\byrule{oplus-frame}
\whichentails
  \cond{h}{H}{
  \bigl(
    \sure{\p{h}=h}
    *
    \cond{v}{\set{0,1}}{\sure{\p{v}=v}}
  \bigr)}
\byrule{unit-r}
\end{eqexplain}
In the last step, \ref{rule:oplus-cons} allows us to apply the
\ref{rule:unit-r} rule inside the outer \condition[ing] modality.
Since we now have a handle on each value~$h$ that~\p{h} can take,
we can perform the first step of our proof outline:
we can rewrite $\sure{\p{v} = v}$ to
$\sure{\p{v} = v \land \code{h+v} \equiv_2 h+v}$
which holds in the context of $\sure{\p{h}=h}$.
We can then split the two cases on $h$:
\[
  \cond{h}{H}{\left(
    \sure{\p{h}=h}
    *
    \begin{cases}
      \cond{v}{\set{0,1}}{\sure{\code{h+v} \equiv_2 v}}
        \CASE h \equiv_2 0
      \\
      \cond{v}{\set{0,1}}{\sure{\code{h+v} \equiv_2 1-v}}
        \CASE h \equiv_2 1
    \end{cases}
  \right)}
\]
The next step is showing that the second case is equivalent to the first one,
up to a change of variable, \ie expressing the partition in terms of~$1-v$,
which also ranges over $\set{0,1}$.
Such a change can be made by the \ref{rule:oplus-transf} rule,
also in \cref{fig:violet-ent-rules}.
We can apply it to the second case using $f(v) = 1-v$, obtaining
\[
  \cond{h}{H}{\bigl(
    \sure{\p{h}=h}
    *
    \cond{v}{\set{0,1}}{\sure{\code{h+v} \equiv_2 v}}
  \bigr)}
\]
The application of \ref{rule:unit-r} from right to left
gives us
$
  \cond{h}{H}{\bigl(
    \sure{\p{h}=h}
    *
    \In{\code{(h+v) mod 2}}{\set{0,1}}
  \bigr)}
$.

The final step is the most crucial:
we need to bring the assertion on \code{(h+v) mod 2} outside the \condition[ing]
on~$\p{h}$, essentially applying \ref{rule:oplus-frame} in reverse.
The reverse direction of the rule is however unsound in general!
This is the main safeguard for validity of inferences that the \condition[ing]
modality provides:
Facts that hold for each case of a \condition\ do not necessarily translate
to facts that hold for the whole set of states.
For example, conditional independence $\vcond{\p{z}}{(\own{\p{x}}*\own{\p{y}})}$
does not entail independence $\own{\p{x}}*\own{\p{y}}$: although the assertion inside
the modality is uniform across the partition via \p{z},
the $\own{\hole}$ notation hides an existential quantification which can be
resolved differently for each case.
Luckily, there is a general side condition we can add to the rule to regain
soundness: an assertion is \emph{precise} when there is a
\emph{unique minimal model} which satisfies it.
Precision is a condition which arises in standard separation logic too,
and holds for $\In{\expr}{X}$ and $\sure{\expr}$ facts,
and is closed under separating conjunction and \condition[ing].
The sound rule we need is \ref{rule:oplus-unframe}
in \cref{fig:violet-ent-rules}.
Since $\In{\expr}{X}$ facts are precise,
we can apply the rule to our example and get
$
  \bigl(\cond{h}{H}{\sure{\p{h}=h}}\bigr)
  *
  \In{\code{(h+v) mod 2}}{\set{0,1}}
$
which, by \ref{rule:unit-r}, is equivalent to
$
  \In{\p{h}}{H}
  *
  \In{\code{(h+v) mod 2}}{\set{0,1}}.
$
Clearly, it is possible to summarize these common patterns of reasoning
into high-level derived rules.
But the derivation shows that the fundamental reasoning principles
can be expressed as structural rules that explain the interplay of the
connectives, and that these principles allow for the derivation without
breaking the abstraction.

We have thus proved that $
  \T
  {\In{\p{h}}{H}}
  {\code{v := rnd(\{0,1\})}}
  {\own{\p{h}} * \own{\code{(h+v) mod 2}}}.
$
To complete the proof of our triple \eqref{ideas:walkthrough:triple}
it remains to prove the assignment to \p{l} correct.
The assignment axiom gives us
$ \T { \True }{\code{l:=(h+v) mod 2}}{ \sure{\p{l} = \code{(h+v) mod 2}} } $,
so by framing and sequencing we obtain
\begin{equation}
  \T
  {\In{\p{h}}{H}}
  {\code{v := rnd(\{0,1\}); l:=(h+v) mod 2}}
  {\own{\p{h}} * \own{\code{(h+v) mod 2}} * \sure{\p{l} = \code{(h+v) mod 2}} }.
\end{equation}
From the principles we outlined above, it is easy to derive the entailment
$ \sure{E_1=E_2} * \In{E_1}{V} \entails \In{E_2}{V} $,
which applied to our postcondition gives us the desired
$ \own{\p{h}} * \own{\p{l}}. $

\subsubsection{Verifying Indirect Flows}
\label{sec:key-ideas:verify:indirect-flow}

The versatility of expressing conditional independence as a combination
of \condition[ing] and separation shines when reasoning about indirect flows.
Consider a classic program that satisfies \TINI\ but would be ruled out
by typical syntactic disciplines to enforce noninterference:
\[
  \p{indfl} \is \bigl(\code{if h then log!(low,0) else log!(low,0)}\bigr)
\]
The security of the program hinges on the two branches writing the same
value, something that static typing would not track.
To prove that the $\low$ log stays independent from \p{h}
we can prove the triple:\footnote{Technically, the precondition also needs a token
  giving permission to mutate the log,
  but we omit this detail here for simplicity.
  The full details are explained in \cref{ex:h-assigned-l}.
}\[
  \T
  {\own{\lowOf} * \own{\p{h}}}
  {\p{indfl}}
  {\own{\lowOf} * \own{\p{h}}}
\]
Why is the example intuitively correct?
Because although it branches on a secret,
the effect on the $\low$ log in the two branches is the same.
To see how the reasoning in \ourlogic\ is exactly mirroring this intuition,
we need to introduce
the if-then-else rules in \cref{fig:violet-rules}.
At first sight,
\labelcref{rule:if-then,rule:if-else} seem hopelessly restrictive:
While the rules make perfect sense for a determined guard
(\ie one that has a constant truth value across all
states in the input state set),
they just do not apply to a precondition that does not fix whether~$\bexpr=0$,
like $\In{\bexpr}{\set{0,1}}$.
Luckily, \ourlogic\ validates an important rule, \ref{rule:oplus-triple},
that allows us to lift a triple which assumes some determined expression~$\expr$,
to an appropriate triple where~$\expr$ is undetermined.
The triple in the premise of \ref{rule:oplus-triple}
analyzes the program~$\prog$ by case analysis on
each possible value of $\expr$, and, from this, the rule
concludes that the program's behaviour on the union of all input cases
is the union of all output cases.

In our example the precondition is $\own{\lowOf}*\own{\p{h}}$
which implies
$ \In{\lowOf}{L} * \In{\p{h}}{H} $ for some sets~$L,H$.
Then, via \labelcref{rule:unit-r,rule:oplus-frame},
we can condition on both expressions and obtain
$
  \cond{h}{H}{\cond{\obsL}{L}{
    \bigl(\sure{\p{h}=h} * \sure{\lowOf=\obsL}\bigr)
  }}
$ as precondition.
Thus we can build the following derivation:
\begin{derivation}
  \infer*{
  \infer*{
    \T {
        \sure{\p{h}=h} * \sure{\lowOf=\obsL}
    }{\code{log!(low,0)}}{
\sure{\p{h}=h} * \sure{\lowOf=\obsL\app 0}
    }
  }{\forall h\in H,\obsL\in L\st
    \T {
        \sure{\p{h}=h} * \sure{\lowOf=\obsL}
}{\code{if h then}\ \dots}{
\sure{\p{h}=h} * \sure{\lowOf=\obsL\app 0}
    }
  }}{\T {
      \cond{h}{H}{\cond{\obsL}{L}{
        \bigl(\sure{\p{h}=h} * \sure{\lowOf=\obsL}\bigr)
      }}
    }{\p{indfl}}{
      \cond{h}{H}{\cond{\obsL}{L}{
        \bigl(\sure{\p{h}=h} * \sure{\lowOf=\obsL\app 0}\bigr)
      }}
    }
  }
\end{derivation}
The top triple captures the effect of the two (identical) branches
on determined (but arbitrary) \p{h} and $\lowOf$.
The if-then-else rules can be applied (to both cases) to obtain
the effect of the if-then-else on determined state.
Finally, the \ref{rule:oplus-triple} rule infers
the effect on a set of states by bundling together the determined cases
with a \condition[ing] modality.
The \ref{rule:oplus-triple} rule requires the program to have \emph{some}
terminating trace from any input ($\mayterm{c}$).
We discuss the side condition and ways to weaken it or remove it in
\cref{sec:logic} (see \cref{ex:weak-swap-ti}).

These steps are routine ``symbolic execution'' steps combined with
a way to lift them to assertions on sets of states.
What remains is to prove that the overall postcondition indeed implies
independence of \p{h} and $\lowOf$,
the crucial step of the proof.
The main driver of this step is \ref{rule:oplus-unframe}:
\begin{eqexplain}
&
  \cond{h}{H}{\cond{\obsL}{L}{
    \bigl(\sure{\p{h}=h} * \sure{\lowOf=\obsL\app 0}\bigr)
  }}
\whichentails
  \cond{h}{H}{\bigl(
    \sure{\p{h}=h} *
    \cond{\obsL}{L}{
      \sure{\lowOf=\obsL\app 0}
    }
  \bigr)}
  \explain{\ref{rule:oplus-unframe}}
\whichentails
  \cond{h}{H}{\bigl(
    \sure{\p{h}=h} *
    \cond{\obsL'}{L'}{
      \sure{\lowOf=\obsL'}
    }
  \bigr)}
  \explain{\ref{rule:oplus-transf} with $L'=L\app 0$}
\whichentails
  \cond{h}{H}{\bigl(
    \sure{\p{h}=h} *
    \In{\lowOf}{L'}
  \bigr)}
  \explain{\ref{rule:unit-r}}
\whichentails
  \In{\lowOf}{L'}
   *
  \bigl(
  \cond{h}{H}{
    \sure{\p{h}=h}
  }\bigr)
  \explain{\ref{rule:oplus-unframe}}
\whichentails
  \In{\lowOf}{L'} *
  \In{\p{h}}{H}
  \explain{\ref{rule:unit-r}}
\whichentails
  \own{\lowOf} *
  \own{\p{h}}
\end{eqexplain}
The \ref{rule:oplus-unframe} rule
only applies if the unframed assertion is precise and does not
refer to the variable bound by the modality it is escaping.
In both applications above, these conditions are met.
The \ref{rule:oplus-unframe} rule is the main tool to derive
independence facts from conditionally stated assertions.
The rule's checks are the crucial safeguards for soundness of the derivation.
For example, if we repeated the derivation for the leaky
program \code{if h then log!(low,1) else log!(low,0)}
we would end up with the assertion
$
  \cond{h}{H}{\bigl(
    \sure{\p{h}=h} *
\In{\lowOf}{L'(h)}
  }\bigr)
$.
In this case the possible values $L'(h) = L \app (h\ne0)$ of~$\lowOf$
depend on~$h$, so the assertion about $\lowOf$ cannot be unframed as is.
To remove the reference to~$h$ one can weaken the information on $\lowOf$
to
$
  \cond{h}{H}{\bigl(
    \sure{\p{h}=h} *
    \exists L'.\In{\lowOf}{L'}
  }\bigr)
$
but $\exists L'.\In{\lowOf}{L'}$ is not precise and thus cannot be unframed.

This line of argument is one of the main proof patterns to derive unconditional
independence from conditional facts in \ourlogic,
and it rests entirely on the fundamental rules that express the interplay
between the connectives of the logic.

\subsection{Abstractions and Semantic Typing}
\label{sec:key-ideas:abstractions}

Doing proofs in \ourlogic\ benefits from the abstractions provided
by independence and \condition[ing], but it can feel manual and low-level
compared to methods based on, say, type systems for enforcing information flow properties.
Indeed, when focusing on a specific security notion or domain of application,
it is often possible to create higher-level abstractions that are easier
to manipulate.
These two approaches---\ourlogic's general assertions and bespoke abstractions---are not in conflict. Rather, they strengthen
and complement each other:
\ourlogic's triples can be used to define the semantics of type judgments
in such a way that the rules of the type system are valid lemmas about
the corresponding triples.
This approach, called \emph{semantic typing}, has a distinctive
advantage:
  it allows verifying, within a \emph{program logic},
  programs that cannot be typed in
  the type system rules but actually comply with the semantics of their
  type signature.
Moreover, such manually verified components can be seamlessly integrated
in a typing derivation,
allowing all the parts of a larger program that can be (automatically) typed using the rules to be handled that way, and to only resort to manual
verification in \ourlogic\ for truly tricky components.

As a demonstrator,
we encode a nondeterministic variant
of the classic Volpano-Irvine-Smith type system in \ourlogic\ by:
\begin{enumerate*}
  \item compiling the security-label-annotated programs
        into \ourlogic\ programs where labelling occurs by logging,
  \item encoding the typing judgment as a triple, and
  \item proving the rules as lemmas in the \ourlogic\ program logic.
\end{enumerate*}
Furthermore, we demonstrate semantic typing by showing,
in \cref{sec:abstractions}, that classic
programs that are secure but ill-typed can be proven to satisfy
the semantic typing judgment in \ourlogic.
 \section{A Possibilistic Model of Conditional Independence}
\label{sec:model}

\paragraph{Programming Language}
\label{sec:progs}

We consider a generic imperative programming language.
Since our results do not depend on the specific choices of primitives,
we assume countable sets of
  \emph{(program) variables} $ \p{x} \in \Var $,
  \emph{values} $ \val \in \Val \sups \Int $,
  and \emph{atomic expressions} $ \aexpr,\bexpr \in \AExpr \sups \Val $.
For brevity, we represent false with the value~$0$, and any other value
is treated as true.
We use $\code{rnd}$ to inject explicit nondeterminism.
Commands $\prog \in \Prog$ are formed according to the grammar:
\begin{grammar}
    \prog \is \code{skip}
           | \code{x:=$\aexpr$}
           | \code{x:=rnd($V$)}
           | \code{log!($\aexpr$)} \\
           &| \prog_1\p;\prog_2
           | \code{if $\;\bexpr\;$ then $\;\prog_1\;$ else $\;\prog_2$}
           | \code{repeat $\;\aexpr$: $\;\prog_1$}
           | \code{while $\;\bexpr\;$ do $\;\prog_1$}
\end{grammar}
The language is entirely standard with the exception of the
\code{log!($\aexpr$)} construct, which we use as a means to record an
\emph{``event''} (\ie a value) in a log. \code{repeat $\;\aexpr$: $\;\prog_1$} evaluates $\aexpr$ to a number $n$ and then executes $\prog_1$ $n$ times in sequence ($0$ if negative).
To give semantics to the language, we define
\emph{variable stores} $ \store \in \Store \is \Var \to \Val $ and we let \emph{logs} be sequences of values $\obs \in \Obs$.
The state of a program is a pair of
a store and a log:
$\state \in \State \is \Store \times \Obs $.

We assume expressions come equipped with a semantic function
$ \Sem[E]{\aexpr} \from \State \to \Val $.
For non-total functions (\eg division, or reading from the log) we assume the existence of a value $\bot \in \Val$.
The command \code{x:=rnd($V$)} nondeterministically assigns to \p{x} any value in the set~$V$.

The small-step transition relation
$ \smallstep{\prog_1}{\state_1}{\prog_2}{\state_2} $
is standard.
For example:
\[
  \infer{}{
    \smallstep
      {\code{log!($\aexpr$)}}{\store,\obs}
      {\pskip}{\store, \obs \app {\Sem[E]{\aexpr}(\store,\obs)}}
  }
  \qquad
  \infer{}{
    \smallstep
    {\code{x:=$\aexpr$}}{\store,\obs}
    {\pskip}{\store\upd{\p{x} -> {\Sem[E]{\aexpr}(\store,\obs)}}, \obs}
  }
\]
We let $\smallstep*{\prog_1}{\state_1}{\prog_2}{\state_2}$
be the reflexive-transitive closure of the small-step transition relation.

From the small-step semantics we can derive the collecting semantics
$\sem{\prog} \from \State \to \powset(\State)$
by letting
$
  \sem{\prog}(\state) \is
    \set{ \state' \in \State | \smallstep*{\prog}{\state}{\pskip}{\state'} }
$,
which captures all the terminating behaviour of the program.
We extend the semantics by union to $\sem{\prog} \from \powset(\State) \to \powset(\State)$: $\sem{\prog}(S) \is \Union_{\state\in S} \sem{\prog}(\state)$.

Note that our semantics treats the log as an append-only buffer:
\begin{equation*}
    \All (\store_0,\obs_0),(\store_1,\obs_1) \in \State.
      \smallstep*{\prog_1}{\store_0,\obs_0}{\prog_2}{\store_1,\obs_1}
        \implies \Exists \obs \in \Obs. \obs_1  = \obs_0 \app \obs
  \tag{ObsApp}
\end{equation*}

The programming language given above is meant to be used as an intermediate
modelling language: it does not have any intrinsic notion of flows,
nor of security levels.
Instead, it provides the \code{log!} command which declares which events of the execution of the program are
observable.
Then, security notions can be defined on top of the observable log of events.

For example, to encode the standard security labels $\low$ and $\high$
(for low/high confidentiality respectively),
one would include in the values pairs $ (\lvl, \val) $
where $\lvl \in \set{\low,\high}$;
then \code{log!(low, $\val$)} and \code{log!(high, $\val$)} can be used
to declare (at runtime) a value~$\val$ as having been observed
at $\low$ or $\high$ confidentiality respectively.
From a state $(\store, \obs) \in \State$,
one can reconstruct what was leaked to the low confidentiality channel
as $\lowOf(\obs)$, where $\lowOf$ is the function that returns
the substring of $\obs$ consisting of values of the form $(\low, \val)$.
By abuse of notation we let $\lowOf(\store,\obs) = \lowOf(\obs)$.
We define $\highOf(\hole)$ extracting $\high$ observations analogously.
As notation, we write functions of the log in $\obsfun{purple}$: for instance, $\obsfun{\length}$ is the function that computes the length of the log.

Notice that, by default, every command is seen as unobservable,
thus at the modelling/compilation stage the program needs
to be carefully annotated with \code{log!} commands
so that the desired leakage assumptions are modelled correctly.

\paragraph{Possibility Spaces}
\label{sec:possibility_spaces}

We now define \emph{possibility spaces}, a structure that can represent
both sets of states and the level of precision at which we can describe them.
Note that while the following definitions are given using~$\State$,
the set of program states, they would be well-defined on any set.

\begin{definition}[Possibility Space]
    A set $\psp \subs \powset(\State)$ is a \emph{possibility space}
    over~$\State$,
    if it is a non-empty set of non-empty pairwise disjoint subsets of $\State$,
    \ie if
    \begin{enumerate*}[(i)]
        \item $\psp \ne \emptyset$,
        \item $\All \evA \in \psp. \evA \ne \emptyset$,
        \item $\All \evA,\evB \in \psp. \evA \neq \evB \implies \evA \inters \evB = \emptyset$.
    \end{enumerate*}
The elements of $\psp$ are called \emph{events} and they represent the coarseness of observations
    which a model is afforded.
    We call the set $ \Union \psp $ the ``possible'' states of $\psp$,
    and $ \State \setminus \Union \psp$ the ``impossible'' states.
We write $\PSp$ for the domain of all possibility spaces over $\State$.
\end{definition}

A possibility space $\psp$ is isomorphic to the complete Boolean Algebra generated via union by $\psp \union \set{ \set{\state} | \state \notin \Union \psp }$.
The states in $\Union \psp$ are the ones that $\psp$ declares are possible.
Possibility spaces can be seen as a representation of the Partial Equivalence Relation~(PER) model of information flow~\citep{DBLP:journals/lisp/SabelfeldS01}.
A PER $R \subs X \times X$ is a relation which is an equivalence over
a subset of $X$ called its domain $\card{R}$ (the ``possible'' states).
The relation describes what states are indistinguishable from the attacker's point of view.
A possibility space represents this information by recording the equivalence
classes of the domain of a PER.
What our work crucially adds to this structure is the notion of independent product, a knowledge-preserving refinement order, and the interpretation of partitioning as disjoint union, all defined below.

\begin{definition}[Independent Product]
    Given $\pspA,\pspB \in \PSp$,
    we say $\pspA$ and $\pspB$ are \emph{compatible},
    written $\pspA \compat \pspB$,
    if $ \All A\in\pspA,B\in\pspB. A \inters B \ne \emptyset$.
    The \emph{independent product}
    $ {\iprod} \from \PSp \times \PSp \to \PSpInvalid$, where $\PSpInvalid \is \PSp \dunion \set{ \invalid }$ contains a placeholder value $\invalid$ for when the independent product is undefined, is defined as follows:
    \[
    \pspA \iprod \pspB \is
    \begin{cases}
        \set{ A \inters B | A\in\pspA, B\in\pspB  }
        \CASE \pspA \compat \pspB
        \\
        \invalid \OTHERWISE
    \end{cases}
    \]
\end{definition}
\begin{example}
    Consider states with two variables \p{h} and \p{v}, inspired by the examples in \cref{sec:key-ideas:spec}. We will use \p{(h+v)\%2} instead of \p{l}. Let $B = \set{ 0, 1 }$ for this example, and
    \[
    \pspA \is \set{ \set{ \state | \state(\p{h}) = h } | h \in H }
    \quad
    \pspB \is \set{ \set{ \state | \state(\p{(h+v)\%2}) = l } | l \in B }
    \quad
    \pspC \is \set{ \set{ \state | \state(\p{v}) = v } | v \in B }
    \]
    In $\pspA$ it is possible to know the value of \p{h}, so it is a model of $\In{\p{h}}{H}$. Similarly $\pspB$ is a model of $\In{\p{(h+v)\%2}}{B}$, and $\pspC$ of $\In{\p{v}}{B}$.
    Any two of the three possibility spaces above are compatible: fixing the value of any one of the three expressions does not constrain the value of either of the other two.
    For instance $\pspA \compat \pspB$: each pair of events from $\pspA$ and $\pspB$ has non-empty intersection:
    \begin{align*}
        \pspA \iprod \pspB
        &= \set[\big]{ \set{ \state | \state(\p{h}) = h, \state(\p{(h+v)\%2}) = l } | (h, l) \in H \times B }
    \end{align*}
    The resulting possibility space $\pspA \iprod \pspB$ is therefore a model of $\In{\p{h}}{H} * \In{\p{(h+v)\%2}}{B}$.
    However, if we consider all three $\pspA$, $\pspB$, $\pspC$ at the same time, they are not compatible: $\pspA \iprod \pspB$ fixes the value of \p{v} and therefore is not compatible with $\pspC$. For instance, for $h = 0$, $l = 0$, $v = 1$ there is no state in the intersection of the three events.
\end{example}

\begin{definition}[Expressible]
  Let $\pspA\in \PSp$.
  We say a set $\ev \subs \State$ is \emph{expressible in $\pspA$},
  written $ \ev \ein \pspA $,
  if it is a union of events of $\pspA$ up to impossible states.
  Formally:
  \[
    \ev \ein \pspA \is
    \Exists \pspB \subs \pspA.
      \ev \inters \Union \pspA = \Union \pspB
  \]
  Notice that $\pspB$ is not required to be a possibility space,
  and thus can be empty, implying that sets of impossible states
  are always expressible.
\end{definition}

\begin{definition}[Refinement order]
    Given $\pspA,\pspB \in \PSp$,
    we define the \emph{refinement order} $\pspA \pLeq \pspB$ to hold if and only if
    \begin{enumerate*}[(i)]
        \item $\Union \pspB \subs \Union \pspA$,
        \item $
          \All \evA \in \pspA. \evA \ein \pspB
$, and
        \item $
          \All \evA \in \pspA.
              \evA \inters \Union\pspB \ne \emptyset
          $.
    \end{enumerate*}
    We overload the refinement order so that for $\psp \in \PSp$ and $S\subs\State$,
$ \psp \pLeq S \is \psp \pLeq \set{ \set{\state} | \state \in S }. $
\end{definition}

\begin{example}
\label{ex:psp-ref}
    Below is a graphical representation of some examples of the refinement order.
    {\small
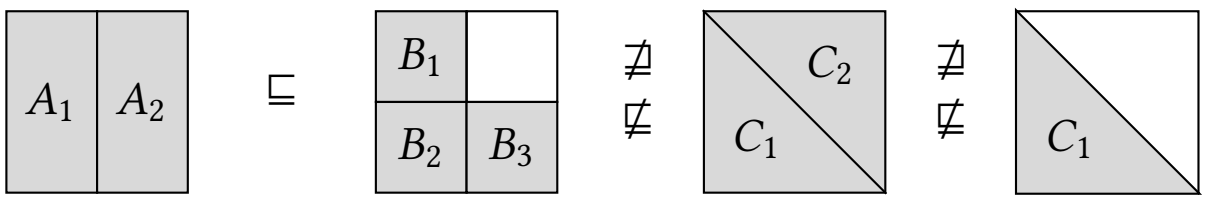
\[
\def\cellw{.6cm}
\tikz[baseline]{\matrix[matrix of nodes,
          nodes={draw, minimum width=\cellw, minimum height=2*\cellw, anchor=center},
          column sep=-\pgflinewidth,
          row sep=-\pgflinewidth] (m) {
    |[fill=gray!30]| $\evA_1$ & |[fill=gray!30]| $\evA_2$ \\
  };}
\quad\pLeq\quad
\tikz[baseline]{\matrix[matrix of nodes,
      nodes={draw, minimum width=\cellw, minimum height=\cellw, anchor=center},
      column sep=-\pgflinewidth,
      row sep=-\pgflinewidth] (m) {
    |[fill=gray!30]| $\evB_1$ & {}  \\
    |[fill=gray!30]| $\evB_2$     &  |[fill=gray!30]|   $\evB_3$ \\
  };}
\quad
  \begin{matrix}
    \not\pGeq
    \\
    \not\pLeq
  \end{matrix}
  \quad
\tikz[baseline]{
  \node[minimum size=2*\cellw,draw,fill=gray!30,draw=black] (m) {};
  \draw (m.north west) -- (m.south east);
  \path (m.north east) --
    node[pos=.7] {$\evC_1$}
    node[pos=.3] {$\evC_2$}
  (m.south west);
  }
\quad
  \begin{matrix}
  \not\pGeq
  \\
  \not\pLeq
  \end{matrix}
  \quad
\tikz[baseline]{
  \node[minimum size=2*\cellw,draw] (m) {};
  \draw[fill=gray!30] (m.north west) --
    (m.south east) -- (m.south west) -- cycle;
  \path (m.north east) --
    node[pos=.7] {$\evC_1$}
  (m.south west);
  }
\]
}     Here $\pspB = \set{ \evB_1, \evB_2, \evB_3 }$ is a refinement of $\pspA = \set{ \evA_1, \evA_2 }$: intuitively, every event in $\pspA$ has been partitioned, some of the parts have been declared impossible, and the others appear as events in $\pspB$.

    On the contrary, $\pspB$ is \emph{not} a refinement of $\pspC = \set{ \evC_1, \evC_2 }$: while every state that is possible in $\pspB$ is also possible in $\pspC$, the events in $\pspC$ cannot be partitioned in such a way that each part is either impossible or an event in $\pspB$.
For a different reason, $\pspC' = \set{ \evC_1 }$ is not a refinement of $\pspC$ either: every event in $\pspC$ is ``possible'', which means at least one of its states is possible. However, in $\pspC'$ this is not the case: all states in $\evC_2$ are impossible.
\end{example}

To support the \condition[ing] modality, we want to split a given possibility space into disjoint slices. \begin{definition}[Disjoint possibility spaces]
    Let $\set{ \pspA_v }_{v \in X}$ be a set of possibility spaces.
    We say that $\set{ \pspA_v }_{v \in X}$ are \emph{disjoint} if
    $
    \All v, w \in X . v \neq w \implies \Union \pspA_v \cap \Union \pspA_w = \emptyset .
    $
    If an expression $\expr$ is such that $\expr(\Union \pspA_v) = \set{ v }$, we say that $\expr$ is a \emph{witness} of the disjointness.
\end{definition}
\begin{proposition}\label{lemma:disjoint-union-pspace}
	Given a set of disjoint possibility spaces $\set{\pspA_v}_{v\in X}$, their (disjoint) union $\pspA \is \Dunion_{v\in X} \pspA_v$ is a possibility space.
\end{proposition}

Note that, just like $\pspC$ and $\pspC'$ in \cref{ex:psp-ref}, $\pspA_v \not\pLeq \Dunion_{v\in X} \pspA_v$: this is not a refinement, but a case analysis that requires checking all the cases.

To build a model of Bunched Implications~(BI)---the assertion language of separation logic---it suffices to show that possibility spaces form a Resource Algebra~\citep{DBLP:journals/pacmpl/KrebbersJ0TKTCD18}.

\begin{definition}[Ordered Unital Resource Algebra]
\label{def:ra}
  An \emph{ordered unital resource algebra}~(RA) is a tuple
  $
    (\raM, \raLeq, \raValid, \raOp, \raUnit)
  $
  where
  $ {\raLeq} \subs \raM \times \raM $ is a
    reflexive and transitive relation
    called \emph{resource order},
    $\raValid \from \raM \to \Prop$ is the \emph{validity predicate},
  $ (\raOp) \from \raM \times \raM \to \raM $ is a
    commutative and associative binary operator
    called \emph{resource composition},
  and
  $ \raUnit \in \raM $ is called the \emph{unit} of $\raM$,
  satisfying, for all $a,b,c \in \raM$:
  \begin{align*}
  \raValid(\raUnit)
  	&&
  \raUnit \raOp a = a
    &&
  \infer{ \raValid(a \raOp b) }{ \raValid(a) }
    &&
  \infer{ a \raLeq b \quad \raValid(b) }{ \raValid(a) }
    &&
  \infer{ a \raLeq b }{ a \raOp c \raLeq b \raOp c }
  \end{align*}
\end{definition}

Given any instance of an RA~$\raM$ one can build a separation logic
where assertions are the upward-closed sets of elements of~$\raM$,
and separation is interpreted via resource composition.
Possibility spaces form an RA.

\begin{proposition}\label{lemma:psp-ra}
    The structure $(\PSpInvalid, \pLeq, \raValid, \iprod, \set{\State})$ is an
    ordered unital resource algebra,
    where
    $\pLeq$ is the refinement order with $\invalid$ added as the top element;
    $\raValid = \fun a . (a \neq \invalid)$;
    and $\iprod$ is the independent product extended to return $\invalid$ whenever one of the arguments is $\invalid$.
\end{proposition}

While this structure would suffice for a (say, functional) language without
mutable state, to allow for a side-condition-free Frame rule it is necessary
to add to the resource algebra a notion of permission on variables
to specify in assertions whether a program is allowed to write to a variable.
To do this we adapt the solution based on permissions proposed by \Bluebell.

\begin{definition}Let $\VarLog \is \Var \dunion \set{ \logs }$.
	The permissions RA is defined as $(\Perm, \leq, \raValid, +, \funconst 0)$
	where
	$\Perm \is \VarLog \to \PosReal$;
	$\permap \leq \permaq \is \All \p{x} \in \VarLog . \permap(\p{x}) \le \permaq(\p{x})$ is the point-wise lifting of the usual $\le$ on real numbers;
	$\raValid(\permap) \is (\All \p{x} \in \VarLog . \permap(\p{x}) \le 1)$; and
	$\permap + \permaq \is \fun \p{x} . \permap(\p{x}) + \permaq(\p{x})$ is the point-wise sum.
\end{definition}

Intuitively, a permission of~$1$ on a regular variable is ``read-write'', a permission strictly between~0 and~1 is ``read-only'' and a permission of~$0$ means the assertion cannot read the variable.
For the log we use a different interpretation: since logs are append-only, even with permission~$0$ the assertion can talk about the past log, because it cannot be mutated by the program. However, only a permission strictly greater than~$0$ allows an assertion to fix any upper bound on the length of the log.
We formalize this intuition via compatibility of a possibility space and a permission map:

\begin{definition}[Compatibility]\label{def:permission-compatibility}
	Recall that $\prefixOf$ is the prefix order on sequences.

	Given a permission map $\permap \in \Perm$, define the relation:
	$
	(\store_1, \obs_1) \preccurlyeq_{\permap} (\store_2, \obs_2)
	$
	to hold iff
	\begin{itemize}
		\item $
		\forall \p{x}\in\Var\st
		\permap(\p{x}) > 0 \implies \store_1(\p{x}) = \store_2(\p{x})
		$, and
		\item $
		\permap(\logs) > 0 \implies \obs_1 = \obs_2
		$, and
		\item $
		\permap(\logs) = 0 \implies \obs_1 \prefixOf \obs_2
		$
	\end{itemize}

	Given a possibility space $\pspA \in \PSp$ and a permission map $\permap \in \Perm$, we say they are \textit{compatible}, written $\pspA \compat \permap$, if:
	$
		\forall \evA\in \pspA\st
		\forall \state \in \evA\st
		\forall \state' \in\State\st
		\state \preccurlyeq_{\permap} \state' \implies \state' \in \evA.
	$
\end{definition}

We can then define the RA that is the model of our assertions:
\begin{definition}
	Let $\PSpPm \is \set{ (\pspA, \permap) | \pspA \in \PSpInvalid, \permap \in \Perm, \pspA \compat \permap }$.
	We define the \textit{possibility spaces with permissions} resource algebra to be $(\PSpPm, \raLeq, \raValid, \raOp, \raUnit)$ where
	$\raLeq$ is defined point-wise from the resource orders on $\PSpInvalid$ and $\Perm$;
	$\raValid$ is defined combining validity of the two resources separately;
	$\raOp$ is defined point-wise; and
	$\raUnit$ is the pair of the two underlying units:
	\begin{align*}
		(\pspA, \permap) \raLeq (\pspB, \permaq) &\is \pspA \pLeq \pspB \land \permap \leq \permaq
		& \raValid(\pspA, \permap) &\is (\pspA \neq \invalid) \land (\permap \leq \funconst 1)
		\\
		(\pspA, \permap) \raOp (\pspB, \permaq) &\is (\pspA \iprod \pspB, \permap + \permaq)
		&\raUnit &\is (\set{ \State }, \funconst 0)
	\end{align*}
\end{definition}

\begin{proposition}\label{lemma:psppm-ra}
	$(\PSpPm, \raLeq, \raValid, \raOp, \raUnit)$ is an ordered unital resource algebra.
\end{proposition}

\begin{remark}
  \label{rm:bluebell-model}
  The model of \thelogic\ can be derived from the one of \Bluebell\ by
  ``forgetting'' the probability measure from resources and only keeping its support (here represented as $\Union \psp$).
  The definition of a new model is not just more self-contained; it is necessary:
  defining \thelogic\ directly on \Bluebell's model would not validate
  the crucial~\cref{rule:oplus-unframe} rule,
  as we discuss in \cref{rm:bluebell-unframe}.
  In addition, not having a measure allows \thelogic\ to support
  uncountable domains, allowing, for example, a nondeterministic choice
  among all permutations of the natural numbers (useful for the so-called shuffling model of security).
\end{remark} \section{A Logic for Possibilistic Conditional Independence}
\label{sec:logic}

We now define the logic of \thelogic\ assertions
following the so-called \emph{semantic approach}:
the logic manipulates the \emph{interpretation} of assertions
as the sets of resources that satisfy them.
Syntax is introduced as notation to express these sets.

\begin{definition}[Assertions]
  \label{def:assrt}
  We consider as an assertion any element $P,Q \in \Assrt \is \PSpPm \ucto \Prop$,
  that is, functions $P \from \PSpPm \to \Prop$ such that
  for any $a, b \in \PSpPm$ with $a \raLeq b$,
  if $P(a)$ holds, then $P(b)$ holds.
  We write $a \models P$ to mean $P(a)$ holds.
\end{definition}
\begin{definition}[Entailment]
  \label{def:entails}
  The \emph{entailment relation}~$\entails$ is defined as:
  \[
      P \entails Q \is \forall a \in \PSpPm \st
      \raValid(a) \implies
      (a \models P) \implies (a \models Q)
  \]
\end{definition}

Since $\PSpPm$ is an RA, all the standard rules of affine BI are valid for it;
see \cite{DBLP:journals/pacmpl/KrebbersJ0TKTCD18}.

In \cref{fig:assert-sem} we define notation for the main assertions we use,
\ie the ones we consider to be the primitives to build formulas.
The main atomic proposition is
$\In{\expr}{X}$: it does not only mean that $\expr$ evaluates to some $d \in X$ in all possible states of $\psp$, but also that for each $d \in X$ there is a possible state $\state \in \Union \psp$ such that $\expr(\state) = d$.
The \condition[ing] modality $\cond{v}{X}{P(v)}$ holds when the possibility space
can be partitioned into one space~$\psp_v$ for each value~$v$ in~$X$ such that
$P(v)$ holds in $\psp_v$; notice that the modality binds~$v$.
The $P\withp{\permaq}$ assertion states that $P$ holds and we hold permissions~$\permaq$.
The other assertions are standard BI assertions.
In $\Exists x \of X . P(x)$ we omit the type of $x$ when clear from context.
In the context of assertions, we implicitly treat atomic expressions $\aexpr$ as the function $\Sem[E]{\aexpr}$.

We remark that assertions about a prefix of the log can be supported by permission $0$: since the log is append-only, everything that happened in the past is immutable and does not need permissions. For instance, $\lowOf \obspointsto[t] \obsL$ needs permission $0$ on the log. This assertion fixes the sequence of $\low$ observations among the first $t$ entries of the log to be $\obsL$.
On the contrary, $\lowOf \obspointsto \obsL$ specifies that the \emph{entire} low log is equal to $\obsL$, and therefore does require non-null permission on it.

We call $\own{\expr}$ \emph{ownership} of $\expr$.
Assertions $\sure{\bexpr}$ are called \emph{sure assertions}, and they prescribe that $\bexpr$ holds on all possible states.
Because they effectively operate on single states, they never entail any ownership and therefore can be
duplicated: $
  \sure{\bexpr}
  \lequiv
  \sure{\bexpr} * \sure{\bexpr}
$.

\begin{figure}
  \adjustfigure[\small]
  \begin{alignat*}{3}
(\pspA, \permap) &\models \In{\expr}{X}
      &\quad\tIff\quad&
      \All d\in D.
        \inv{\expr}(d) \ein \pspA
        \text{ and }
          \expr(\Union\pspA) = X
      \qquad & (\expr \from \State \to D)
    \\
    (\pspA, \permap) &\models \cond{v}{X}{P(v)}
      &\tIff\quad&
\Exists \set{\pspA_v}_{v\in X}.
            \set{\pspA_v}_{v\in X} \text{ are disjoint,}
            \qquad &(P \from X \to \Assrt)
            \\
            &&& \qquad
            \pspA \pGeq \Dunion_{v\in X} \pspA_v
            \text{ and }
            \All v \in X. (\pspA_v, \permap) \models P(v)
    \\
    (\pspA, \permap) &\models P\withp{\permaq}
    &\quad\tIff\quad&
        (\pspA, \permap) \models P
        \text{ and }
        \permap \geq \permaq
    \\
    a &\models P_1 * P_2
      &\tIff\quad&
      \Exists b_1, b_2 \in \PSpPm.
        a \raGeq b_1 \raOp b_2
        \text{ and }
        b_1 \models P_1
        \text{ and }
        b_2 \models P_2
    \\
    a &\models \Exists x \of X . P(x)
        &\tIff\quad&
        \Exists x \of X . a \models P(x).
    \end{alignat*}

    \raggedright\medskip
    \textbf{Abbreviations}
    (for $ \bexpr \from \State \to \Val $,
    $\expr \from \State \to D$,
    $\obsfun{f} \from \Obs \to D$,
    $i, t \in \Nat$
    ):\medskip
    \begin{align*}
        \sure{\bexpr} &\is \In{(\fun\state.\bexpr(\state) \ne 0)}{\set{\true}}
        &
        \own{\expr} &\is \Exists S. \In{\expr}{S}
        &
        \obsfun{f} \obspointsto x &\is \sure{ {\obsfun{f}(\logs)} = x }
        \\
        \vcond[v]{\expr}{P(v)} &\is \Exists X . \cond{v}{X}{\sure{\expr = v} * P(v)}
        &
        \vcond{\expr}{P} &\is \vcond[v]{\expr}{P}
        &
        \obsfun{f} \obspointsto[t] x &\is \sure{ {\obsfun{f}(\restr{\logs}{t})} = x }
        \\
        \auth{t} &\is \sure{\obsfun{\length} = t}\withperm{\logs : 1}
        &
        &
        &
        i \obspointsto \val &\is \sure{\logs[i] = \val}
    \end{align*}
    \caption{Assertion semantics.
        The function $\restr{\hole}{t}$ cuts a sequence to its prefix of length~$t$.}
    \label{fig:assert-sem}
\end{figure}

\paragraph{Weakest Precondition}
To define a program logic
we define a weakest precondition \emph{modality}
and encode Hoare triples in terms of it.\footnote{Notice this is a \emph{modality}, not a predicate transformer.
}
Intuitively, the assertion $\WP{\prog}{Q}$ holds on the current set of possible states if, after executing $\prog$, we end up in a set of possible states on which $Q$ holds.
\begin{definition}[Weakest precondition]
    For program $\prog$ and resource $a \in \PSpPm$,
    \begin{multline*}
        a \models \WP{\prog}{Q}
        \quad \tIff \quad
\All S \subs \State, f . (a \raOp f \raLeq S \land \sem{\prog}(S) \neq \emptyset) \implies (\Exists b . b \raOp f \raLeq \sem{\prog}(S) \land b \models Q)
    \end{multline*}
    where $(\pspA, \permap) \raLeq S$ means that both $\raValid(\pspA, \permap)$ and $\pspA \pLeq S$.
\end{definition}

Since we work with a separation logic, we want updates to be frame-preserving \citep{DBLP:journals/jfp/JungKJBBD18}.
We bake this condition into the weakest precondition:
$f\in \PSpPm$ describes a frame that is unchanged by execution of~$\prog$.
The key idea is that~$S$ is the set of possible states,
and $a \raOp f$ doesn't need to describe it exactly.
The program~$\prog$ then executes from each possible state in~$S$,
and the results are collected by union in $\sem{\prog}(S)$.
This corresponds to execution in the nondeterminism monad
\citep{DBLP:journals/iandc/Moggi91}.
$a \raOp f \raLeq S$ means that $a \raOp f$ offers a \emph{view}
on the possible states $S$.
It is this view that imposes the level of granularity
on the possible observations
(the events in the possibility space of $a \raOp f$).
Therefore, the model $b\in \PSpPm$ of the postcondition need not be directly related to~$a$ through $\sem{\prog}$:
it can be any resource that satisfies the postcondition~$Q$
and that can be combined with the (unchanged) frame~$f$
to produce a view of the set of possible states after execution:
$b \raOp f \raLeq \sem{\prog}(S)$.
For~$a$ to be a model of $\wpsymb$, there must exist one such~$b$ for every set of possible states~$S$ that is compatible with~$a$ and the frame~$f$.
The condition $\sem{\prog}(S) \neq \emptyset$ makes the~$\wpsymb$ termination-insensitive: we ignore any collection of states that only exhibits non-terminating behaviour.

With $\wpsymb$, we define Hoare triples as notation:
\[
\T{P}{\prog}{Q} \;\is\; P \entails \WP{\prog}{Q} .
\]

\begin{figure}[t]
    \begin{proofrules}
        \infer*[lab={triple-log}]{
	\forall \p{y} \in \pvar(\aexpr) . \permap(\p{y}) > 0
    \\
    \permap(\log) = 0
    \\
    \log \notin \pvar(\aexpr)
}{
	\T{\auth{t}\withp{\permap}}
	  {[\code{log!($\aexpr$)}]}
    {\auth{(t+1)} * ((t+1) \obspointsto \aexpr)\withp{\permap}}
}
         \label{rule:triple-log}

        \infer*[lab={$\oplus$-own-merge}]{
    \expr = f(\expr')
}{
    \vcond{\expr}{\own{\expr'}}
    \lequiv
    \own{\expr'}
}
         \label{rule:oplus-own-merge}

        \infer*[lab={triple-while}]{
    \All n, i . P_i(n) \entails \sure{\bexpr = i} \\
    \All n . \T{P_{\true}(n)}{\prog}{P(n+1) } \\
    \All n . P(n) \entails \Exists X . \cond{i}{X}{P_i(n)} \\
    \partitions{n}{\Nat}{P_{\false}}
}{
    \T{P(0)}
    {[\code{while $\;\bexpr\;$ do $\;\prog$}]}
    {{ \Exists N . \cond{n}{N}{P_{\false}(n)} }}
}
         \label{rule:triple-while}

        \infer*[lab=$\oplus$-triple-weak]{
    \forall v\in V\st
    \T
    {\sure{\expr=v}*P(v)}
    {\prog}
    {\sure{\expr=v}*Q(v)}
}{
    \T
    {\cond{v}{V}{\bigl(\sure{\expr=v}*P(v)\bigr)}}
    {\prog}
    {\Exists V' . \pure{\emptyset \subsetneq V' \subseteq V} * \cond{v}{V'}{\bigl(\sure{\expr=v}*Q(v)\bigr)}}
}
         \label{rule:oplus-triple-weak}
    \end{proofrules}
    \caption{Other notable rules of \ourlogic{}.}
    \label{fig:violet-rules-sec-logic}
\end{figure}

The full list of rules that hold for our $\wpsymb$ is in \appendixref{app:rules}.
\Cref{fig:violet-rules,fig:violet-rules-sec-logic} show the relevant derived rules for triples.
The log-specific \ref{rule:triple-log} rule
says that appending to the log only needs the token $\auth{t}$,
that fixes the length of the log and ensures permission~$1$ on the log.

\paragraph{\Condition[ing] Modality}

For non-branching programs, the rules above are sufficient: we can use the relevant axioms for primitive commands and chain them with \ifappendix\cref{rule:wp-seq}\else the standard sequencing rule\fi. However, \cref{rule:if-then,rule:if-else} for conditionals require the guard to be a deterministic value.
When this is not the case, we follow the pattern in \cref{sec:key-ideas:verify:indirect-flow}. We use \cref{rule:unit-r,rule:oplus-frame} to introduce a \condition[ing] modality $\condmod$ that ``determinizes'' the value of the guard and proceed by cases with \cref{rule:oplus-cons}, and then put together the result on each fiber with \cref{rule:oplus-triple}.
Note that requiring the existence of an expression $\expr$ unchanged by $\prog$ to decide the branch of the \condition[ing] is not very restrictive. In fact, the value $v$ can be written in the log, so that it appears in the history and becomes immutable. This logging can happen without any security label: it is exclusively a proof tool, rather than an event observable by the adversary.

The use of \condition[ing] to determinize a value is not just useful for conditionals: for instance, \cref{rule:triple-log} requires a deterministic length of the log. This can be extracted from $\own{\logs}$ using the (derived) \cref{rule:oplus-own-merge}: $\obsfun{\length}$ is a function of the log so
$
\own{\logs}\withperm{\logs : 1} \lequiv \tcond[t]{ (\auth{t} * \own{\restr{\logs}{t}}) }
$ .
Then, we can proceed by cases on the length $\auth{t}$ within the \condition[ing] on $\obsfun{\length}$ and put the results together with \cref{rule:oplus-triple}.

\begin{example}\label{ex:h-assigned-l}
    Consider the code \code{h := l;log!(high,h)} and the precondition from \eqref{triple:tini}:
    \[
    (\lowOf \obspointsto \obsL * \sure{\p{l} = v_l} * \sure{\p{h} = v_h} * \own{\logs})\withperm{\logs : 1}
    \]

    Using axioms for atomic commands, sequential composition and frame we can prove (we omit permissions on everything but the log because they can be handled mechanically, see \appendixref{ex:app:h-assigned-l})
    {\small
    \begin{equation*}
        \T
        { \auth{t} * \sure{\p{l} = v_l} * \sure{\p{h} = v_h}  }
        { \code{h := l;log!(high,h)} }
        {  \auth{t + 1} * (t+1) \obspointsto (\high, \p{h}) * \sure{\p{l} = v_l} * \sure{\p{h} = \p{l}}  }
    \end{equation*} }

    Using the equivalence above, we can lift the triple to the noninterference precondition:
    \begin{equation*}
        \begin{proofoutline}
            \assert{ (\lowOf \obspointsto \obsL * \sure{\p{l} = v_l} * \sure{\p{h} = v_h} * \own{\logs})\withperm{\logs : 1} } \\
            $\entails$ \assert{ \exists T \st \cond{t}{T}{ \bigl(\auth{t} * \sure{\p{l} = v_l} * \sure{\p{h} = v_h} * \lowOf \obspointsto[t] \obsL * \own{\restr{\logs}{t}} \bigr) } } \\
            \begin{proofjump}[rule:oplus-triple]
            \begin{proofjump}[rule:frame]
                \assert{ \auth{t} * \sure{\p{l} = v_l} * \sure{\p{h} = v_h}  } \\
                \quad\text{ \code{h := l;log!(high,h)} } \\
                \assert{ \auth{t + 1} * (t+1) \obspointsto (\high, \p{h}) * \sure{\p{l} = v_l} * \sure{\p{h} = \p{l}} } \\
            \end{proofjump} \\
            \end{proofjump} \\
            \phantom{$\entails$} \assert{ \exists T \st \cond{t}{T}{\bigl(  \auth{t + 1} * (t+1) \obspointsto (\high, \p{h}) * \sure{\p{l} = v_l} * \sure{\p{h} = \p{l}} * \lowOf \obspointsto[t] \obsL * \own{\restr{\logs}{t}}  \bigr)} }
        \end{proofoutline}
    \end{equation*}

    As typical of separation logics, a small-footprint triple is lifted to the entire context by the structural rules of \thelogic{}.
    We remark that in \thelogic{} there are two levels of locality: with respect to independent facts (locality with respect to $*$, supported by \cref{rule:frame}) and to nondeterministic values (locality with respect to $\condmod$, supported by \cref{rule:oplus-triple}).
\end{example}

A crucial characteristic of \condition[ing] is that it requires \emph{every branch} to be inhabited: the $\pspA_v$ are all possibility spaces, which cannot be empty.
This introduces a subtlety with our termination-insensitive $\wpsymb$:
applying \cref{rule:oplus-triple} to a program that (surely) diverges in some branches would be unsound.

\begin{example}\label{ex:weak-swap-ti}
    Consider the code
    $
    \code{while (h = 0) do skip;log!(0)}
    $
    and the precondition $(\auth{t-1} * \In{\p{h}}{\set{ 0, 1 }})$. By \cref{rule:unit-r} we can work on the two branches $\p{h} = 0$ and $\p{h} = 1$ separately. On the latter the program terminates with the postcondition $\sure{\p{h} = 1} * \auth{t} * t \obspointsto 0$. On the former, the program always diverges, so in our termination-insensitive interpretation of $\wpsymb$ it also satisfies the postcondition $\sure{\p{h} = 0} * \auth{t} * t \obspointsto 0$. However, putting these together with \cref{rule:oplus-triple} is unsound: in the combined postcondition $\cond{v}{\set{0, 1}}{\bigl( \sure{\p{h} = v} * \auth{t} * t \obspointsto 0 \bigr)}$ the branch for $v = 0$ is not satisfied by any final state.
\end{example}

A solution to this issue is requiring that the program has some terminating trace from every initial state:
the $\mayterm{\prog}$ side condition of \cref{rule:oplus-triple} is defined as $\mayterm{\prog} \is \left( \All \state . \sem{\prog} ( \state ) \neq \emptyset \right)$.
An alternative solution is to weaken the postcondition: \cref{rule:oplus-triple-weak} ensures a \condition[ing] on a subset of the branches, namely only those that may terminate. In the rule, these are picked by the existential quantification on $V'$: there must exist some branch that survives ($\emptyset \subsetneq V'$), but not necessarily all of them ($V' \subseteq V$). While weaker, this rule can be sufficient for proofs: as shown in \appendixref{ex:app:weak-swap-ti}, it is enough to prove TINI of the program in the above \cref{ex:weak-swap-ti} through triple \eqref{triple:tini}.
By switching to a termination-sensitive WP, the rule could determine exactly which branches of~$V$ survive;
we leave this for future work.

Another use for \condition[ing] is in \cref{rule:triple-while} for \p{while}-loops.
Just like for \p{if}, we need to partition $P(n)$ into two cases depending on $\bexpr$. Interestingly, we also use \condition[ing] in the postcondition to put together the infinitely many cases on the number of iterations $n$: there, the set $N$ contains exactly those iteration counts on which the \p{while}-loop may terminate.

\paragraph{Unframing and Precision}

Thanks to \cref{rule:oplus-triple} we are able to perform ``case analysis'' on some expression. However, as discussed in \cref{sec:key-ideas:verif}, this only proves \emph{conditional} independence.
To recover unconditional independence, we need to ``unframe'' facts outside of $\condmod$,
but unguarded unframing is in general unsound (cf. \cref{sec:key-ideas:verify:indirect-flow}).
Therefore, we guard this step in \cref{rule:oplus-unframe} with \emph{precision}, as already done by, \eg \citet{DBLP:journals/pacmpl/MatsushitaHWD26,DBLP:journals/pacmpl/ZilbersteinST26}.
\begin{definition}[Precision]
    An assertion $P \in \Assrt$ is \emph{precise}, written $P : \precise$, if it has either no valid model or a least model, \ie a model that is smaller than all other models:
    \[
    P : \precise
    \quad\tIff\quad
    \exists a \st (\raValid(a) \implies P(a)) \land \forall b \st \raValid(b) \implies P(b) \implies a \raLeq b
    \]
\end{definition}

Crucially, $(\In{\expr}{X})\withp{\permap}$
(and consequently~$\sure{\bexpr}\withp{\permap}$) is precise,
and precision is closed under separating conjunction and partitioning (see \appendixref{fig:app:precise-rules-all}).
Existential quantification (and thus $\own{\hole}$)
is the key connective that is not precise.

\begin{remark}
    \label{rm:bluebell-unframe}
    It is possible, in principle, to encode \thelogic's assertions as \Bluebell\ assertions.
    For example,
    $(\In{\expr}{X})\withp{\permap}$
    would be modelled in \Bluebell\ with
    $(\exists \mu. \expr \sim \mu \land \psupp(\mu)=X)\withp{\permap}$
    which says that there is \emph{some} distribution of values
    for $\expr$ and the non-zero probability elements ($\psupp(\mu)$)
    are exactly~$X$.
    This encoding is \emph{not precise} due to the existential quantification
    (which is essential since we cannot assume any specific distribution)
    and thus it disables unframing of $(\In{\expr}{X})\withp{\permap}$
    resources completely, breaking every non-trivial \thelogic\ proof,
    including all our examples.
\end{remark}

\paragraph{Soundness}
Following the semantic approach, all the rules of the \thelogic\ program logic
are implications from the (semantic interpretation of the) premises to the conclusion.
\appendixref{app:rules} shows the omitted rules and their proof sketches.
The rules about triples (\eg \cref{fig:violet-rules}) are all derived
on top of the rules for entailment
(\eg \cref{fig:violet-ent-rules} and the ones in \appendixref{app:rules}).
All the rules (primitive and derived) are proven sound in the Rocq mechanization~\cite{artifact}. \section{Modelling Security}
\label{sec:modelling-security}

In this section we show the versatility of \ourlogic\ as a specification tool,
by encoding various security notions as triples.
All results in this section have been mechanized in Rocq~\cite{artifact}.

\subsection{Noninterference}
For a program with security label annotations,
one can state the standard termination-insensitive noninterference
(for the nondeterministic case).
Usually, in noninterference there is a notion of secret input that must be protected. Our programming language does not admit input, but as discussed in \cref{sec:key-ideas:spec} we can model it with the log.
For noninterference, we assume the secret input is written in the first entry in the log and we read it with the following function:
\begin{equation*}
    \highIn(\store, \obs) \is \text{if } \obs[1] = (\high, v) \text{ then } v \text{ else } \bot
    \tag{HighIn}
    \label{eq:high-in-def}
\end{equation*}

Our noninterference is parametric over an arbitrary set $H \subs \Val$ of valid high inputs with $\bot \notin H$.
For instance, for the programs in \cref{fig:running-ex},
one would take $H = \Int$.
We define Termination-Insensitive Noninterference with Declassification
as follows:
\begin{definition}\label{def:tini-rel}
    Let $\declOf \from \Obs \to D$ be a function of the log and $H \subs \Val$ with $\bot \notin H$.

    A program~$\prog$ satisfies
    \emph{Termination-Insensitive Noninterference with Declassification}~(\TINIrel{\declOf})
    if:
    \[
    \TINIrel{\declOf}(\prog) \is
    \All \state_1,\state_1',\state_2, \state_2'.
    \begin{conj}
        \state_1' \in \sem{\prog}(\state_1)
        \text{, } \quad
        \state_2' \in\sem{\prog}(\state_2)
        \land
        \lowOf(\state_1) = \lowOf(\state_2)
        \text{, } \quad
        \pv{x}(\state_1) = \pv{x}(\state_2)
        \land
        \highIn(\state_1), \highIn(\state_2) \in H
        \land
        \highIn(\state_1) \neq \highIn(\state_2)
        \land
        \declOf(\state_1') = \declOf(\state_2')
    \end{conj}
    \implies
    \Exists \state_2''.
    \begin{conj}
        \state_2'' \in \sem{\prog}(\state_2)
        \land
        \lowOf(\state_1') = \lowOf(\state_2'')
        \land
        \declOf(\state_1') = \declOf(\state_2'')
    \end{conj}
    \]
\end{definition}

\begin{wrapfigure}[7]{R}{23ex}\vspace{-\baselineskip -0.9ex}
  \centering \begin{tikzpicture}[
  eq/.style={
    shorten = 2pt,
  },
  obs/.style={
    sloped,
    font=\footnotesize,
  },
  trace/.style={
    decorate, decoration={snake, amplitude=1pt, segment length=4pt,pre length=2pt, post length=#1},
    shorten = 1pt,
    -{Latex[scale=.7]},
  },
  trace/.default=4pt,
  state/.style={
    circle, inner sep=0pt,
    minimum size=2pt,
    fill,
  },
  st/.style={strut sized,inner sep=2pt},
  exists/.style={blindblue},
thick,
  node distance=1cm and 1.2cm,
]

\node[state,label={[st]$s_1$}]   (s1)   {};
\node[state,label={[st,below]$s_2$},below=of s1]   (s2)   {};
\node[state,label={[st]$s_1'$},right=of s1]  (s1')  {};
\node[state,label={[st,below,gray]$s_2'$},right=of s2,gray]  (s2')  {};
\path (s1') --
        node[state,label={[st,right]$s_2''$},pos=.67,right=1] (s2'') {}
      (s2');

\draw (s1) edge[trace] (s1');
\draw (s2) edge[trace,gray] (s2');

\draw (s1) edge[eq] node[obs,below]{$\lowOf$} (s2);
\draw (s1') edge[eq,gray] node[obs,above,gray]{$\mathsf{decl}$} (s2');

\draw (s1') edge[eq,exists] node[obs,above]{$\lowOf$} (s2'');
\draw (s2') edge[eq,exists] node[obs,below]{$\declOf$} (s2'');
\fill[white] (s2'') ++(-2pt,0) rectangle +(-8pt,-3.1pt);
\draw (s2) edge[trace=5pt,exists] (s2'');

\end{tikzpicture}
   \vspace{-1ex}
  \caption{\TINIrel{\declOf}.
  }
  \label{fig:tini-decl}
\end{wrapfigure}
In \cref{fig:tini-decl} we show a schematic representation of the \TINIrel{\declOf} definition: starting from low-equivalent states~$\state_1,\state_2$,
if there are terminating traces from them declassifying the same information,
then for every trace from $\state_1$ there is a trace from $\state_2$
which reaches a low-equivalent state with the same declassified information.
If we fix $\declOf = (\fun \state . \top)$ nothing is declassified and we recover standard termination-insensitive noninterference as in equation (\ref{triple:tini}).

We remark that the special case when $\prog$ is deterministic corresponds to the usual definition from the literature \citep{DBLP:journals/jsac/SabelfeldM03}: for any two initial $\low$-equivalent but not $\highIn$-equivalent states $\state_1$, $\state_2$, if $\prog$ terminates on both then the final states are also low equivalent.

We can prove that $\TINIrel{\declOf}$ is \emph{characterized} by a triple like \eqref{triple:tini} of \cref{sec:key-ideas:spec}.

\begin{theorem}[Adequacy]\label{theorem:adequacy}
    Fix any program $\prog$, set of valid high inputs $H \subs \Val$ with $\bot \notin H$ and function $\declOf$.
    Then $\TINIrel{\declOf}(\prog)$ holds if and only if
    \begin{align*}
        \All \obsL, \vec{v} .
        \T {(\sure{\lowOf = \obsL} * \sure{\pv{x} = \vec{v}} * \sure{\highIn \in H} * \own{\logs})\withperm{\_ : 1}}{\prog}{ \vcond{\declOf}{\own{\lowOf} * \own{\highIn}} }
    \end{align*}
\end{theorem}

\subsection{`When' Declassification}

The triple for $\TINIrel{\declOf}$ states that everything that is declassified is completely fixed (by the \condition[ing] on $\declOf$).
This effectively allows releasing a secret \emph{before} its declassification.
As we discussed in \cref{sec:key-ideas:spec-beyond},
in \thelogic{} it is possible to encode that declassifications
are only valid from when they happen onward (a `when' declassification policy).
We do so by requiring the security policy
(\eg noninterference with declassification)
to hold every time a relevant event is emitted.
Formally, we consider the log to contain all relevant events. Therefore, `when' declassification is described by requiring $\TINIrel{\declOf}$ for all $m$ at which the log can be cut during the execution of the program: replace $\declOf$ with $\restr{\declOf}{m}$ and $\lowOf$ with $\restr{\lowOf}{m}$.
Such a definition is then characterized by a triple like \eqref{ideas:triple:prefix-close}, with the universal quantification on $m \geq n_0$. For space reasons, the formal correspondence is given only in \appendixref{sec:app:modelling:when-decl}.

\subsection{Timing Side Channel}

Timing attacks~\citep{DBLP:conf/uss/BrumleyB03} are a class of side-channel attacks on cryptographic procedures. They extract information about a secret input from the duration of the computation.
In \thelogic{} we can model this kind of attack by writing a tick in the log every time the program performs an expensive operation. In the end, the number of ticks in the log is proportional to the time taken by the execution.
Let $\obsfun{ticks} \from \Obs \to \Nat$ be the function that counts the number of \code{tick} entries in the log. Then we can express constant-time in \thelogic{} as independence between $\highIn$ and $\obsfun{ticks}$ (for a fixed number of ticks in the initial log):
\[
\forall n_0 \st
\T
{(\own{\logs} * \sure{\obsfun{ticks} = n_0} * \sure{\highIn \in H})\withperm{\_ : 1}}
{\prog}
{\own{\highIn} * \own{\obsfun{ticks}}}
\]

Notice that this triple is exactly the one characterizing $\TINIrel{\declOf}$ when $\lowOf = \obsfun{ticks}$ and $\declOf = (\fun \state . \top)$: a program is constant-time exactly if the (secret) input does not influence the execution time (the number of ticks). \Cref{theorem:adequacy} directly proves the adequacy of the above triple.

Consider the two variants of the code in \cref{fig:timing-ex},
which perform some operation for each bit of \p{h} that is 1 (for instance, exponentiation-by-squaring). We assume \p{h} to have a 64-bit representation.
\begin{figure}
  \adjustfigure[\small]
  \begin{codecols}
  \lstset{gobble=4}
  \begin{sourcecode*}
    def timevuln:
      h := $\highIn$;
      while h > 0:
        if h mod 2 = 1: log!(tick);  // work
        else: skip
        h = h/2;
  \end{sourcecode*}
  \begin{sourcecode*}
    def timeconst:
      h := $\highIn$;
      repeat 64:
        if h mod 2 = 1: log!(tick);  // work
        else: log!(tick);  // waste time
        h = h/2;
  \end{sourcecode*}
  \end{codecols}
  \caption{Timing side channel examples.}
  \label{fig:timing-ex}
\end{figure}
The constant-time triple does not hold for \p{timevuln} because in the end the number of ticks is $n_0$ plus the number of 1s in the binary representation of \p{h}. In \p{timeconst} it is always $n_0 + 64$, and therefore it is independent of \p{h}.

\subsection{Paralocks}

To illustrate the fact that \ourlogic\ can robustly represent very
dynamic notions of security, we show how to encode Paralocks~\cite{DBLP:conf/popl/BrobergS10} as a \ourlogic\ specification.
Paralocks is a model where the applicable policy can vary dynamically
during the execution of a program.\footnote{The full Paralocks model also allows policies to be parametric on roles, but we omit this orthogonal aspect in our discussion.}
The idea is to express policies by fixing a set of \emph{actors}
(modelling the roles of principals of the system)
and expressing the allowed flows on each location by a set of clauses $\set{c_1;\dots;c_n}$.
Each clause~$c_i$ is of the form $\Sigma \implies a$
where $a$ is an actor, and $\Sigma$ is a (possibly empty) set of \emph{locks}.
Locks~$\sigma$ are boolean flags that
can be set (with the \code{open($\sigma$)} instruction)
or unset (with the \code{close($\sigma$)} instruction)
by the program at runtime.
A clause $ \set{\sigma} \implies a $ associated with some value means that
the value is allowed to flow to $a$ if the lock $\sigma$ is set.
For example, a situation with high/low-confidentiality data can be modelled
by having two actors $\high$ and $\low$;
a value with policy $\set{\emptyset \implies \high}$ (written $\set{\high}$)
is the policy for secrets, $\set{\low}$ for public values.
A policy $\set{\high; \sigma \implies \low}$ is a secret that
can be declassified when~$\sigma$ is set.

The security notion used by Paralocks~\cite[Def.~5.4]{DBLP:conf/popl/BrobergS10}
is progress-insensitive, and formulated in terms of ``attacker knowledge'',
\ie the set of high inputs that are still plausible after some observations.
Specifically, the notion quantifies over all attackers $A = (a, \Sigma)$,
\ie a pair of an actor~$a$ (the role of the attacker) and
a set of locks~$\Sigma$, which can be thought of as the set of locks that~$A$
can \emph{force} open.
Every attacker $A=(a, \Sigma)$ induces a split
between ``high'' and ``low'' observations,
\ie all the values with policy including a clause
    $\set{\Sigma'\implies a}$ with $\Sigma' \subs \Sigma$
    are considered \emph{\pre A-low} and the others \emph{\pre A-high}.
Paralocks security checks all flows are allowed by considering every attacker.
Given an attacker~$A=(a,\Sigma)$, and arbitrary
\pre A-low observations~$\vec{u}$ (the ``declassified'' values),
Paralocks compares any trace with \pre A-low observations $\vec{u}u$,
with any trace with \pre A-low observation $\vec{u}u'$.
Let~$\Delta$ be a possible set of open locks after the observations~$\vec{u}$.
If $\Delta \subs \Sigma$ then no flow from \pre A-high inputs to \pre A-low observations is allowed.
The ``no flow'' condition is expressed by saying that the
set of plausible \pre A-high inputs given $\vec{u}u$ is the same
as the one plausible given $\vec{u}u'$.

The Paralocks security notion translates directly to our setting,
via a simple compilation $\sem{\prog}$
from programs~$\prog$ with locks and policies to our language.
Every assignment logs a pair of the assigned value and the policy
associated with the variable assigned to.
The \code{open($\sigma$)}/\code{close($\sigma$)} operations
log the current state of the lock, the identity of the lock $\sigma$ and
the policy associated with the lock.
Any initial input is stored in the log as for \TINI.
Then Paralocks security is encoded as:
\begin{align}
    \label{triple:paralocks}
    \begin{split}
    &
    \text{Init}(A, \obsL, \vec{v}) \is (\sure{\obsfun{view}^A = \obsL} * \sure{\pv{x} = \vec{v}} * \sure{\highIn^A \in H^A} * \own{\logs})\withperm{\_ : 1}
    \\
    &\forall A=(a,\Sigma) \st
    \forall n\in \Nat \st
    \forall \obsL, \vec{v} \st
    \\
    &\qquad
    \T
    { \text{Init}(A, \obsL, \vec{v}) }
    {\sem{\prog}}
    {\vcond{\obsfun{view}^A_n}{
        \vcond[b]{\obsfun{hasNxt}^A_n}{
        \bigl(
            \pure{b} \implies
                \bigl(
                    \sure{\obsfun{open}^A_n \not\subs \Sigma}
                    \lor
                    (
                        \own{\obsfun{nxt}^A_n} * \own{\highIn^A}
                    )
                \bigr)
        \bigr)
        }
    }}
    \end{split}
\end{align}
where
    $\obsfun{view}^A_n$ extracts from the log the values visible to~$A$ according to their policies, truncated at~$n$;
    $\obsfun{hasNxt}^A_n$ returns a boolean indicating whether the log has a next event in the view of~$A$ at~$n$;
    $\obsfun{open}^A_n$ is the set of open locks at $\obsfun{view}^A_n$ (including locks invisible to $A$, the dependency is only to compute when $n$ $A$-visible events have happened);
    $\obsfun{nxt}^A_n$ is the next \pre A-visible event after~$n$;
    $\highIn^A$ is the \pre A-high input.

The precondition \text{Init} is analogous to \TINIrel{\declOf}.
The postcondition first conditions on the first $n$ \pre A-visible events:
this amounts, like for \TINIrel{\declOf}, to declassifying these events.
For example, for the attacker $A=(\low, \set{\sigma})$
all the events with policy including $\set{\low}$ or $\set{\sigma \implies \low}$
are disclosed to~$A$.
Then the conditioning on $\obsfun{hasNxt}^A_n$ expresses the
progress-insensitivity of the original Paralocks definition,
which only considers pairs of traces which have some next \pre A-visible event,
\ie absence of an event is not considered to leak information.
The postcondition expresses this by only imposing constraints on the subset
of the traces that have a \pre A-visible event ahead,
that is, when $b$---the result of $\obsfun{hasNxt}^A_n$---is true.
On this subset of cases,
if even one possible lock state sees as open only locks that $A$ could have forced open, then no flow from \pre A-high to \pre A-low is allowed.
Only if every possible lock state sees an open lock that is not in $\Sigma$
are additional flows allowed.
This is represented by the inner disjunction in the postcondition:
either in every state some lock not in $\Sigma$ is open (allowing flows),
or the next \pre A-visible event is independent of the \pre A-high input
(\ie does not refine the attacker knowledge).

Notice that by removing the conditioning on $\obsfun{hasNxt}^A_n$
we would obtain an ``occurrence-sensitive'' variant where absence of a low
output upon termination would be considered a leak.
For instance,
  \code{if h then log!(\{low\},1) else skip}
would be rejected by the revised notion.

Our triple implies a knowledge-based definition analogous to \citet[Def.~5.4]{DBLP:conf/popl/BrobergS10}. For space reasons, we only detail this correspondence in \appendixref{sec:app:modelling:paralocks}.

\newcommand{\bdBclass}{\ensuremath{\obsfun{bclass}}}
\newcommand{\bdB}{\ensuremath{\mathsf{B}}}
\newcommand{\bdT}{\ensuremath{\obsfun{T}}}

\subsection{Other Definitions of Information Flow Security}
Definitions of security that combine different forms
of declassification can also be modelled as conditional independence
with further effort.
For example, bounded-deducibility
(BD)~\cite{DBLP:conf/itp/0001BL21}, which combines `when' and `what'
forms of declassification, says that \emph{unless} the log has a
declassification trigger, the adversary learns information only up to
a bound. The declassification trigger and the bound are characterized
as a predicate~$\bdT$ on log events and a relation $\bdB$ on lists of
secrets, respectively. BD can be approximated in \ourlogic{} as
conditional independence, following the template of
\cref{theorem:adequacy} together with the prefix quantification of
`when' declassification, \cref{ideas:triple:prefix-close}. For the specific
case where $\bdB$ is an equivalence relation, BD morally takes the following
form, where $\restr{\highOf}{n}$ extracts the secrets from the first
$n$ log entries, $\bdBclass$ maps them to their $\bdB$-equivalence
classes, and $\restr{\bdT}{n}$ is true iff at least one event among
them satisfies $\bdT$.
$$\forall n, \obsL, \vec{v} \st \T{\sure{\lowOf = \obsL} *
  \sure{\pv{x} = \vec{v}} * \own{\logs}} {\prog}
{\vcond[b]{\restr{\bdT}{n}}{ \bigl(\pure{\neg b} \implies
    \vcond{\restr{\bdBclass}{n}}{ \own{\restr{\lowOf}{n}} *
      \own{\restr{\highOf}{n}}} \bigr)}} $$

A formal precise encoding of BD in \thelogic\ is left for future work.
Explicit logs can also model reactive programs and, therefore,
{\ourlogic} should be able to model information flow security on such
programs, e.g., reactive
noninterference~\cite{DBLP:conf/ccs/BohannonPSWZ09}, which applies to
deterministic reactive programs, as well as suitable generalizations
to nondeterministic reactive programs.

 \section{Verification}
\label{sec:verification}

In \cref{sec:key-ideas:verif} we have already shown the critical proof patterns.
For reasons of space,
we detail here only one example that exploits advanced features of \ourlogic;
further examples are in \appendixref{sec:app:more-case-studies};
all the examples are mechanized in \cite{artifact}.

\begin{figure*}
    \adjustfigure[\small]
    \begin{sourcecode}[gobble=8,frame=tb]
        def auction:
            n := rnd($\Nat$);
            repeat n:
                i := i+1; log!(start,i);  // announce the start of auction i
                bid := rnd($U$);           // user input
                log!(high, bid);        // the bid is dynamically declared secret
                k := rnd($\Int_p$);            // nondeterministic choice of key
                // sending on an insecure channel is modelled as low log
                log!(low, bid+k mod $p$);    // sending the encrypted bid to the server
                log!(decl, bid);        // result announced, declassifies bid
                log!(low, k);           // releases the session key
    \end{sourcecode}
    \caption{An example program using dynamic secrets. $U \subs \Int$ is a set of values the user can input.}
    \label{fig:auction-case-study}
\end{figure*}

The program in \cref{fig:auction-case-study} models a secret-bidding auction client. For each auction (identified by index $i$), it first reads the bid from the user.
To model dynamic input, we use nondeterministic values from some arbitrary set $U$ of possible values. The bid is then \emph{dynamically} declared secret by logging it on the $\high$ log.
To send it to the server, the client mints a session key \p{k}, a new internal secret for encryption.
Using \p{k}, it sends the bid encrypted on an insecure channel (the $\low$ log).
After the server announces the result, the \emph{current} bid is dynamically declassified. This does not declassify previous or future bids, only the current one.
At this point, there is no longer any need to keep the session key private. By writing it on the public channel $\low$, we model a possible leak that happens in the future, but we still forbid leaks in the past.

Recall the prefix-closed specification from \eqref{ideas:triple:prefix-close}. Since we are dealing with a closed program, we assume for simplicity that the initial log is empty and that the initial value for the auction identifier \p{i} is $0$. Since the number \p{n} of auctions is public, we condition on it: this could be achieved by logging it as $\decl$, but we keep it split here for clarity. As usual, we omit permissions for simplicity.
Thus, the target security triple is
\begin{equation*}
    \All m\in\Nat .
    \T
    { \sure{\logs = \emptylog} * \sure{\p{i} = 0} }
    {\p{auction}}
    {
\vcond{\p{n}}{
            \vcond{\restr{\declOf}{m}}{\bigl(
                \own{\restr{\lowOf}{m}} * \own{\restr{\highOf}{m}}
            \bigr)}
        \bigr)}
    }
\end{equation*}

To prove this specification in \thelogic{}, we define the following loop invariant:
\begin{align*}
    A(i) &\is \begin{grp}
        (5i + 1) \obspointsto (\p{start}, i + 1)
        * \In{\logs[5i + 2]}{(\set{ \high } \times U)}
        * \In{\logs[5i + 5]}{(\set{ \low } \times \Int_p)} \\
        {} * (5i + 3) \obspointsto (\low, \logs[5i + 2] + \logs[5i + 5] \% p)
        * (5i + 4) \obspointsto (\decl, \logs[5i + 2])
    \end{grp} \\
    P(i) &\is \sure{\obsfun{\length} = 5i} * \sure{\p{i} = i} * \Sep_{j = 0}^{i-1} A(j)
\end{align*}

Intuitively, the loop keeps adding the same block of five observations described by $A(i)$ every iteration.
As in \appendixref{ex:app:h-assigned-l}, we can prove the triple
$\T{P(i)}{\p{body}}{P(i+1)}$
for the body of the loop.
To lift this to a triple for the whole program, we first use \appendixref{rule:wp-rnd} to obtain $\In{\p{n}}{\Nat}$, and then \cref{rule:unit-r} to introduce the \condition[ing] $\cond{n}{\Nat}\sure{\p{n} = n}$ on it. The initial precondition is
$
\bigl( \sure{\logs = \emptylog} * \sure{\p{i} = 0} \bigr)
\entails
\bigl( \sure{\obsfun{\length} = 5 \cdot 0} * \sure{\p{i} = 0} \bigr) = P(0)
$
so, by duplicating the sure fact $\sure{\p{n} = n}$, we obtain $\vcond{\p{n}}\sure{\p{n} = n} * P(0)$.
Using rules \appendixref{rule:wp-repeat,rule:wp-bind} we obtain
\[\T{\sure{\p{n} = n} * P(0)}{\code{repeat n: body}}{P(n)}, \]
which we can apply inside the \condition[ing] (\cref{rule:oplus-triple}) to get the postcondition $\vcond{\p{n}}{ P(n) }$.
Then, using \cref{rule:oplus-cons}, we can focus on a fixed $\p{n} = n$ and reduce the thesis to the entailment below.
\[
\sure{\obsfun{\length} = 5n} * \sure{\p{i} = n} * \Sep_{j = 0}^{n-1} A(j)
\;\stackrel{\smash{?}}{\entails}\;
\vcond{\restr{\declOf}{m}}{\bigl(
    \own{\restr{\lowOf}{m}} * \own{\restr{\highOf}{m}}
\bigr)}
\]

Notice that $\Sep_{i = 0}^{n-1} A(i)$ gives an exact description of the log up to index $5 n$, which is the entire length of the log. So we know exactly the shape of the entire log.
Let $m = 5k + r$. We observe that $\restr{\declOf}{5k + r}$ contains all entries $\logs[5i + 2]$ for $0 \leq 5i + 2 \leq 5k + r$, which includes the entire $\restr{\highOf}{5k}$. We then proceed by case analysis on the remainder $r$ of the division.

If $r = 0$ or $r = 1$, $\restr{\declOf}{m} = \restr{\highOf}{5k} = \restr{\highOf}{m}$, that is, the entire $\high$ log is declassified. This makes it trivially independent of the low log inside the \condition[ing] on $\restr{\declOf}{m}$.
If $r = 2$, $\restr{\highOf}{m} = \restr{\highOf}{5k} \app \logs[5k + 2]$. Independence from the low log follows from the fact that $\restr{\highOf}{5k} = \restr{\declOf}{m}$ is \condition[ed] on, and $\logs[5k + 2]$ is an independent sample from $U$.
If $r = 4$, $\restr{\declOf}{m} = \restr{\highOf}{m}$, so again the entire $\high$ log is declassified and thus independent under the \condition[ing].
The last case is when $r = 3$. Here, $A(k - 1)$ gives all the ingredients to redo the exact same proof shown in \cref{sec:key-ideas:verif} but modulo $p$ instead of modulo $2$: independence of $\logs[5k + 2]$ and $(\logs[5k + 2] + \logs[5k + 5] \% p)$ follows from $\logs[5k + 5]$ being a value in $\Int_p$ independent of $\logs[5k + 2]$.

 \section{Abstractions}
\label{sec:abstractions}

So far, we have shown how to prove programs secure in \thelogic{}. However, proofs for larger programs involve simple bookkeeping for most of the code, and only a few points where actual ingenuity is required. This suggests that it may be possible to devise abstractions on top of \thelogic{} to ease proofs.
As anticipated in \cref{sec:key-ideas:abstractions}, we believe logical semantic typing~\citep{DBLP:journals/jacm/TimanyKDB24} to be the solution to this issue. The idea is to define a syntactic type system to deal with most of the program, but define its semantics in terms of a program logic. This way, for complex components that cannot be typed syntactically it is possible to prove the type judgment in the logic, and then use it in a syntactic type derivation for the whole program.

Developing a realistic type system for security is a challenging task that requires searching the space of possible abstractions. Such a task is outside the scope of this paper.
Our goal here is to demonstrate the viability of building such abstractions on top of \thelogic{}, with a proof of concept which we have mechanized in \cite{artifact}.
We build a simple type system for proving \TINI{} inspired by the one in \citet{DBLP:journals/jcs/VolpanoIS96}, with two main deviations. To stress-test the nondeterministic nature of \thelogic{}, we add a nondeterministic assignment in the language, analogous to \code{rnd}. However, it is known that the Volpano-Irvine-Smith type system is unsound for termination-insensitive noninterference in the presence of nondeterminism~\citep{DBLP:conf/popl/SmithV98}; therefore, we replace unbounded \p{while}-loops with the bounded \p{repeat}.
An alternative solution that keeps both is to make termination depend only on low inputs~\citep{DBLP:conf/popl/SmithV98}. This makes the system termination-\emph{sensitive}: termination is allowed to leak secret information, therefore it is constrained by the type system. In this paper we choose to remain termination-insensitive to have a coherent presentation of the novel aspects of \thelogic{}; the termination-sensitive variant is left as future work.

In this section, we follow the presentation with a security context $\pclabel$ from \citet{DBLP:journals/jsac/SabelfeldM03}, generalized to a security lattice of labels $\secl \in \Lvl$, with the two changes described above: \p{while} replaced by \p{repeat} and the addition of \code{rnd}.
Let $\seclabelof \from \Var \to \Lvl$ be the function that associates to each variable its security label.
We show as examples the rules for assignments, conditionals and the new \code{rnd}, where $\aexpr : \secl$ means $\All \p{x} \in \pvar(\aexpr) . \seclabelof(\p{x}) \le \secl$. The full type system is in \appendixref{fig:app:vs-type-system}.
\begin{proofrules}
    \infer*[lab=vs-assign]{
    \seclabelof(\p{x}) \ge \pclabel \\
    \aexpr : {\seclabelof(\p{x})}
}{
    \VSJ{\pclabel}|-{\code{x := $\aexpr$}}
}
     \label{rule:vs-assign}

    \infer*[lab=vs-if]{
    \bexpr : \pclabel \\
    \VSJ{\pclabel}|-{\prog_1} \\
    \VSJ{\pclabel}|-{\prog_2}
}{
    \VSJ{\pclabel}|-{\code{if $\;\bexpr\;$ then $\;\prog_1\;$ else $\;\prog_2$}}
}
     \label{rule:vs-if}

    \infer*[lab=vs-rnd]{
    \seclabelof(\p{x}) \ge \pclabel \\ V \neq \emptyset
}{
    \VSJ{\pclabel}|-{\code{x := rnd($V$)}}
}     \label{rule:vs-rnd}
\end{proofrules}

As discussed, to encode a security notion in \thelogic{} we need to write security-relevant events in the log. To this end, we define a compilation $\Comp{\hole}$ of programs
that annotates them carefully to model the desired information leakage. As an example, we show below the compilation of assignments and conditionals:
\begin{align*}
    \Comp{\code{x:=$\aexpr$}} &\is \code{log!($\seclabelof$(x),'x', $\aexpr$); x:=$\aexpr$} \\
    \Comp{\code{if $\;\bexpr\;$ then $\;\prog_1\;$ else $\;\prog_2$}} &\is \code{if $\;\bexpr\;$ then } \Comp{\code{$\prog_1$}}\code{ else }\Comp{\code{$\prog_2$}}
\end{align*}

This compilation ensures that every time a variable is changed, a corresponding entry appears in the log. Therefore, compiled programs preserve ``state-log consistency'', the fact that variables contain the last value associated with them in the log.
Let $\lastOf(\p{x}) \from \Obs \to \Val$ be the function that reads the last entry of shape $(\seclabelof(\p{x}),\code{'x'}, \val)$ in the log and returns $\val$, the value of $\p{x}$ according to the log.
To avoid the need for permissions on the log, we add a superscript $t \in \Nat$ to truncate the log to length $t$: $\lastOf^t(\p{x}) \is \fun \obs . \lastOf(\p{x})(\restr{\obs}{t})$.
Given the list $\pv{x}$ of all program variables, we define the ``state-log consistency invariant'' (parametric on $t$) as
$I^t \is \sure{t \leq \obsfun{\length}} * \sure{\LAnd_{\p{x} \in \pv{x}} \p{x} = \lastOf^t(\p{x})}\withp{\pv{x} : 1}$.

We generalize $\lowOf$ and $\highOf$ for the security lattice $\Lvl$. For $\secl \in \Lvl$, let $\secbelow{\secl} \is \set{ \secl' \in \Lvl | \secl' \le \secl }$ be the set of security labels below $\secl$ (inclusive).
We overload notation and write, for instance, $\lastOf(\secnotbelow{\secl})$ for the sequence of values $\lastOf(\p{x})$ for all variables $\p{x}$ such that $\seclabelof(\p{x}) \in \secnotbelow{\secl}$ (the complement of $\secbelow{\secl}$).
We let $\obsSecl(N) \from \Obs \to \Obs$ be the function extracting observations with label in $N$ from the log. For instance, $\sure{\obsSecl(\secbelow{\secl}) = \obsL}$ means the sequence of events with security label in $\secbelow{\secl}$ (the view of $\secl$) is $\obsL$. As above, we use a superscript $\obsSecl^t(N)$ to denote application of $\obsSecl(N)$ to the prefix of length $t$.

For a security label $\secl$, we define
\[
\Pre^t_{\secl}(\obsL) \is (\auth{t} * I^t * \sure{\obsSecl^t(\secbelow{\secl}) = \obsL} * \own{\restr{\logs}{t}}) .
\]
This assertion fixes the the initial length of the log to be $t$ and the low input to be $\obsL$, then assumes state-log consistency and ownership of the log.
Fixing the initial length of the log to be $t$ also allows to talk about the initial high input in the postcondition: we can use the values of high variables stored in the log at the prefix $t$, before executing the command, that is $\lastOf^{t}(\secnotbelow{\secl})$.

\begin{definition}
    Let $\prog$ be a program on variables $\pv{x}$.
    A type judgment is \emph{valid}, written $\VSJ{\pclabel}|-{\prog}$, if the following formula holds.

    $\All \secl \in \Lvl, \obsL \in \Obs . \Exists \Slow \subs \Obs . (\secl \ngeq \pclabel \implies \Slow = \set{ \obsL })$ and
    \begin{equation*}
        \forall t \in \Nat \st
        \T
        { \Pre^t_{\secl}(\obsL) }
        { \Comp{\prog} }
        {
            \vcond{\lastOf^{t}(\secnotbelow{\secl})}{
                \cond{\obsL'}{\Slow}{
                    \vcond[t']{\finalposOf_{\prog}(t)}{
                        \Pre^{t'}_{\secl}(\obsL')
                    }
                }
            }
        }
    \end{equation*}
\end{definition}

Let us digest the formula step by step. Just like the soundness in \citet[Theorem~6.8]{DBLP:journals/jcs/VolpanoIS96}, our triple also holds for all security labels $\secl \in \Lvl$, corresponding to security with respect to each label. We \condition a priori with the outside quantification on the length of the log $t$ and the low input $\obsL$ (with respect to the fixed $\secl$). The set $\Slow$ is the set of possible final low logs.
Crucially, $\Slow$ depends \emph{only} on the initial low log $\obsL$, not the rest of the state, which yields independence from the high input.
Moreover, if the security label $\secl$ cannot view the current program counter $\pclabel$, the final low log must be precisely the input low log. This means that no write happens on a variable that cannot see the program counter. Equivalently, all writes happen on variables above the program counter \citep[Lemma~6.4]{DBLP:journals/jcs/VolpanoIS96}.

In the postcondition, we re-establish the precondition for the next program step, but inside a triple \condition[ing]. This allows us, at the same time, to also ensure independence of the high input $\lastOf^{t}(\secnotbelow{\secl})$ and the low output $\obsSecl^{t'}(\secbelow{\secl})$.
To continue execution of the subsequent program, the precondition $\Pre^{t'}_{\secl}(\obsL')$ is established at the log length $t'$ at which the execution of $\prog$ ended. This value can depend on the execution trace, so it is extracted from the final log by the function $\finalposOf_{\prog}(t)$. This function simply inspects the log in the final state to determine where the execution of command $\prog$ ended, provided it started at length $t$.
The following derivation, where for clarity we let $\obsfun{lowOut} \is \obsfun{obs}^{\finalposOf_{\prog}(t)}(\secbelow{\secl})$, formally shows that high input and low output are independent in the postcondition:
\begin{eqexplain}
    \vcond{\lastOf^{t}(\secnotbelow{\secl})}{
        \cond{\obsL'}{\Slow}{
            \vcond[t']{\finalposOf_{\prog}(t)}{
                \Pre^{t'}_{\secl}(\obsL')
    }}}
    \entails{}&
    \vcond{\lastOf^{t}(\secnotbelow{\secl})}{
        \cond{\obsL'}{\Slow}{
            \vcond[t']{\finalposOf_{\prog}(t)}{
                \sure{\obsfun{lowOut} = \obsL'}
    }}}
    \whichentails
    \vcond{\lastOf^{t}(\secnotbelow{\secl})}{
        \cond{\obsL'}{\Slow}{
            \sure{\obsfun{lowOut} = \obsL'}
    }}
    \byrule{oplus-unframe}
    \whichentails
    \vcond{\lastOf^{t}(\secnotbelow{\secl})}{
        \In{\obsfun{lowOut}}{\Slow}
    }
    \byrule{unit-r}
    \whichentails
    \In{\obsfun{lowOut}}{\Slow} * \vcond{\lastOf^{t}(\secnotbelow{\secl})}{ \True }
    \byrule{oplus-unframe}
    \whichentails
    \In{\obsfun{lowOut}}{\Slow} * \own{\lastOf^{t}(\secnotbelow{\secl})}
    \byrule{unit-r}
\end{eqexplain}

The (logical) semantics of a judgment is adequate, in the sense that it implies \TINI{}. To play the role of $\highIn$, we use the \emph{first} occurrence of each variable in the log (extracted by $\firstOf$): since \TINI{} is established on closed programs, we can assume that the initial log is empty.
\begin{theorem}[Adequacy of the type system]
    Let $\VSJ{\pclabel}|-{\prog}$ be a valid judgment and $\secl \in \Lvl$ a security label. Let \p{z} be a variable not appearing in $\prog$, and define $\lowOf = \obsSecl(\secbelow{\secl})$ and $\highIn = \firstOf(\secnotbelow{\secl})$. Then \TINI{} holds for the program
    \[
    \code{ z := $\length(\logs)$; for x in $\ \pv{x}\ $ do x:=$\firstOf$(x); log!($\seclabelof(\p{x})$,'x',x) done; } \Comp{\prog}
    \]
\end{theorem}
\begin{proof}[Proof sketch]
    By \cref{theorem:adequacy} we can prove \TINI{} via a triple in \thelogic{}. Starting from
    \[(\sure{\lowOf = \obsL} * \sure{\highIn \in H} * \own{\logs})\withperm{\_ : 1} ,\]
    \condition on the length $t$ of the log. After the preamble, it is straightforward to prove that $I^t$ holds and $\sure{\highIn = \lastOf^{\p{z}}(\secnotbelow{\secl})}$. By semantic validity of the judgment we get
    \[
    \vcond{\lastOf^{\p{z}}(\secnotbelow{\secl})}{
        \cond{\obsL'}{\Slow}{
            \vcond[t']{\finalposOf_{\prog}(t)}{
                \Pre^{t'}_{\secl}(\obsL')
            }
        }
    }
    \]
    As shown above, this entails $\In{\obsfun{lowOut}}{\Slow} * \own{\lastOf^{\p{z}}(\secnotbelow{\secl})}$, the postcondition of \TINI{}.
\end{proof}

\begin{theorem}[Soundness]
    \label{theorem:vs-soundness}
    All rules in the type system are valid lemmas in \thelogic{}.
\end{theorem}
We remark that the soundness proofs of all the rules are carried out entirely in \thelogic, without ever needing to drop down to the level of the model.

In the following, we show the key advantage of semantic typing: we type in the logic some components that are syntactically ill-typed. This allows using them in syntactic type derivations just like any other syntactically typed component.

\begin{example}[Semantic typing]
    We start from the known example of a syntactic implicit flow that does not exist semantically. Namely, we prove the following judgment valid in \thelogic{}.
    \[
    \VSJ{\low}|-{\code{if h then l := l+1 else l := l+1}}
    \]

    For a fixed $\secl$, the precondition of the type judgment is $\bigl( \auth{t} * I^t * \sure{\obsSecl^t(\secbelow{\secl}) = \obsL} * \own{\restr{\logs}{t}} \bigr)$. Thanks to the state-log consistency invariant, all variables are a function of the $\logs$, so we can use \cref{rule:oplus-own-merge} to condition on \p{h}. With slight modifications to the proof in
\cref{sec:key-ideas:verify:indirect-flow},
    we re-establish state-log consistency $I^{t'}$ in the post. The \condition[ing] on $\lastOf^t(\secnotbelow{\secl})$ follows from the \condition[ing] on \p{h}.
    The last important bit is to define $\Slow$. If $\low = \seclabelof(\p{l}) \leq \secl$ then $\p{l} = \lastOf(\p{l})(\obsL)$, so we can define $\Slow = \set{ \obsL \app (\lastOf(\p{l})(\obsL) + 1) }$. Else $\Slow = \set{ \obsL }$. The only update happens on \p{l}, so the final possible values for $\obsSecl(\secbelow{\secl})$ only depend on the initial $\obsL$.

    The proof is almost identical for the program that assigns to \p{l} a nondeterministic value from $V$:
    \[
    \VSJ{\low}|-{\code{if h then l := rnd($V$) else l := rnd($V$)}}
    \]
    The only notable difference is that in the postcondition we obtain $\In{\p{l}}{V}$, which leads to the definition $\Slow = \set{ \obsL \app v | v \in V }$ if $\low \leq \secl$.
    One can imagine that this code is modelling some protocol where a nonce is sampled, used during the protocol and then discarded (\ie released). Such a protocol can be typed in the logic, and then used in typing a whole program.

    As a last example, we consider the program \code{v:=rnd(\{0,1\}); l := (h + v) mod 2} from \cref{fig:running-ex}. The typing for this program can be established essentially following the pattern described in \cref{sec:key-ideas:verif}. There, we eventually proved the postcondition $\In{\p{h}}{H} * \In{\code{(h+v) mod 2}}{\set{0,1}}$.
    Letting $\Slow = \set{ \obsL \app 0, \obsL \app 1 }$ if $\low \leq \secl$ yields validity of the judgment
    $
    \VSJ{\low}|-{\code{v:=rnd(\{0,1\}); l := (h + v) mod 2}}.
    $
\end{example}
 \section{Related Work}

\paragraph{Definitions of Information Security}
Definitions of (perfect) information security as independence or
segregation of low outputs from high information date back to the
1980s. Notably, the specific formulation of security as independence
that we use was first presented by \citet{sutherland1986} and was
later named \emph{non-deducibility} by the community. Related
definitions of information security are the immensely popular
\emph{noninterference}~\cite{DBLP:journals/jcs/VolpanoIS96,DBLP:conf/sp/GoguenM82a},
which says that two runs that differ only in high inputs must produce
equal low outputs, \emph{knowledge-based} definitions first introduced
by \citet{DBLP:conf/sp/AskarovS07}, which represent the same concept
in terms of adversary knowledge of high state and are used widely now
(e.g., \cite{DBLP:conf/sp/AskarovS07,DBLP:conf/popl/BrobergS10}), and
their many instantiations using logical relations, bisimulations and
partial-equivalence relations
(PERs)~\cite{DBLP:conf/popl/HeintzeR98,DBLP:conf/csfw/Rajani018,DBLP:journals/pacmpl/MenzHLG23,DBLP:journals/toplas/NanevskiBG13,DBLP:journals/lisp/SabelfeldS01,DBLP:conf/csfw/Mantel00,DBLP:conf/esorics/Mantel00,DBLP:conf/isss2/SabelfeldM03,DBLP:journals/jcs/VolpanoIS96,DBLP:conf/ccs/ChongM04,DBLP:conf/ccs/CecchettiMA17,DBLP:conf/popl/PottierS02,DBLP:conf/ifip1-7/Boudol08,DBLP:conf/icfp/BuirasVR15}. While
the three definitions coincide on deterministic programs,
noninterference as defined above ceases to be a useful definition on
nondeterministic programs as the low outputs of a nondeterministic
program may not agree even in pairs of runs starting from the exact
same state (the other two definitions remain relevant). An adaptation
of noninterference with a third mediating run, called generalized
noninterference, applies to nondeterministic
programs~\cite{McCullough88}.

Declassification can be specified in many different ways, each
necessitating a different security definition and different
verification rules. \citet{DBLP:journals/jcs/SabelfeldS09} classify
declassification policies by ``what'' can be declassified, ``where'' in
the program it can be declassified and ``when'' (under which dynamic
conditions) it can be declassified.  Literature abounds in security
definitions and verification techniques for these declassification
classes~\cite{DBLP:conf/isss2/SabelfeldM03,DBLP:conf/pldi/AskarovS07,DBLP:conf/sp/AskarovS07,DBLP:conf/popl/BrobergS10,DBLP:conf/esop/BrobergS06,DBLP:conf/ccs/ChongM04,DBLP:conf/csfw/MyersSZ04,DBLP:conf/ccs/CecchettiMA17,DBLP:journals/pacmpl/MenzHLG23,DBLP:conf/ifip1-7/Boudol08}.

Our work argues that conditional independence over possibility spaces
is a foundational primitive that can be used to represent and verify
all the notions of perfect security and several notions of declassification
security in both the deterministic and nondeterministic settings.

\paragraph{Verification of Information Security}
Static techniques for verifying information security span type
systems, relational program logics, traditional static analyses and
hyperlogics~\cite{DBLP:journals/mscs/BartheDR11,DBLP:journals/pacmpl/LambaTBBBL24,DBLP:journals/jcs/VolpanoIS96,DBLP:conf/isss2/SabelfeldM03,DBLP:conf/pldi/AskarovS07,DBLP:conf/csfw/Rajani018,DBLP:conf/popl/Myers99,DBLP:conf/ccs/ChongM04,DBLP:conf/ccs/CecchettiMA17,DBLP:conf/popl/AbadiBHR99,DBLP:conf/popl/PottierS02,DBLP:journals/pacmpl/MenzHLG23,DBLP:conf/ifip1-7/Boudol08,DBLP:conf/icfp/BuirasVR15,DBLP:conf/popl/HuntS06,DBLP:journals/scp/AmtoftB07,GregersenBTB21,DBLP:journals/toplas/NanevskiBG13,DBLP:conf/sas/AmtoftB04,DBLP:conf/popl/AssafNSTT17,DBLP:journals/sLogica/GradelV13,DBLP:conf/mfcs/KrebsMV018,DBLP:conf/fsttcs/VirtemaHFK021,DBLP:journals/iandc/KontinenSV25,hammer09ijis,DBLP:conf/sac/MastroeniP19}.
We compare to the most closely related of these lines of work below, but
note upfront that we are unaware of any other work that internalizes
independence into the reasoning framework as a connective to directly
establish (conditional) non-deducibility for security as we do. Most
work targets noninterference or definitions derived from it.

\paragraph{Program Logics for Information Security}
Program logics for information flow (IFC) security trace their lineage to
self-composition, the idea of encoding two runs of a program by
duplicating variables and using standard (unary) Hoare logic to
establish information flow
properties~\cite{DBLP:journals/mscs/BartheDR11}. Subsequent work in
this area has been largely based on relational program logics,
pioneered by RHL~\cite{Benton04}.  \citet{DBLP:journals/scp/AmtoftB07}
share similarities with our work in that they propose a relational Hoare
logic for IFC with a predicate that specifies the independence of two
variables. However, they do not internalize independence as a
connective in the logic and do not consider
separation. \citet{DBLP:conf/sp/BanerjeeNR08} present a relational
Hoare logic with an ``agreement'' predicate to reason about
information security with declassification. This work influenced
\citet{DBLP:journals/toplas/NanevskiBG13}, who present relational
Hoare Type Theory, a monadic dependent type theory for relational
reasoning in higher-order programs, which can also establish
information flow properties (among other examples). Neither work
considers separation or independence as internalized
connectives. \citet{GregersenBTB21} show how to carry out proofs of
information security, specifically, those targeting TINI, in terms of
relational separation logic specifications in Iris, enabling semantic
typing. They rely on separation, but only of heaps; they do not
consider separation of information as we do. \citet{ErnstM19}
represent a related security property, observational determinism, in
SecCSL, a concurrent separation logic where standard separation logic
assertions are annotated with security levels and given relational
meaning but, again, separation is limited to heaps.

\paragraph{Program Logics over Sets of States}
\citet{ZilbersteinSS24} proposed Outcome Logic as a Hoare-like logic
able to unify over/under-approximating specifications by moving to
assertions on \emph{outcomes}, which in the case of nondeterministic
computation are sets of states.  The main new connective of the logic
is the outcome conjunction, internalizing union of sets in the
assertion language, similarly to our \condition[ing] modality.
\citet{DardinierM24} make a similar move by defining a Hoare-like
logic over the whole set of possible states.  The logic can be used to
represent properties that include over/under-approximate triples, but
also hyperproperties.  Later, both logics have been extended to
support heaps via the integration of separation, resulting in Outcome
Separation Logic~\cite{ZilbersteinSS24} and in Hyper Separation
Logic~\cite{gospodinov2026hyper}.  None of these proposals internalize
(conditional) independence with connectives in the logic, and thus
they cannot express any of our reasoning principles.

Another relevant direction is separation logics for probabilistic
programs.  PSL~\cite{BartheHL20} pioneered the idea of internalizing
\emph{probabilistic independence} as separating conjunction, an idea
later refined by Lilac~\cite{Li0H23} which added a \emph{conditioning
  modality} to express conditional independence.
\Bluebell~\cite{DBLP:journals/pacmpl/BaoDF25} built on Lilac's
independence model
and proposed a different modality for conditioning,
which enjoys a number of extremely effective reasoning principles.
\Bluebell's model is the main inspiration behind \ourlogic's connectives.
In particular, the rules that govern \condition[ing] in \ourlogic\
are modelled after \Bluebell's conditioning rules.
\Bluebell's connectives were also shown to be adequate to explain, extend, and make interoperable unary principles (like conditional independence) and relational ones (like probabilistic couplings) in a uniform logic.
The uniform treatment of unary and relational reasoning
is obtained in \Bluebell\ by adopting the approach of LHC~\cite{DOsualdoFD22},
which is a Hoare logic for proving hypersafety properties
for a non-probabilistic language with nondeterminism.
It would be interesting to explore what new reasoning techniques for IFC
would be enabled by making \ourlogic\ relational in the LHC style.
DIBI~\cite{BaoDH021} extended PSL with a connective to express conditioning (in a less expressive way than the more recent modalities).
Notably, the DIBI logic has a possibilistic instantiation, which the authors
linked to notions from relational algebra and databases (but not to IFC).
The assertions and rules of \ourlogic\ are strictly more expressive than possibilistic DIBI.
\citet{DBLP:journals/pacmpl/ZilbersteinST26} proposed \textsc{pcOL}, a PSL-style
logic for concurrent probabilistic programs with conditioning; \textsc{pcOL} proposed precision as a suitable side condition
for the probabilistic version of \ref{rule:oplus-unframe}.
None of these logics linked the use of conditional independence to IFC.

\citet{DBLP:conf/popl/AssafNSTT17} define a logic over a
hypercollecting semantics to validate static analyses for IFC. Like
{\ourlogic}, the subjects of predicates in their logic are sets, but
they use the logic to establish refinement properties, not conditional
independence. Separation does not arise in their setting.

\paragraph{Team Semantics and Hyperproperties}
Our assertions are naturally suited to expressing \emph{hyperproperties}~\cite{DBLP:journals/jcs/ClarksonS10}, as illustrated by the triples we present for \TINI\ and its extension with declassification.
Within the landscape of logics for hyperproperties,
our use of independence is closest in spirit to
\emph{team semantics}~\cite{DBLP:journals/sLogica/GradelV13},
where formulas are evaluated over sets of states
and (conditional) independence is available as
a primitive atomic proposition~$\vec{x} \mathrel{\bot_{\vec{z}}} \vec{y}$,
and in particular to the recent \emph{temporal team logics},
which use teams of traces as specification languages
for hyperproperties~\cite{DBLP:conf/mfcs/KrebsMV018,DBLP:conf/fsttcs/VirtemaHFK021,DBLP:journals/iandc/KontinenSV25}.
This line of work focuses on studying the expressivity and the complexity of the satisfiability and the model checking problems.
To the best of our knowledge, team semantics has not been used as a reasoning primitive for program-level information-flow control, and richer noninterference conditions such as \TINI, or declassification, have not been explicitly addressed in this setting.
By contrast, we internalize independence as a \emph{connective} between assertions and express conditional independence as the combination
of the \condition[ing] modality and separation.
Moreover, we build a deductive, Hoare-style program logic out of the connectives, expressive enough to verify \TINI\ with declassification.

\begin{acks}
This work was partially supported by the Austrian Science Fund through the SFB project SpyCoDe 10.55776/F85.
\end{acks}

\setlabel{LAST}
\label{paper-last-page}

\section*{Data Availability Statement}

All the results in the paper are mechanized in Rocq and available as \citet{artifact}.
The development only relies on standard axioms: informative excluded middle, functional choice, functional and propositional extensionality plus the classical axioms of Stdlib's Reals (for permissions).
The few deviations from the paper are cosmetic and made to streamline the Rocq development. They are fully documented in the artifact's README file, which includes a table indicating which Rocq identifier corresponds to which Definition/Theorem in this paper.

\section*{Disclosure of Use of AI in Research}
The conception of this paper and a first thorough draft of this paper,
including all main results and their proof sketches,
was prepared \emph{without} AI assistance.
Claude Code (Anthropic), was used extensively
in the Rocq mechanization of this paper.
We prevalently used the Opus 5/5.5 and Fable 5 models.
After a first complete version of the paper had been drafted,
Claude was tasked with an initial mechanization of the meta theory.
After several iterations with extensive human direction,
the logic was mechanized and proven correct.
Then examples (including the type system) were carried out by Claude
following the outlines of the original paper draft.
The whole development underwent extensive review and refinements
by the authors with the aim of ensuring the top-level results
and the structure of the proofs matched the original paper versions
(modulo few minor imprecisions that emerged during the mechanization effort).
In particular, the adequacy theorems (\eg for \TINI) were exercised on the examples to show that the logic can be used to establish end-to-end verification results (and thus the logic is not in the TCB for those results).
The artifact includes a mapping between the results in this paper and their mechanized version to facilitate external audits.

\appendix

Below we provide the proofs of the most interesting statements. The remaining proofs can be found in the Rocq mechanization~\citep{artifact}.

\allowdisplaybreaks
\section{Programs and Semantics}
\label{app:programs}

We define $\pvar \from \AExpr \to \powset(\VarLog)$ to be the set of program variables read by the semantic function $ \Sem[E]{\aexpr}$. The definition is standard, but we remark that it can also include the observation log, if the expression reads from it.

Evaluation contexts are defined as
\begin{grammar}
    \Ectxt[\hole] \is
      \p{x := } [\hole]
    | \code{log!($[\hole]$)}
    | \code{if $\;[\hole]\;$ then $\;\prog_1\;$ else $\;\prog_2$}
    | \code{repeat $\;[\hole]$: $\;\prog_1$}
\end{grammar}

\begin{proposition}\label{lemma:prog-semantics-subst}
    $\sem{\Ectxt[\expr]}(\state) = \sem{\Ectxt[\expr(\state)]}(\state)$
\end{proposition}
\begin{proof}
    By cases on the evaluation context.
\end{proof}
 \section{Possibility Spaces and Assertions}
\label{app:model}

\subsection{Possibility Spaces}
The following lemmas describe facts about the structure of possibility spaces in relation to the RA structure.

\begin{lemma}\label{lemma:set-states-expressible}
	Let $S \subs \State$ and $\pspA \is \set{ \set{ \state } | \state \in S }$. Then, for every $B \subs \State$, $B$ is expressible in $\pspA$: $B \ein \pspA$.
\end{lemma}

\begin{corollary}\label{lemma:pleq-psp-set}
	Let $S\subs\State$ and $\pspA \in \PSp$. $\pspA \pLeq S$ if and only if
	\begin{enumerate}[(i)]
		\item $S \subs \Union \pspA$
		\item $\All \evA \in \pspA. \evA \inters S \ne \emptyset$
	\end{enumerate}
\end{corollary}

\begin{corollary}\label{lemma:pleq-psp-set-union}
    Let $\pspA \in \PSp$. Then $\pspA \pLeq \Union \pspA$.
\end{corollary}

\begin{lemma}\label{lemma:pleq-union-psp}
    Let $\pspA \in \PSp$. Then $\set{ \Union \pspA } \pLeq \pspA$.
\end{lemma}

\begin{lemma}\label{lemma:pleq-events-contained}
	Let $\pspA,\pspB \in \PSp$ such that $\pspA \pLeq \pspB$. Then $\All \evB \in \pspB . \Exists \evA \in \pspA . \evB \subs \evA$.
\end{lemma}

\begin{lemma}\label{lemma:union-iprod}
	Let $\pspA,\pspB \in \PSp$ such that $\pspA \compat \pspB$. Then $\Union (\pspA \iprod \pspB) = (\Union \pspA) \inters (\Union \pspB)$.
\end{lemma}
\begin{proof}
	By definition, $\Union (\pspA \iprod \pspB) = \Union \set{ A \inters B| A \in \pspA, B \in \pspB }$. Fix a state $\state \in \State$.
	$\state \in \Union \set{ A \inters B| A \in \pspA, B \in \pspB }$ if and only if there exists $A \in \pspA, B \in \pspB$ such that $\state \in A \inters B$.
	That is, if and only if there exists $A \in \pspA$ such that $\state \in A$ and there exists $B \in \pspB$ such that $\state \in B$.
	That is, if and only if $\state \in \Union \set{ A | A \in \pspA }$ and $\state \in \Union \set{ B | B \in \pspB }$.
	That is, if and only if $\state \in (\Union \pspA) \inters (\Union \pspB)$.
\end{proof}

\begin{lemma}\label{lemma:union-partitioning-monotone}
    Let $\set{ \pspA_v }_{v \in X}$ and $\set{ \pspB_v }_{v \in X}$ be two families of possibility spaces. Suppose that $\set{ \pspA_v }_{v \in X}$ is disjoint and for every $v$ we have $\pspA_v \pLeq \pspB_v$. Then $\set{ \pspB_v }_{v \in X}$ is disjoint and $\Union_{v \in X} \pspA_v \pLeq \Union_{v \in X} \pspB_v$.
\end{lemma}
\begin{proof}
    Fix $v, w \in X$ such that $v \neq w$. $\pspA_v \pLeq \pspB_v$ means $\Union \pspB_v \subseteq \Union \pspA_v$ and analogously for $w$. Since $\set{ \pspA_v }_{v \in X}$ is disjoint, $\Union \pspA_v \cap \Union \pspA_w = \emptyset$. Therefore $\set{ \pspB_v }_{v \in X}$ is disjoint.

    To prove $\Union_{v \in X} \pspA_v \pLeq \Union_{v \in X} \pspB_v$ we have to show three facts.

    (i) $\Union (\Union_{v \in X} \pspB_v) \subseteq \Union (\Union_{v \in X} \pspA_v)$.

    \[
    \Union (\Union_{v \in X} \pspB_v) = \Union_{v \in X} (\Union \pspB_v) \subseteq \Union_{v \in X} (\Union \pspA_v)  = \Union (\Union_{v \in X} \pspA_v)
    \]

    (ii) $\forall A \in (\Union_{v \in X} \pspA_v) . A \ein (\Union_{v \in X} \pspB_v)$.

    Fix $A \in \Union_{v \in X} \pspA_v$. Then there must exist $v \in X$ such that $A \in \pspA_v$. This means $A \ein \pspB_v$, that is, there exists $\pspB \subseteq \pspB_v$ such that $A \cap \Union \pspB_v = \Union \pspB$. But because $\set{ \pspA_v }_{v \in X}$ is disjoint, $A \in \pspA_v$ means that $A \cap \Union \pspB_w \subseteq A \cap \Union \pspA_w = \emptyset$ for every $w \neq v$. Thus, $A \cap \Union (\Union_{v \in X} \pspB_v) = A \cap \Union \pspB_v = \Union \pspB$ and $\pspB \subseteq \pspB_v \subseteq \Union_{v \in X} \pspB_v$.

    (iii) $\forall A \in (\Union_{v \in X} \pspA_v) . A \cap \Union (\Union_{v \in X} \pspB_v) \neq \emptyset$.

    Fix $A \in \Union_{v \in X} \pspA_v$. Then there must exist $v \in X$ such that $A \in \pspA_v$. This means $\emptyset \neq A \cap \Union \pspB_v \subseteq A \cap \Union (\Union_{v \in X} \pspB_v)$.
\end{proof}

\begin{lemma}\label{lemma:union-iprod-id}
    For every $\psp \in \PSpInvalid$ and $\evA \sups \Union \psp$, $\set{ \evA } \iprod \psp = \psp$.
\end{lemma}
\begin{proof}
    Obvious by definition of $\iprod$.
\end{proof}

\begin{corollary}\label{lemma:unit-psp-iprod-id}
    For every $\psp \in \PSpInvalid$, $\set{ \State } \iprod \psp = \psp$.
\end{corollary}

\begin{lemma}\label{lemma:unit-psp-minimal}
    For every $\psp \in \PSpInvalid$, $\set{ \State } \pLeq \psp$.
\end{lemma}
\begin{proof}
    If $\psp = \invalid$ the thesis follows because $\invalid$ is the top element of $\pLeq$. Otherwise, we have to check three conditions.

    (i) $\Union \psp \subseteq \Union \set{ \State }$.

    Trivial.

    (ii) $\forall B \in \set{ \State } . B \ein \psp$.

    $B \in \set{ \State }$ means $B = \State$. $\State$ is expressible in $\psp$ because $\State \cap \Union \psp = \Union \psp$.

    (iii) $\forall B \in \set{ \State } . B \cap \Union \psp \neq \emptyset$.

    $B \in \set{ \State }$ means $B = \State$. Then $B \cap \Union \psp = \Union \psp \neq \emptyset$, where the last inequality is because $\psp$ is a non-empty set of non-empty subsets of $\State$.
\end{proof}

\begin{proof}[Proof of \cref{lemma:psp-ra}]
    We check the five conditions one by one, but in a different order.
    \begin{casesplit}
        \case*[$\raValid(\raUnit)$] Obvious by definition of $\raValid$ because $\raUnit \neq \invalid$.

        \case*[$\raUnit \iprod \psp = \psp$] \Cref{lemma:unit-psp-iprod-id}.

        \case*[$\pspA \pLeq \pspB$ and $\raValid(\pspB)$ implies $\raValid(\pspA)$] Obvious because $\invalid$ is the only invalid element and it is the top of the order $\pLeq$.

        \case*[$\pspA \pLeq \pspB$ implies $\pspA \iprod \pspC \pLeq \pspB \iprod \pspC$]
        If $\pspB \iprod \pspC = \invalid$, then $\pspA \iprod \pspC \pLeq \pspB \iprod \pspC$ because $\invalid$ is the top element of $\pLeq$. So, without loss of generality, we assume $\pspB \iprod \pspC \neq \invalid$, which means $\pspB \compat \pspC$.
        We have to prove three facts.

        (i) From \cref{lemma:union-iprod} and the hypothesis $\pspA \pLeq \pspB$, point (i), we have
        \[
        \Union (\pspB \iprod \pspC) = \Union \pspB \inters \Union \pspC \subs \Union \pspA \inters \Union \pspC = \Union (\pspA \iprod \pspC)
        \]

        (ii) Recall that an event of $\pspA \iprod \pspC$ is $\evA \inters \evC$ for $\evA \in \pspA$, $\evC \in \pspC$. Fix an arbitrary event of $\pspA \iprod \pspC$, that is fix $\evA \in \pspA$, $\evC \in \pspC$. Then by the hypothesis $\pspA \pLeq \pspB$, point (ii), we have $\evA \ein \pspB$, which means there exists a $\pspB' \subs \pspB$ such that $\evA \inters \Union \pspB = \Union \pspB'$. Then
        \begin{align*}
            &(\evA \inters \evC) \inters \Union (\pspB \iprod \pspC) \\
            =\; &\evA \inters \evC \inters \Union \pspB \inters \Union \pspC \\
            =\; &(\evA \inters \Union \pspB) \inters (\evC \inters \Union \pspC) \\
            =\; &\Union \pspB' \inters \evC \\
            =\; &\Union_{\evB \in \pspB'} (\evB \inters \evC)
        \end{align*}
        So the set $\set{ \evB \inters \evC | \evB \in \pspB' } \subs \pspB \iprod \pspC$ witnesses $\evA \inters \evC \ein \pspB \iprod \pspC$.

        (iii) Fix an arbitrary event of $\evA \inters \evC \in \pspA \iprod \pspC$. Then by the hypothesis $\pspA \pLeq \pspB$, point (iii), $\evA \inters \Union \pspB \neq \emptyset$. This means there exists $\evB \in \pspB$ such that $\evA \inters \evB \neq \emptyset$. By \cref{lemma:pleq-events-contained} (and the fact that events in $\pspA$ are disjoint) this means $\evB \subs \evA$. Therefore $\evB \inters \evC \subs \evA \inters \evC$, and because $\pspB \compat \pspC$ we have $\evB \inters \evC \neq \emptyset$.

        \case*[$\raValid(\pspA \iprod \pspB)$ implies $\raValid(\pspA)$] From \cref{lemma:unit-psp-minimal} $\raUnit \pLeq \pspB$. From the point above $\raUnit \iprod \pspA \pLeq \pspB \iprod \pspA$. From the point two above, $\raValid(\pspA \iprod \pspB)$ implies $\raValid(\pspA \iprod \raUnit) = \raValid(\pspA)$.
    \end{casesplit}
\end{proof}

\subsection{Permissions and $\PSpPm$}
Lemmas about the resource algebra $\PSpPm$.

The intuitive meaning of the compatibility order $\preccurlyeq_{\permap}$ defined in \cref{def:permission-compatibility} is as follows.
If $s \preccurlyeq_{\permap} s'$ then $s'$ must be indistinguishable from $s$ in any compatible possibility space. For instance, if $\permap(\p{x}) = 0$, any state $s'$ that only differs from $s$ in the value of $\p{x}$ is greater than or equal according to $\preccurlyeq_{\permap}$, and therefore must be indistinguishable from $s$: any possibility space compatible with $\permap$ cannot predicate on the value of $\p{x}$. More constructively:
\begin{proposition}\label{lemma:compatibility-constructive}
    A possibility space $\pspA \in \PSp$ and a permission map $\permap \in \Perm$ are compatible if and only if, for every $\evA \in \pspA$:
    \begin{enumerate}
        \item for every $\p{x} \in \Var$, if $\permap(\p{x}) = 0$ then $\forall (\store, \obs) \in \evA, \val \in \Val . (\store\upd{\p{x} -> \val}, \obs) \in \evA$;
        and
        \item if $\permap(\logs) = 0$ then $\forall (\store, \obs) \in \evA, \obs' \in \Obs . (\store, \obs \app \obs') \in \evA$.
    \end{enumerate}
\end{proposition}

\begin{lemma}\label{lemma:unit-minimal}
	For every $a \in \PSpPm$, $\raUnit \raLeq a$.
\end{lemma}
\begin{proof}
    Fix $a = (\psp, \permap) \in \PSpPm$. We want to show that $\raUnit \raLeq (\psp, \permap)$, that is
    \[
    \set{\State} \pLeq \psp \qquad \funconst 0 \leq \permap
    \]
    The latter is trivially true. The former is \cref{lemma:unit-psp-minimal}.
\end{proof}

\begin{proof}[Proof of \cref{lemma:psppm-ra}]
    Recall that $\raUnit = (\set{ \State }, \funconst 0)$. Below, when we say that ``$\PSpInvalid$ is a resource algebra'' we mean that $(\PSpInvalid, \pLeq, \raValid, \iprod, \set{\State})$ is an ordered unital resource algebra (\cref{lemma:psp-ra}).
    We check the five conditions one by one, but in a different order.
    \begin{casesplit}
        \case*[$\raValid(\raUnit)$] Obvious by definition of $\raValid$ and $\raUnit$.

        \case*[$\raUnit \raOp a = a$] Obvious by definition of $\raOp$ and $\raUnit$, using that $\PSpInvalid$ is a resource algebra.

        \case*[$a \raLeq b$ and $\raValid(b)$ implies $\raValid(a)$]
        $\raValid(b)$ implies $b = (\pspB, \permaq)$ for some $\pspB$, $\permaq$ such that $\pspB \neq \invalid$ and $\permaq \leq \funconst 1$.
        Then $a \raLeq b$ implies $a = (\pspA, \permap)$ for some $\pspA$, $\permap$ such that $\pspA \pLeq \pspB$ and $\permap \leq \permaq \leq \funconst 1$.
        $\pspA \neq \invalid$ follows from the fact that $\PSpInvalid$ is a resource algebra.

        \case*[$a \raLeq b$ implies $a \raOp c \raLeq b \raOp c$]
        Let $a = (\pspA, \permap)$, $b = (\pspB, \permaq)$, $c = (\pspC, \permar)$.
        By $a \raLeq b$ we have $\pspA \pLeq \pspB$ and $\permap \leq \permaq$.
        The latter implies that $\permap + \permar \leq \permaq + \permar$.
        The former implies $\pspA \iprod \pspC \pLeq \pspB \iprod \pspC$ because $\PSpInvalid$ is a resource algebra.

        \case*[$\raValid(a \raOp b)$ implies $\raValid(a)$] From \cref{lemma:unit-minimal} $\raUnit \raLeq b$. From the point above $\raUnit \raOp a \raLeq b \raOp a$. From the point two above, $\raValid(a \raOp b)$ implies $\raValid(a \raOp \raUnit) = \raValid(a)$.
    \end{casesplit}
\end{proof}
\begin{corollary}\label{lemma:raop-increasing}
    For every $a, b \in \PSpPm$, $a \raLeq a \raOp b$.
\end{corollary}

\begin{proof}[Proof of \cref{lemma:disjoint-union-pspace}]
    The only condition to check is that two different events in $\Dunion_{v\in X} \pspA_v$ are disjoint.
    If the two events come from the same $\pspA_v$ then they are disjoint by the hypothesis that $\pspA_v$ is a possibility space.
    If they come from different $\pspA_v, \pspA_w$ then they are disjoint by disjointness of $\Union \pspA_v$ and $\Union \pspA_w$.
\end{proof}

\subsection{Possibility Spaces and Assertions}
A very common pattern is to \condition on the value of an expression $\expr$. When this happens, expression $\expr$ witnesses the disjointness of the family of possibility spaces $\set{\pspA_v}_{v\in X}$. At the assertion level, we describe this situation with the following definition.

\begin{definition}[Partitioning assertions]\label{def:partitioning-assertions}
    Let $P \from X \to \Assrt$ be a family of assertions.
    We say that $P$ \emph{partitions} (equivalently, it is \emph{partitioning}) if there exists an expression $\expr$ such that
    \[
    \forall \val \in X. P(\val) \entails \sure{\expr = \val}
    \]
    We say that $\expr$ is the \emph{witness} of the partitioning.
\end{definition}

We define notation $\pure{\phi}$, where $\phi$ is a proposition (\ie, a statement that does not depend on the state) to describe the assertion
\[
(\psp, \permap) \models \pure{\phi} \quad\tIff\quad \phi \text{ holds}
\]
For brevity, we write $\True \is \pure{\True}$ and $\False \is \pure{\False}$.

The following lemmas tell us something about the structure of possibility spaces satisfying certain assertions.

\begin{lemma}\label{lemma:model-in-characterization}
    $\pspA \models \In{\expr}{X}$ if and only if
    \begin{enumerate}
        \item for every $\evA \in \pspA$ there exists $v \in X$ such that $\expr(\evA) = \set{ v }$; and
        \item $\expr(\Union \pspA) = X$.
    \end{enumerate}
\end{lemma}
\begin{proof}
    We prove the two implications separately.
    \begin{casesplit}
        \case[$\Rightarrow$] (2) is part of the definition of $\pspA \models \In{\expr}{X}$. For (1), fix $\evA \in \pspA$ and suppose $\expr(\evA) = \set{ v, w }$ for two distinct values $v \neq w$. Then there exists $\state, \state' \in \evA$ such that $\expr(\state) = v$, $\expr(\state') = w$.
        Because $\inv{\expr}(v) \ein \pspA$, there exists $\pspB \subs \pspA$ such that $\inv{\expr}(v) = \Union \pspB$. But $\state \in \inv{\expr}(v)$ and $\state' \not\in \inv{\expr}(v)$, so $\evA \not\subs \inv{\expr}(v)$ and therefore $\evA \not\in \pspB$. But also $\state \in \inv{\expr}(v) \inters \Union \pspA$, so $\evA \in \pspB$, which is a contradiction.
        \case[$\Leftarrow$] We have to prove two conditions, the second of which is exactly (2). So fix $v \in X$ and let us prove that $\inv{\expr}(v) \ein \pspA$. Let $\pspB = \set{ \evA \in \pspA | \expr(\evA) = \set{ v } }$. By definition of $\inv{\expr}(v)$ we have $\Union \pspB \subs \inv{\expr}(v)$, and obviously $\Union \pspB \subs \Union \pspA$, so $\Union \pspB \subs \inv{\expr}(v) \inters \Union \pspA$. For the other inclusion, fix a state $\state \in \inv{\expr}(v) \inters \Union \pspA$. Then $\expr(\state) = v$ and there exists $\evA \in \pspA$ such that $\state \in \evA$. But this means $\expr(\evA) \sups \set{ v }$, and by hypothesis this implies $\expr(\evA) = \set{ v }$, so $\evA \in \pspB$.
    \end{casesplit}
\end{proof}

\begin{corollary}\label{lemma:model-in-exists-value}
    Let $\pspA \models \In{\expr}{X}$. Then for every $\evA \in \pspA$ there exists $v \in X$ such that $\expr(\evA) = \set{ v }$.
\end{corollary}

\begin{corollary}\label{lemma:model-e-equal-v}
	$\pspA \models \sure{\expr = v}$ if and only if \ $\All \state \in \Union \pspA. \expr(\state) = v$.
\end{corollary}

\begin{proposition}\label{lemma:model-cond-partition}
	Let $D$ be a set of values, $P \from D \to \Assrt$ a $D$-indexed family of assertions such that $P$ partitions with witness $\expr$.
	Let $\pspA \in \PSp$ be such that $\pspA \models \cond{v}{D}{P(v)}$, and let $\set{ \pspA_v }_{v \in D}$ be the family of possibility spaces witnessing that $\pspA \models \cond{v}{D}{P(v)}$, and define $\restr{\pspA}{\expr = v} \is \set{ \evA \in \pspA | \expr(\evA) = \set{ v } }$.

	Then:
	\begin{propenum}
\item \label{lemma:model-cond-partition:split-partitions-states} for any state $\state \in \Union \pspA$, $\state \in \Union \restr{\pspA}{\expr = \expr(\state)}$;
		\item\label{lemma:model-cond-partition:split-value} for all $v$, $\pspA_v \pLeq \restr{\pspA}{\expr = v}$.
	\end{propenum}
\end{proposition}

\begin{proposition}\label{lemma:p-partitions-disjoint}
    Let $P \from X \to \Assrt$ be a family of assertions, $\permap \from X \to\Perm$ a family of permission maps, and $\set{\pspA_\val}_{\val \in X}$ a family of possibility spaces. If $P$ partitions with witness $\expr$ and $\forall \val \in X . (\pspA_\val, \permap_\val) \models P(\val)$, then $\set{\pspA_\val}_{\val \in X}$ is disjoint with witness $\expr$.
\end{proposition}

\begin{lemma}\label{lemma:model-own-characterization}
    $\pspA \models \own{\expr}$ if and only if $\All A \in \pspA. \expr(A) \text{ is a singleton}$.
\end{lemma}

\begin{lemma}\label{lemma:wp-state-subset-model}
    Let $S \subseteq \State$ and $a = (\pspA, \permap), f = (\pspF, \permar) \in \PSpPm$ be such that $a \raOp f \raLeq S$. Then
    \[
    S \subs \Union \pspA \qquad \text{and} \qquad S \subs \Union \pspF
    \]
\end{lemma}
\begin{proof}
    Corollary of \cref{lemma:union-iprod}.
\end{proof}

\begin{definition}[Permission reduction]
    Let $P : \Assrt$, $\p{x} \in \VarLog$ and $n \in \Nat$. We define the assertion $\halveperm(P, \p{x}) : \Assrt$ as
    \[
    (\pspA, \permap) \models \halveperm(P, \p{x}) \;\tIff\quad \Exists \epsilon \leq 2 \cdot \permap(\p{x}). (\pspA, \permap\upd{\p{x} -> \epsilon}) \models P
    \]
    We say $P$ has scalable permission on \p{x} if $P \lequiv \halveperm(P, \p{x})$.
\end{definition}

Intuitively, if $(\pspA, \permap) \models P$ then $\pspA$ with half the permission on \p{x} is a model of $\halveperm(P, \p{x})$.
Roughly, an assertion has scalable permission on \p{x} if it does not contain $\withperm{\p{x} : q}$: for a possibility space to be compatible with a permission map $\permap$ it is only important whether $\permap(\p{x})$ is $0$, not its exact value.
For instance, $\sure{x = 0}$ has scalable permissions on \p{x}, but $\sure{x = 0}\withperm{\p{x} : \onehalf}$ does not.

\begin{lemma}\label{lemma:model-sure-and-to-star}
    For every possibility space $\pspA$ and assertion $P$ that does not depend on permissions (\ie if two resources $a$, $b$ have the same possibility space then $a \models P$ if and only if $b \models P$), we have that
    \[
    \pspA \models \sure{\expr = v} * P \qquad\tIff\qquad \pspA \models \sure{\expr = v} \land P
    \]
\end{lemma}

 \section{Soundness}
\label{app:rules}

\begin{figure}[t]
    \begin{proofrules}
        \infer*[lab={precise-in}]{
    \forall \p{x} \in \pvar(\expr) \st \permap(\p{x}) > 0
}{
    (\In{\expr}{X})\withp{\permap} : \precise
}
         \label{rule:precise-in}

        \infer*[lab={precise-perm}]{
    P : \precise
}{
    P\withp{\permap} : \precise
}
         \label{rule:precise-perm}

        \infer*[lab={precise-sure}]{
    \forall \p{x} \in \pvar(\bexpr) \st \permap(\p{x}) > 0
}{
    \sure{\bexpr}\withp{\permap} : \precise
}
         \label{rule:precise-sure}

        \infer*[lab={precise-star}]{
    P, Q : \precise
}{
    P * Q : \precise
}
         \label{rule:precise-star}

        \infer*[lab={precise-$\oplus$}]{
    \forall \val \in X . P(\val) : \precise \\
    \partitions[\expr]{\val}{X}{P}
}{
    \cond{\val}{X}{P(\val)} : \precise
}
         \label{rule:precise-oplus}
    \end{proofrules}
    \caption{Rules to prove assertions precise.}
    \label{fig:app:precise-rules-all}
\end{figure}

\begin{figure}[t]
    \begin{proofrules}
        \infer*[lab=sure-and]{}{
    \sure{\bexpr_1} * \sure{\bexpr_2}
    \lequiv
    \sure{\bexpr_1 \land \bexpr_2}
}
         \label{rule:sure-and}

        \infer*[lab=sure-persistent]{}{
    \sure{\bexpr}
    \lequiv
    \sure{\bexpr} * \sure{\bexpr}
}
         \label{rule:sure-persistent}

        \infer*[lab=sure-and-to-star]{
    \All \p{x} \in \pvar(\bexpr) . P \lequiv \halveperm(P, \p{x})
}{
    \sure{\bexpr} \land P
    \entails
    \sure{\bexpr} * P
}
         \label{rule:sure-and-to-star}
    \end{proofrules}

    \caption{\thelogic{} rules not about $\condmod$ or $\wpsymb$.}
    \label{fig:app:rest-rules}
\end{figure}

\begin{figure}[t]
    \rulesection*{Primitive rules about $\condmod$}
    \begin{proofrules}
        \infer*[lab=$\oplus$-unit-r]{}{
  \In{\expr}{X}
  \lequiv
  \cond{\val}{X} \sure{\expr=\val}
}

        \infer*[lab=$\oplus$-unit-l]{}{
    P(\val_0)
    \lequiv
    \cond{\val}{\set{ \val_0 }} P(\val)
}

        \infer*[lab={$\oplus$-cons}]{
    \All v \in X. P(\val) \entails Q(\val)
}{
    \cond{\val}{X}{P(\val)}
    \entails
    \cond{\val}{X}{Q(\val)}
}

        \infer*[lab={$\oplus$-fuse}]{
    Y_{\hole} \from X \to \powset(D)
    \qquad Y = \Union_{\val \in X} (\set{ \val } \times Y_\val)
}{
    \cond{\val}{X} \cond{w}{Y_\val} P(v, w)
    \entails
    \cond{(v, w)}{Y} P(v, w)
}
         \label{rule:oplus-fuse}

        \infer*[lab={$\oplus$-unfuse}]{
    Y_{\hole} \from X \to \powset(D)
    \\
    Y = \Union_{\val \in X} Y_\val
    \\
    \forall v, w \in X \st v \neq w \implies Y_v \inters Y_w = \emptyset
    \\
    \forall v \in X \st Y_v \neq \emptyset
}{
    \cond{w}{Y} P(w)
    \entails
    \cond{\val}{X} \cond{w}{Y_\val} P(w)
}
         \label{rule:oplus-unfuse}

        \infer*[lab={$\oplus$-skolem}]{}{
    \cond{\val}{X}{\Exists y \in Y . P(\val, y)}
    \lequiv
    \Exists f \from X \to Y . \cond{\val}{X}{P(\val, f(\val))}
}
         \label{rule:oplus-skolem}

        \infer*[lab={$\oplus$-witness}]{}{
    \cond{\val}{X} P(\val)
    \entails
    \Exists \expr . \cond{\val}{X} (\sure{\expr = \val} \land P(\val))
}
         \label{rule:oplus-witness}

        \infer*[lab={$\oplus$-frame}]{}{
  F * \cond{\val}{X} P(\val)
  \entails
  \cond{\val}{X} (F * P(\val))
}

        \infer*[lab={$\oplus$-unframe}]{
    F : \precise \\
    \partitions[\expr]{\val}{X}{P}
}{
    \cond{v}{X} (F * P(v))
    \entails
    F * \cond{v}{X} P(v)
}

        \infer*[lab={$\oplus$-pure}]{
    \forall \val \in X \st \phi(\val)
}{
    \cond{\val}{X}{Q(\val)}
    \entails
    \cond{\val}{X}{\bigl( \pure{\phi(\val)} * Q(\val) \bigr)}
}
         \label{rule:oplus-pure}
    \end{proofrules}

    \rulesection*{Derived rules about $\condmod$}
    \begin{proofrules}
        \infer*[lab={$\oplus$-own-merge}]{
    \expr = f(\expr')
}{
    \vcond{\expr}{\own{\expr'}}
    \lequiv
    \own{\expr'}
}

        \infer*[lab={$\oplus$-transf}]{
    f \from Y \to X \text{ bijective}
}{
    \cond{v}{X}{ P(v) }
    \entails
    \cond{w}{Y}{ P(f(w)) }
}

\end{proofrules}

    \caption{\thelogic{} rules about $\condmod$.}
    \label{fig:app:oplus-rules-all}
\end{figure}

\begin{figure}[t]
    \rulesection*{Structural rules about $\wpsymb$}
    \begin{proofrules}
        \infer*[lab={wp-cons}]{
    Q \entails Q'
}{
    \WP{\prog}{Q}
    \entails
    \WP{\prog}{Q'}
}
         \label{rule:wp-cons}

        \infer*[lab={wp-exist}]{}{
    \exists x \of X . \WP{\prog}{Q(x)}
    \entails
    \WP{\prog}{\exists x \of X . Q(x)}
}
         \label{rule:wp-exist}

        \infer*[lab={wp-frame}]{}{
	P * \WP{\prog}{Q}
	\entails
	\WP{\prog}{P * Q}
}
         \label{rule:wp-frame}

        \infer*[lab=wp-bind]{
    P \entails \sure{\expr = \val} \\
    P \entails \WP{[\Ectxt[\val]]}{Q}
}{
    P \entails \WP{[\Ectxt[\expr]]}{Q}
}
         \label{rule:wp-bind}

        \infer*[lab={$\oplus$-wp-swap}]{
    \mayterm{\prog}
}{
	\cond{v}{X} (\sure{\expr = v}@\permaq * \WP{\prog}{Q(v)})
	\entails
	\WP{\prog}{\cond{v}{X} (\sure{\expr = v}@\permaq * Q(v))}
}
         \label{rule:oplus-wp-swap}

        \infer*[lab={$\oplus$-wp-swap-weak}]{}{
    \cond{v}{X} (\sure{\expr = v}@\permaq * \WP{\prog}{Q(v)})
    \entails
    \WP{\prog}{\Exists X' . \pure{\emptyset \subsetneq X' \subseteq X} * \cond{v}{X'} (\sure{\expr = v}@\permaq * Q(v))}
}
         \label{rule:oplus-wp-swap-weak}

        \infer*[lab=wp-sure-and]{}{
    \WP{\prog}{\sure{\bexpr}} \land \WP{\prog}{Q} \entails \WP{\prog}{\sure{\bexpr} \land Q}
}
         \label{rule:wp-sure-and}
    \end{proofrules}
    \caption{Primitive structural rules about $\wpsymb$.}
    \label{fig:app:wp-rules-struct-prim}
\end{figure}

\begin{figure}[t]
    \rulesection*{Program rules about $\wpsymb$}
    \begin{proofrules}
        \infer*[lab={wp-skip}]{}{
    P
    \entails
    \WP{[\code{skip}]}{P}
}
         \label{rule:wp-skip}

        \infer*[lab={wp-assign}]{
	\forall \p{y} \in \pvar(\aexpr) . \permap(\p{y}) > 0
	\\
	\permap(\p{x}) = 1
}{
	\sure{\varphi\upd{\p{x}->\aexpr}}\withp{\permap}
\entails
	\WP{[\code{x:=$\aexpr$}]}
		 {\sure{\varphi}\withp{\permap}}
}
         \label{rule:wp-assign}

        \infer*[lab={wp-rnd}]{
}{
    \perm{\p{x} : 1}
    \entails
    \WP{[\code{x:=rnd($V$)}]}{\In{\p{x}}{V}\withperm{\p{x} : 1} }
}
         \label{rule:wp-rnd}

        \infer*[lab={wp-log}]{
	\forall \p{y} \in \pvar(\aexpr) . \permap(\p{y}) > 0
    \\
    \permap(\log) = 0
    \\
    \log \notin \pvar(\aexpr)
}{
	\auth{t}\withp{\permap}
	\entails
	\WP{[\code{log!($\aexpr$)}]}{\auth{(t+1)} * ((t+1) \obspointsto \aexpr)\withp{\permap}}
}
        \label{rule:wp-log}

        \infer*[lab={wp-seq}]{}{
    \WP{[\prog_1]}{ \WP{[\prog_2]}{Q} }
    \entails
    \WP{[\prog_1\p;\prog_2]}{Q}
}
         \label{rule:wp-seq}

        \infer*[lab={wp-if-prim}]{}{
    \text{if $v$  then } \WP{[\prog_1]}{Q(1)} \text{ else } \WP{[\prog_0]}{Q(0)}
    \entails
    \WP{[\code{if $\;v\;$ then $\;\prog_1\;$ else $\;\prog_0$}]}{Q(v)}
}
         \label{rule:wp-if-prim}

        \infer*[lab={wp-repeat}]{
    \All k < n . P(k) \entails \WP{\prog}{ P(k+1) } \\
    \sure{\aexpr = n}\withp{\permaq} * R \entails P(0)
}{
    \sure{\aexpr = n}\withp{\permaq} * R \entails
    \WP{[\code{repeat $\;\aexpr\;$ : $\;\prog$}]}{ P(n) }
}
         \label{rule:wp-repeat}

\infer*[lab={wp-while}]{
    \All n, i . P_i(n) \entails \sure{\bexpr = i} \\
    \All n . P_{\true}(n) \entails \WP{\prog}{P(n+1) } \\
    \All n . P(n) \entails \Exists X . \cond{i}{X}{P_i(n)} \\
    \partitions{n}{\Nat}{P_{\false}}
}{
    P(0) \entails
    \WP{[\code{while $\;\bexpr\;$ do $\;\prog$}]}{{ \Exists N . \cond{n}{N}{P_{\false}(n)} }}
}
        \label{rule:wp-while}
    \end{proofrules}

    \caption{Primitive program rules about $\wpsymb$.}
    \label{fig:app:wp-rules-prog-prim}
\end{figure}

\begin{figure}[t]
    \rulesection*{Derived program rules about $\wpsymb$}
    \begin{proofrules}
        \infer*[lab={wp-log-val}]{}{
    \auth{(t-1)}
    \entails
    \WP{[\code{log!($\val$)}]}{\auth{t} * t \obspointsto \val}
}
         \label{rule:wp-log-val}

        \infer*[lab={wp-assign-val}]{}{
    \perm{\p{x} : 1}
    \entails
    \WP{[\code{x:=$\val$}]}{\sure{\p{x} = \val}\withperm{\p{x} : 1}}
}
         \label{rule:wp-assign-val}

        \infer*[lab={wp-if-det}]{
    P * \sure{\bexpr = i} \entails \WP{[\prog_i]}{Q * \sure{\bexpr = i}}
}{
    P * \sure{\bexpr = i}
    \entails
    \WP{[\code{if $\;\bexpr\;$ then $\;\prog_1\;$ else $\;\prog_0$}]}{Q * \sure{\bexpr = i}}
}
         \label{rule:wp-if-det}
    \end{proofrules}

    \caption{Derived rules about $\wpsymb$.}
    \label{fig:app:wp-rules-derived}
\end{figure}

\subsection{Precision}

Soundness of the rules for precision in \cref{fig:app:precise-rules-all}.

\begin{lemma}
    \Cref{rule:precise-in} is sound.
\end{lemma}
\begin{proof}
    Let $\evA_\val \is \set{ \state \in \State | \expr(\state) = \val } = \inv{\expr}(\val)$. Define $\pspA \is \set{ \evA_\val | \val \in X }$ and $a \is (\pspA, \permap)$. We show \begin{enumerate*}
        \item $a \models (\In{\expr}{X})\withp{\permap}$, and
        \item $a$ is the least model of $(\In{\expr}{X})\withp{\permap}$
    \end{enumerate*}.

    (1)
    Observe that, since $\forall \p{x} \in \pvar(\expr) \st \permap(\p{x}) > 0$, $(\pspA, \permap)$ is valid. Therefore, since $\permap \geq \permap$, item (1) follows from $a \models \In{\expr}{X}$.
    First fix $d \in D$. By definition $\inv{\expr}(d) = \evA_d$. If $d \in X$ we have $\inv{\expr}(d) \in \pspA$ so it is expressible. If $d \notin X$ then $\inv{\expr}(d) \cap \Union \pspA = \emptyset$, so it is expressible.
    Then we observe that
    \[
    \expr(\Union \set{ \evA_\val | \val \in X }) = \Union_{\val \in X} \expr(\evA_\val) = \Union_{\val \in X} \set{ \val } = X
    \]

    (2)
    Fix any model $(\pspB, \permaq) \models (\In{\expr}{X})\withp{\permap}$. By definition of $\withp{\permap}$ we have $\permaq \geq \permap$ and $(\pspB, \permaq) \models \In{\expr}{X}$.
    To prove $(\pspA, \permap) \raLeq (\pspB, \permaq)$ we have to show $\pspA \pLeq \pspB$, which amounts to proving three things.

    (i) $\Union \pspB \subseteq \Union \pspA$.

    By definition $\expr (\Union \pspB) = X$, therefore $\Union \pspB \subseteq \inv{\expr}(X) = \Union \pspA$.

    (ii) $\forall A \in \pspA . A \ein \pspB$.

    Fix $A \in \pspA$. Then there exists $\val \in X \subseteq D$ such that $A = \inv{\expr}(\val)$. Thus $A \ein \pspB$ because $(\pspB, \permaq) \models \In{\expr}{X}$.

    (iii) $\forall A \in \pspA . A \cap \Union \pspB \neq \emptyset$.

    Fix $A \in \pspA$. Then there exists $\val \in X \subseteq D$ such that $A = \inv{\expr}(\val)$. By definition $(\pspB, \permaq) \models \In{\expr}{X}$ implies $\expr(\Union \pspB) = X$, therefore there must be at least one $\state \in \Union \pspB$ such that $\expr(\state) = \val$. By definition, $\state \in A$, so $\state \in A \cap \Union \pspB$.
\end{proof}

\begin{lemma}
    \Cref{rule:precise-perm} is sound.
\end{lemma}
\begin{proof}
    If $P$ has no valid model, also $P\withperm{\permap}$ does not have any.
    So assume there exists a minimal model $a = (\pspA, \permaq) \models P$. Let $\permar = \fun \p{x} . \max(\permap(\p{x}), \permaq(\p{x}))$. We claim $(\pspA, \permar)$ is the minimal model of $P\withp{\permap}$.

    By definition of $P\withp{\permap}$, $(\pspA, \permaq) \models P$ and upwards closure of assertions, $(\pspA, \permar) \models P\withp{\permap}$.
    Any other model $(\pspB, \permap') \models P\withp{\permap}$ implies, by definition, $(\pspB, \permap') \models P$ and $\permap' \geq \permap$. By minimality of $(\pspA, \permaq) \models P$ this means $\pspA \pLeq \pspB$, so $(\pspA, \permar) \raLeq (\pspB, \permap')$.
\end{proof}

\begin{lemma}
    \Cref{rule:precise-star} is sound.
\end{lemma}
\begin{proof}
    If either $P$ or $Q$ has no (valid) minimal model, then also $P * Q$ has none, so it is trivially precise.
    So assume there are minimal models $a_P \models P$, $a_Q \models Q$. We claim $a_P \raOp a_Q \models P * Q$ is the minimal model of $P * Q$.
    Fix any model $b \models P * Q$. By definition there exist $b_P \models P$, $b_Q \models Q$ such that $b_P \raOp b_Q \raLeq b$. Then, by properties of resource algebras and minimality of $a_P$, $a_Q$
    \[
    b \raGeq b_P \raOp b_Q \raGeq a_P \raOp a_Q .
    \]
\end{proof}

\begin{lemma}
    \Cref{rule:precise-oplus} is sound.
\end{lemma}
\begin{proof}
    Since $P(v)$ is precise for all $\val \in X$, let $a_\val = (\psp_\val, \permap_\val) \models P(\val)$ be the least model of $P(\val)$. If for any $v$ there is no minimal model, then $\cond{\val}{X}{P(\val)}$ has no valid model.
    Define $\psp \is \Union_{\val \in X} \psp_\val$ and $\permap \is \fun \p{x} . \sup \set{ \permap_\val(\p{x}) | \val \in X }$. We claim that $(\psp, \permap)$ is the least model of $\cond{\val}{X}{P(\val)}$.

    By \cref{lemma:p-partitions-disjoint}, we know that $\set{ \psp_\val }_{\val \in X}$ is disjoint, therefore $\psp$ is a possibility space. Moreover, by upwards closure of assertions and $\permap \geq \permap_\val$, we have that $(\psp_\val, \permap) \models P(\val)$. Thus, $(\psp, \permap) \models \cond{\val}{X}{P(\val)}$.

    Fix a model $(\pspB, \permaq) \models \cond{\val}{X}{P(\val)}$. Then there exists a disjoint family $\set{ \pspB_\val }_{\val \in X}$ such that $\pspB \pGeq \Dunion_{\val \in X} \pspB_\val$ and $(\pspB_\val, \permaq) \models P(\val)$. Because for all $\val$ $(\psp_\val, \permap_\val)$ is the minimal model of $P(\val)$ we have $\psp_\val \pLeq \pspB_\val$ and $\permap_\val \leq \permaq$.
    We also know that $\set{ \psp_\val }_{\val \in X}$ is disjoint, so by \cref{lemma:union-partitioning-monotone} we have $\psp = \Dunion_{\val \in X} \psp_\val \pLeq \Union_{\val \in X} \pspB_\val \pLeq \pspB$.
    Lastly, observe that for any fixed $\p{x} \in \VarLog$ we have $\permap(\p{x}) = \sup \set{ \permap_\val(\p{x}) | \val \in X } \leq \sup \set{ \permaq(\p{x}) | \val \in X } = \permaq(\p{x})$, so $\permap \leq \permaq$.
    Therefore, $a = (\psp, \permap) \raLeq (\pspB, \permaq)$.
\end{proof}

\subsection{Rules on Sure Assertions}

\begin{lemma}
    \Cref{rule:sure-and} is sound.
\end{lemma}
\begin{proof}
    We prove the two entailments separately.
    \begin{casesplit}
        \case[$\entails$] Fix a model $a = (\pspA, \permap) \models \sure{\bexpr_1} * \sure{\bexpr_2}$. By definition of $*$ there exists $a_i = (\pspA_i, \permap_i) \models \sure{\bexpr_i}$ for $i \in \set{ 1, 2 }$ and $a_1 \raOp a_2 \raLeq a$.
        By \cref{lemma:model-e-equal-v}, we have $\bexpr_i(\Union \pspA_i) = \set{ \true }$.
        \Cref{lemma:raop-increasing} prescribes that $a_1 \raOp a_2 \raLeq a$ implies $a_i \raLeq a$, so that, by definition of $\pLeq$:
        \[
        \bexpr_i(\Union \pspA) \subs \bexpr_i(\Union \pspA_i) = \set{ \true }
        \]
        Thus, again by \cref{lemma:model-e-equal-v}, $a = (\pspA, \permap) \models \sure{\bexpr_1 \land \bexpr_2}$.
        \case[$\entailedby$] Fix a model $a = (\pspA, \permap) \models \sure{\bexpr_1 \land \bexpr_2}$. By \cref{lemma:model-e-equal-v} $\bexpr_1(\Union \pspA) = \bexpr_2(\Union \pspA) = \set{ \true }$. Define $\evA = \Union \pspA$. Again by \cref{lemma:model-e-equal-v} we get $(\set{ \evA }, \permap) \models \sure{\bexpr_i}$ for $i \in \set{ 1, 2 }$. Moreover, $\set{ \evA } \iprod \set{ \evA } = \set{ \evA }$. Lastly, $\set{ \evA } \pLeq \pspA$ by \cref{lemma:pleq-union-psp}. This proves $a = (\pspA, \permap) \models \sure{\bexpr_1} * \sure{\bexpr_2}$.
    \end{casesplit}
\end{proof}

\begin{lemma}
    \Cref{rule:sure-persistent} is sound.
\end{lemma}
\begin{proof}
    We derive this rule from \cref{rule:sure-and}:
    \[
    \sure{\bexpr}
    \lequiv
    \sure{\bexpr \land \bexpr}
    \lequiv
    \sure{\bexpr} * \sure{\bexpr}
    \]
\end{proof}

\begin{lemma}
    \Cref{rule:sure-and-to-star} is sound.
\end{lemma}
\begin{proof}
    Fix a model $a = (\pspA, \permap) \models \sure{\bexpr} \land P$. By \cref{lemma:model-e-equal-v} $\bexpr(\Union \pspA) = \set{ \true }$ and $\pspA \compat \permap$, so $\permap(\p{x}) > 0$ for all $\p{x}$ on which $\bexpr$ depends (which in general is a subset of $\pvar(\bexpr)$).
    Define $\evA = \set{ \state \in \State | \bexpr(\state) = \true }$ and $\permaq = \fun \p{x} . \text{if } \p{x} \in \pvar(\bexpr) \text{ then } \permap(\p{x}) / 2 \text{ else } 0$ and $\permap' = \fun \p{x} . \text{if } \p{x} \in \pvar(\bexpr) \text{ then } \permap(\p{x}) / 2 \text{ else } \permap(\p{x})$. Clearly $\permap = \permaq + \permap'$.
    By \cref{lemma:model-e-equal-v} we get $(\set{ \evA }, \permaq) \models \sure{\bexpr}$ (compatibility of $\set{ \evA }$ and $\permaq$ follows from definition of $\permaq$ and $\pvar(\bexpr)$). Moreover, $\set{ \evA } \iprod \pspA = \pspA$ by \cref{lemma:union-iprod-id}.
    Since $P$ has scalable permissions on all $\p{x} \in \pvar(\bexpr)$, we get $(\pspA, \permap') \models P$. Thus
    $
    (\pspA, \permap) = (\set{ \evA }, \permaq) \raOp (\pspA, \permap') \models \sure{\bexpr} * P .
    $
\end{proof}

\subsection{Rules for the \texorpdfstring{\Condition[ing]}{Partitioning} Modality}

Soundness of the primitive rules about $\condmod$.
We show here only a representative subset of the proofs.

\begin{lemma}
    \Cref{rule:unit-r} is sound.
\end{lemma}
\begin{proof}
    We prove the two directions separately.
    \begin{casesplit}
        \case[$\entails$] Fix a valid model $a = (\pspA, \permap) \models \In{\expr}{X}$. For every $\val \in X$ define $\pspA_\val \is \set{ A \in \pspA | \expr(A) = \set{ \val } }$. By definition $\set{ \pspA_\val}_{\val \in X}$ is partitioning. By \cref{lemma:model-in-exists-value}, $\pspA = \Union_{\val \in X} \pspA_\val$. By \cref{lemma:model-e-equal-v} $\pspA_\val \models \sure{\expr=\val}$. Therefore $a \models \cond{\val}{X} \sure{\expr=\val}$.
        \case[$\dashv$] Fix a valid model $a = (\pspA, \permap) \models \cond{\val}{X} \sure{\expr=\val}$. Then there exist $\set{ \pspA_\val}_{\val \in X}$ such that $\pspA_\val \models \sure{\expr = \val}$ and $\pspA \pGeq \Union_{\val \in X} \pspA_\val$. By \cref{lemma:model-e-equal-v} $\expr(\pspA_\val) = \set{ \val }$. To prove that $(\Union_{\val \in X} \pspA_\val, \permap) \models \In{\expr}{X}$, we have to show two things.
        \begin{enumerate}[(i)]
            \item $\All d\in D. \inv{\expr}(d) \ein \Union_{\val \in X} \pspA_\val$. By definition of $\inv{\expr}(d)$ and the fact that $\expr(\pspA_\val) = \set{ \val }$ it follows that if $d = \val$ then $\Union \pspA_\val \subs \inv{\expr}(d)$ and otherwise $\inv{\expr}(d) \inters \Union \pspA_\val = \emptyset$. Therefore:
                    \begin{enumerate}
                        \item if $d \in X$ then $\inv{\expr}(d) \inters (\Union_{\val \in X} \pspA_\val) = \inv{\expr}(d) \inters \Union \pspA_\val = \Union \pspA_\val$;
                        \item if $d \not\in X$ then $\inv{\expr}(d) \inters (\Union_{\val \in X} \pspA_\val) = \emptyset$.
                    \end{enumerate}
                    In both cases, $\inv{\expr}(d) \ein \Union_{\val \in X} \pspA_\val$.
            \item $\expr(\Union_{\val \in X} \pspA_\val) = X$. This follows from $\expr(\Union_{\val \in X} \pspA_\val) = \Union_{\val \in X} \expr(\pspA_\val) = \Union_{\val \in X} \set{ \val } = X$.
        \end{enumerate}
    \end{casesplit}
\end{proof}

\begin{lemma}
    \Cref{rule:oplus-unfuse} is sound.
\end{lemma}
\begin{proof}
    Fix a valid model $a = (\psp, \permap) \models \cond{w}{Y} P(w)$. By definition of $\condmod$ we get a disjoint set $\set{\pspA_w}_{w \in Y}$ such that $\pspA \pGeq \Union_{w \in Y} \pspA_w$ and $\All w \in Y. (\pspA_w, \permap) \models P(w)$.

    Fix $\val \in X$ and define $\pspB_\val = \Union_{w \in Y_\val} \pspA_w$. Since $\set{\pspA_w}_{w \in Y}$ is disjoint, also $\set{\pspA_w}_{w \in Y_\val}$ is, so $\pspB_\val$ is a possibility space. Moreover, $\All w \in Y. (\pspA_w, \permap) \models P(w)$. Therefore, $(\pspB_\val, \permap) \models \cond{w}{Y_\val} P(w)$.
    We observe that
    \begin{align*}
        \Union_{\val \in X} \pspB_\val = \Union_{\val \in X} (\Union_{w \in Y_\val} \pspA_w) = \Union_{w \in Y} \pspA_w
    \end{align*}
    Therefore $\Union_{\val \in X} \pspB_\val$ is a possibility space, thus $\set{ \pspB_\val }_{\val \in X}$ is disjoint.

    We conclude observing that $\psp \pGeq \Union_{w \in Y} \pspA_w = \Union_{\val \in X} \pspB_\val$, so $(\psp, \permap) \models \cond{\val}{X} \cond{w}{Y_\val} P(w)$.
\end{proof}

\begin{lemma}
	\Cref{rule:oplus-unframe} is sound.
\end{lemma}
\begin{proof}
    Fix a model $a = (\psp, \permap) \models \cond{v}{X} (F * P(v))$. By definition of $\condmod$ we get a set $\set{\pspA_v}_{v\in X}$ of disjoint possibility spaces such that $\pspA \pGeq \Union_{v\in X} \pspA_v$ and $\All v \in X. (\pspA_v, \permap) \models F * P(v)$.
    Therefore for all $v$ there exist $(\pspF_v, \permar_v) \models F$ and $(\pspB_v, \permap_v) \models P(v)$ such that $(\pspF_v, \permar_v) \raOp (\pspB_v, \permap_v) \raLeq (\pspA_v, \permap)$.

    Because $F$ is precise and has a valid model, there exists a resource $(\pspF, \permar) \models F$ such that $(\pspF, \permar) \raLeq (\pspF_v, \permar_v)$ for all $v$.
    Let $\permaq \is \permap - \permar$ and fix $v$. By definition $\permar \leq \permar_v$. Then, by properties of the resource algebra:
    \[
    (\pspA_v, \permap) \raGeq (\pspF_v, \permar_v) \raOp (\pspB_v, \permap_v) \raGeq (\pspF, \permar) \raOp (\pspB_v, \permap_v)
    \]
    Therefore, $\pspF \compat \pspB_v$ and $\permap \geq \permar + \permap_v$, which means $\permaq \geq \permap_v$. By upwards closure of assertions, this means that $(\pspB_v, \permaq) \models P(v)$. Since $P$ partitions, by \cref{lemma:p-partitions-disjoint} we have that $\set{\pspB_v}_{v \in X}$ is disjoint. Therefore, $(\Union_{v \in X} \pspB_v, \permaq) \models \cond{v}{X} P(v)$.

    We claim that $(\Union_{v \in X} \pspF \iprod \pspB_v) = \pspF \iprod (\Union_{v \in X} \pspB_v)$.
    Fix $\evA \subseteq \State$.
    \begin{align*}
        &\evA \in (\Union_{v \in X} \pspF \iprod \pspB_v) \\
        \iff &\exists v \in X . \evA \in (\pspF \iprod \pspB_v) \\
        \iff &\exists v \in X , \evF \in \pspF , \evB \in \pspB_v . \evA = \evF \cap \evB \\
        \iff &\exists \evF \in \pspF , v \in X, \evB \in \pspB_v . \evA = \evF \cap \evB \\
        \iff &\exists \evF \in \pspF , \evB \in \Union_{v \in X} \pspB_v . \evA = \evF \cap \evB \\
        \iff &\evA \in \pspF \iprod (\Union_{v \in X} \pspB_v)
    \end{align*}
    In particular, $\raValid(\pspF \iprod (\Union_{v \in X} \pspB_v)) = \raValid(\Union_{v \in X} (\pspF \iprod \pspB_v)) = \true$, so
    \[
    (\pspF \iprod \Union_{v \in X} \pspB_v, \permar + \permaq) = (\pspF, \permar) \raOp (\Union_{v \in X} \pspB_v, \permaq) \models F * \cond{v}{X} P(v)
    \]

    We conclude the proof by showing that $a = (\psp, \permap) \raGeq (\pspF \iprod \Union_{v \in X} \pspB_v, \permar + \permaq)$. By upwards closure of assertions, this implies $a \models F * \cond{v}{X} P(v)$.
    We have $\pspF \iprod \pspB_v \models F * P(v) \entails P(v)$. By \cref{lemma:p-partitions-disjoint} this means $\set{ \pspF \iprod \pspB_v }_{v \in X}$ is disjoint, so by \cref{lemma:union-partitioning-monotone} we have
    \[
    \psp \pGeq \Union_{v\in X} \pspA_v \pGeq \Union_{v\in X} (\pspF \iprod \pspB_v) = \pspF \iprod (\Union_{v \in X} \pspB_v)
    \]
    We conclude using $\permap = \permaq + \permar$:
    \[
    (\psp, \permap) \raGeq (\pspF \iprod (\Union_{v \in X} \pspB_v), \permar + \permaq) .
    \]
\end{proof}

\subsection{Rules for the \texorpdfstring{\Condition[ing]}{Partitioning} Modality: Derived Rules}
Soundness of derived rules about $\condmod$.
We show here only a representative subset of the proofs.

\begin{lemma}
    \Cref{rule:oplus-own-merge} is sound.
\end{lemma}
\begin{proof}
    We start by observing the following equivalences:
    \begin{eqexplain}
        \vcond{\expr} \own{\expr'}
        \whichisequiv*
        \Exists V \subs D. \cond{\val}{V} \sure{{\expr} = \val} * \Exists V'_\val \subs D' . \In{\expr'}{V'_\val}
        \bydef
        \whichisequiv
        \Exists V \subs D, V' \from V \to D'. \cond{\val}{V} \sure{{\expr} = \val} * \In{\expr'}{V'_\val}
        \byrule{oplus-skolem}
        \whichisequiv
        \Exists V \subs D, V' \from V \to D'. \cond{\val}{V}{\cond{\val'}{V'_\val}{\big( \sure{{\expr} = \val} * \sure{\expr' = \val'} \big)}}
        \byrule{unit-r}
        \whichisequiv
        \Exists V \subs D, V' \from V \to D'. \cond{\val}{V}{\cond{\val'}{V'_\val}{\big( \sure{f(\val') = \val} * \sure{\expr' = \val'} \big)}}
        \by{$\expr = f(\expr')$}
    \end{eqexplain}
    and
    \begin{eqexplain}
        \own{\expr'}
        \whichisequiv*
        \Exists V' \subs D'. \cond{\val'}{V'} \sure{{\expr'} = \val'}
        \bydef
    \end{eqexplain}

    To conclude the left-to-right entailment we observe (we omit the types of existentials for brevity)
    \begin{eqexplain}
        &
        \Exists V, V'_{\hole} . \cond{\val}{V}{\cond{\val'}{V'_\val}{\big( \sure{f(\val') = \val} * \sure{\expr' = \val'} \big)}}
        \by{$\expr = f(\expr')$}
        \whichentails
        \Exists V , V'_{\hole}. \cond{(\val, \val')}{\Union_{\val \in V} \big( \set{\val} \times V'_\val \big)}{{\big( \sure{f(\val') = \val} * \sure{\expr' = \val'} \big)}}
        \byrule{oplus-fuse}
        \whichentails
        \Exists V, V'_{\hole}. \cond{(\val, \val')}{\Union_{\val \in V} \big( \set{\val} \times V'_\val \big)}{{\big( \sure{\expr' = \val'} \big)}}
        \whichentails
        \Exists V' . \cond{\val'}{V'}{{\big( \sure{\expr' = \val'} \big)}}
    \end{eqexplain}

    To conclude the right-to-left entailment we observe
    \begin{eqexplain}
        &
        \Exists V'. \cond{\val'}{V'} \sure{{\expr'} = \val'}
        \whichentails
        \Exists V'. \cond{\val}{f(V')}{ \cond{\val'}{\restr{V'}{f = \val}}{\sure{{\expr'} = \val'}} }
        \byrule{oplus-unfuse}
        \whichentails
        \Exists V'. \cond{\val}{f(V')}{ \cond{\val'}{\restr{V'}{f = \val}}{\sure{{\expr'} = \val'} * \sure{f(\val') = \val}} }
        \whichentails
        \Exists V, V' . \cond{\val}{V}{\cond{\val'}{V'_\val}{\big( \sure{f(\val') = \val} * \sure{\expr' = \val'} \big)}}
    \end{eqexplain}
    where the existentials in the last line are witnessed by $f(V')$ and $\fun \val . \restr{V'}{f = \val}$, respectively.
\end{proof}

\begin{lemma}
    \Cref{rule:oplus-transf} is sound.
\end{lemma}
\begin{proof}
    \begin{eqexplain}
        \cond{\val}{X}{P(v)}
        \whichentails*
        \cond{w}{\inv{f}(X)}{\cond{\val}{\set{ {f}(w) }}{P(w)}}
        \byrule{oplus-unfuse}
        \whichentails
        \cond{w}{\inv{f}(X)}{{P({f}(w))}}
        \byrule{unit-l}
    \end{eqexplain}
    By bijectivity of $f$ we have $\inv{f}(X) = Y$.
\end{proof}

\subsection{Structural Rules about $\wpsymb$}

Soundness of structural rules about $\wpsymb$.
We show here only a representative subset of the proofs.

\begin{lemma}
    \Cref{rule:wp-cons} is sound.
\end{lemma}
\begin{proof}
    Fix a model $a \models \WP{\prog}{Q}$.
    The goal is to prove that $a \models \WP{\prog}{Q'}$, that is, by definition of $\wpsymb$:
    \[
    \All S, f. (a \raOp f \raLeq S \land \sem{\prog}(S) \neq \emptyset) \implies (\Exists b . b \raOp f \raLeq \sem{\prog}(S) \land b \models Q')
    \]
    Fix $S \subs \State$ and $f \in \PSpPm$ such that $a \raOp f \raLeq S$ and $\sem{\prog}(S) \neq \emptyset$.
    By $a \models \WP{\prog}{Q}$ we get a resource $b$ such that $b \raOp f \raLeq \sem{\prog}(S)$ and $b \models Q$. By the hypothesis $Q \entails Q'$ we get $b \models Q'$. Thus, $b$ is the witness for the existential.
\end{proof}

\begin{lemma}
    \Cref{rule:wp-exist} is sound.
\end{lemma}
\begin{proof}
    Fix a model $a \models \exists x \of X . \WP{\prog}{Q(x)}$. By definition of $\existsAlt$, there exists a value $\val \of X$ such that $a \models \WP{\prog}{Q(\val)}$.
    The goal is to prove that $a \models \WP{\prog}{\exists x \of X . Q(x)}$, that is, by definition of $\wpsymb$:
    \[
    \All S, f. (a \raOp f \raLeq S \land \sem{\prog}(S) \neq \emptyset) \implies (\Exists b . b \raOp f \raLeq \sem{\prog}(S) \land b \models \exists x \of X . Q(x))
    \]
    Fix $S \subseteq \State$ and $f \in \PSpPm$ such that $a \raOp f \raLeq S$ and $\sem{\prog}(S) \neq \emptyset$.  Because $a \models \WP{\prog}{Q(\val)}$ there exists a $b$ such that $b \raOp f \raLeq \sem{\prog}(S)$ and $b \models Q(\val)$. But $Q(\val) \entails \exists x \of X . Q(x)$, from which the thesis follows.
\end{proof}

\begin{lemma}
	\Cref{rule:wp-frame} is sound.
\end{lemma}
\begin{proof}
	Fix a model $a \models P * \WP{\prog}{Q}$. By definition of $*$, there exist $a_P, a_Q$ such that $a_P \raOp a_Q \raLeq a$, $a_P \models P$, $a_Q \models \WP{\prog}{Q}$. In particular, by \cref{lemma:raop-increasing} we have that $a_P \raLeq a_P \raOp a_Q = a$.

	The goal is to prove that $a \models \WP{\prog}{P * Q}$, that is, by definition of $\wpsymb$:
	\[
	\All S, f. (a \raOp f \raLeq S \land \sem{\prog}(S) \neq \emptyset) \implies (\Exists b . b \raOp f \raLeq \sem{\prog}(S) \land b \models P * Q)
	\]
	Fix $S \subs \State$ and $f \in \PSpPm$ such that $(\pspA, \permap) \raOp f \raLeq S$ and $\sem{\prog}(S) \neq \emptyset$. By monotonicity of $\raOp$ and $a_P \raLeq a$, it follows that $\raValid(a_P \raOp f)$. We then apply the hypothesis that $a_Q \models \WP{\prog}{Q}$ with $S$ and $a_P \raOp f$ to get a $b_Q$ such that $b_Q \raOp (a_P \raOp f) \raLeq \sem{\prog}(S)$ and $b_Q \models Q$. Therefore the thesis is proved by taking $b = b_Q \raOp a_P$.
\end{proof}

\begin{lemma}
	\Cref{rule:oplus-wp-swap} is sound.
\end{lemma}
\begin{proof}
	The idea of the proof is the following. We first use expression $\expr$ to partition a model of the premise. We can then use $\wpsymb$ on each fiber to ``push it forward'' to a fiber of the conclusion that satisfies $Q(v)$. This can only be done if all fibers terminate, whence the totality hypothesis. Moreover, thanks to \cref{rule:wp-frame}, we can also push forward $\sure{\expr = v}\withp{\permaq}$, which allows us to put the fibers back together.

	Fix a valid model $(\pspA, \permap) \models \cond{v}{X} (\sure{\expr = v}\withp{\permaq} * \WP{\prog}{Q(v)})$. By definition of $\cond{v}{X}$, this implies there exists $\set{ \pspA_v }_{v \in X}$ such that $\pspA \pGeq \Union_{v\in X} \pspA_v$ and for all $v \in X$, $(\pspA_v, \permap) \models \sure{\expr = v}\withp{\permaq} * \WP{\prog}{Q(v)}$.
	The goal is to prove that $(\pspA, \permap) \models \WP{\prog}{\cond{v}{X} (\sure{\expr = v}\withp{\permaq} * Q(v))}$, that is, by definition of $\wpsymb$:
	\[
	\All S, f . ((\pspA, \permap) \raOp f \raLeq S \land \sem{\prog}(S) \neq \emptyset) \implies (\Exists b . b \raOp f \raLeq \sem{\prog}(S) \land b \models \cond{v}{X} (\sure{\expr = v}\withp{\permaq} * Q(v))
	\]
	Fix $S \subs \State$ and $f = (\pspF, \permap_f) \in \PSpPm$ such that $(\pspA, \permap) \raOp f \raLeq S$ and $\sem{\prog}(S) \neq \emptyset$.
	Define $S_v \is \set{ \state \in S | \expr(\state) = v }$ and $\restr{\pspA}{\expr = v} \is \set{ \evA \in \pspA | \expr(A) = \set{ v } }$.
	Because we use an affine interpretation of separating conjunction, $\sure{\expr = v}\withp{\permaq} * \WP{\prog}{Q(v)} \entails \sure{\expr = v}$, so by \cref{lemma:model-cond-partition:split-value} we get that $\pspA_v \pLeq \restr{\pspA}{\expr = v}$.

	We now show that $\pspA_v \iprod \pspF \pLeq S_v$. By monotonicity of $\iprod$, it is enough to show that $\restr{\pspA}{\expr = v} \iprod \pspF \pLeq S_v$.
	We first show that $\restr{\pspA}{\expr = v} \compat \pspF$: take any $A \in \restr{\pspA}{\expr = v}$ and $F \in \pspF$. $\restr{\pspA}{\expr = v} \subseteq \pspA$ as sets, therefore $A \in \pspA$ and $\pspA \compat \pspF$, so $A \inters F \ne \emptyset$.
	Now we prove that $S_v \subseteq \Union (\restr{\pspA}{\expr = v} \iprod \pspF)$. We know that $S_v \subseteq S \subseteq \Union (\pspA \iprod \pspF) = (\Union \pspA) \inters (\Union \pspF)$ (where the last equality follows from \cref{lemma:union-iprod}). Moreover, any state $\state \in S_v$ satisfies $\expr(\state) = v$, therefore by \cref{lemma:model-cond-partition:split-partitions-states} $\state \in \Union \restr{\pspA}{\expr = v}$. This means $S_v \subseteq (\Union \restr{\pspA}{\expr = v})$, so that $S_v \subseteq (\Union \restr{\pspA}{\expr = v}) \inters (\Union \pspF) = \Union (\restr{\pspA}{\expr = v} \iprod \pspF)$.
	The fact that any $A \in \restr{\pspA}{\expr = v} \iprod \pspF$ is expressible in $S_v$ follows from \cref{lemma:set-states-expressible}.
	Now we prove that for any $B \in \restr{\pspA}{\expr = v} \iprod \pspF$, $B \inters S_v \ne \emptyset$. This means that there exists $A \in \restr{\pspA}{\expr = v}$ and $F \in \pspF$ such that $B = A \inters F$. But $A \in \restr{\pspA}{\expr = v} \subseteq \pspA$, so $B \in \pspA \iprod \pspF \pLeq S$, so $B \inters S \ne \emptyset$. Since $B \subseteq A \in \restr{\pspA}{\expr = v}$, all states $\state \in B$ are such that $\expr(\state) = v$, so that $\emptyset \ne B \inters S = B \inters S_v$.

	By \cref{rule:wp-frame}, $(\pspA_v, \permap) \models \sure{\expr = v}\withp{\permaq} * \WP{\prog}{Q(v)} \entails \WP{\prog}{\sure{\expr = v}\withp{\permaq} * Q(v)}$. By the $\mayterm{\prog}$ assumption, since $S_v \neq \emptyset$ we get $\sem{\prog}(S_v) \neq \emptyset$. Thus, applying the definition of $\wpsymb$ with $S_v$ and $f = (\pspF, \permap_f)$, we get a $b_v = (\pspB_v, \permar_v)$ such that $b_v \raOp f \raLeq \sem{\prog}(S_v)$ and $b_v \models \sure{\expr = v}\withp{\permaq} * Q(v)$.
	Let $\permar \is \fun x . \max\set{ \permar_v(x) | v \in X }$. By closure of assertions under increased permissions, $(\pspB_v, \permar)  \models \sure{\expr = v}\withp{\permaq} * Q(v)$.

	We want to show that $b = (\Union_{v \in X} \pspB_v, \permar)$ is the witness for the existential, namely that $b \raOp f \raLeq \sem{\prog}(S)$ and $b \models \cond{v}{X} (\sure{\expr = v}\withp{\permaq} * Q(v))$.
	We first show that the $\set{\pspB_v}_{v\in X}$ is disjoint. Again observe that $\sure{\expr = v}\withp{\permaq} * Q(v) \entails \sure{\expr = v}$, which by \cref{lemma:model-e-equal-v} implies $\expr(\Union \pspB_v) = \set{v}$. Therefore it is disjoint, hence by \cref{lemma:disjoint-union-pspace} $\Union_{v \in X} \pspB_v$ is a possibility space. Moreover, since $(\pspB_v, \permar)  \models \sure{\expr = v}\withp{\permaq} * Q(v)$, this implies $b \models \cond{v}{X} (\sure{\expr = v}\withp{\permaq} * Q(v))$.
	To conclude the proof we only need to show that $b \raOp f \raLeq \sem{\prog}(S)$.
	We now show that $\permar + \permap_f \le 1$. Fix an input $\p{x}$ for the permission map: then there exists a $v \in X$ such that $(\permar + \permap_f)(\p{x}) = \permar(\p{x}) + \permap_f(\p{x}) = \permar_v(\p{x}) + \permap_f(\p{x})$. But $b_v \compat f$, so $\permar_v(\p{x}) + \permap_f(\p{x}) \le 1$.
	We now show that, for any $B \in \pspB$, $F \in \pspF$, $B \inters F \inters \sem{\prog}(S) \ne \emptyset$. This implies that $\pspB \compat \pspF$. Since the $\set{\pspB_v}_{v\in X}$ are disjoint, $B \in \pspB$ if and only if there exists a $v \in X$ such that $B \in \pspB_v$. Therefore, from $b_v \raOp f \raLeq \sem{\prog}(S_v)$, it follows that for every $B \in \pspB_v$, $F \in \pspF$ then $B \inters F \inters \sem{\prog}(S_v) \ne \emptyset$ and $\sem{\prog}(S_v) \subseteq \sem{\prog}(S)$.
	Together, these two facts imply that $\raValid(b \raOp f)$. To show that $b \raOp f \raLeq \sem{\prog}(S)$ we are only left to show that $\sem{\prog}(S) \subseteq \Union (\pspB \iprod \pspF)$: any set is expressible in $\sem{\prog}(S)$, and we already proved that $B \inters F \inters \sem{\prog}(S) \ne \emptyset$ for any $B \in \pspB$, $F \in \pspF$.
	$\sem{\prog}(S) = \Union_{v \in X} \sem{\prog}(S_v)$. For a fixed $v$, $b_v \raOp f \raLeq \sem{\prog}(S_v)$ implies that $\sem{\prog}(S_v) \subseteq \Union (\pspB_v \iprod \pspF)$. By taking the union over $v$ on both sides and using \cref{lemma:union-iprod}, we obtain
	\begin{align*}
		\sem{\prog}(S) &= \Union_{v \in X} \sem{\prog}(S_v) \\
        &\subseteq \Union_{v \in X} \Union (\pspB_v \iprod \pspF) \\
		&= \Union_{v \in X} ((\Union \pspB_v) \inters (\Union \pspF)) \\
		&= (\Union_{v \in X} (\Union \pspB_v)) \inters (\Union \pspF) \\
		&= (\Union \pspB) \inters (\Union \pspF) \\
		&= \Union (\pspB \iprod \pspF)
	\end{align*}
\end{proof}

\begin{lemma}
    \Cref{rule:oplus-wp-swap-weak} is sound.
\end{lemma}
\begin{proof}
    The proof proceeds exactly as for the above \cref{rule:oplus-wp-swap}, except for the definition of the resource $b$. Instead of applying the $\wpsymb$ with $S_v$ and $f = (\pspF, \permap_f)$ to get a $b_v = (\pspB_v, \permar_v)$ for every $v \in X$, we can only do so for $v \in X' = \set{ v \in X | \sem{\prog}(S_v) \neq \emptyset }$. This $X'$ is not empty because, by hypothesis, $\sem{\prog}(S) \neq \emptyset$, and can thus serve as the witness for the inner existential. The rest of the proof proceeds exactly as above.
\end{proof}

\begin{lemma}
    \Cref{rule:wp-bind} is sound.
\end{lemma}
\begin{proof}
    Fix a valid model $a = (\psp, \permap) \models P$. Since $P \entails \sure{\expr = \val}$ and $P \entails \WP{[\Ectxt[v]]}{Q}$, we get $a \models \sure{\expr = v}$ and $a \models \WP{[\Ectxt[v]]}{Q}$.

    The goal is to prove that $a \models \WP{[\Ectxt[\expr]]}{Q}$, that is, by definition of $\wpsymb$:
    \[
    \All S, f. (a \raOp f \raLeq S \land \sem{\Ectxt[\expr]}(S) \neq \emptyset) \implies (\Exists b . b \raOp f \raLeq \sem{\Ectxt[\expr]}(S) \land b \models Q)
    \]

    Fix $S \subseteq \State$ and $f = (\pspF, \permar) \in \PSpPm$ such that $a \raOp f \raLeq S$ and $\sem{\Ectxt[\expr]}(S) \neq \emptyset$.
    Since $a \models \WP{[\Ectxt[v]]}{Q}$ we get that there exists a $b$ such that $b \raOp f \raLeq \sem{\Ectxt[\val]}(S)$ and $b \models Q$. We show that $b$ is the witness for the existential. This is implied by $\sem{\Ectxt[\val]}(S) = \sem{\Ectxt[\expr]}(S)$, which we show next.

    By \cref{lemma:wp-state-subset-model}, $S \subseteq \Union \psp$. By \cref{lemma:model-e-equal-v}, $\expr(\state) = \val$ for every state $\state \in \Union \psp \supseteq S$. Therefore, by \cref{lemma:prog-semantics-subst}, $\sem{\Ectxt[\expr]}(S) = \sem{\Ectxt[\val]}(S)$.
\end{proof}

\subsection{Program Rules about $\wpsymb$}
Soundness of program rules about $\wpsymb$.
We show here only a representative subset of the proofs.

\begin{lemma}
    \Cref{rule:wp-rnd} is sound.
\end{lemma}
\begin{proof}
    Fix a valid model $a = (\pspA, \permap) \models \perm{\p{x} : 1}$. This means that $\permap(\p{x}) = 1$.

    The goal is to prove that $a \models \WP{[\code{x:=rnd($V$)}]}{\In{\p{x}}{V}\withperm{\p{x} : 1} }$, that is, by definition of $\wpsymb$:
    \[
    \All S, f. (a \raOp f \raLeq S \land \sem{\code{x:=rnd($V$)}}(S) \neq \emptyset) \implies (\Exists b . b \raOp f \raLeq \sem{\code{x:=rnd($V$)}}(S) \land b \models \In{\p{x}}{V}\withperm{\p{x} : 1}
    \]

    Fix $S \subseteq \State$ and $f = (\pspF, \permar) \in \PSpPm$ such that $a \raOp f \raLeq S$ and $\sem{\code{x:=rnd($V$)}}(S) \neq \emptyset$.

    Define $\evB_v \is \set{ (\store\upd{\p{x} -> v}, \obs) | (\store, \obs) \in \Union \pspA }$, $\pspB \is \set{ \evB_v | v \in V }$ and $b = (\pspB, \permap)$. We show that $b$ is the witness for the existential.

    We first observe that for every state $(\store', \obs') \in \evB_v$ there exists some state $(\store, \obs) \in \Union \pspA$ such that $\obs' = \obs$ and $\store' = \store\upd{\p{x} -> v}$. In other words, $\store = \store'\upd{\p{x} -> w}$ for some $w \in \Val$.
\begin{casesplit}
        \case*[$\pspB \compat \permap$] Fix \p{y} such that $\permap(\p{y}) = 0$, a $\evB_v \in \pspB$ and a state $(\store', \obs) \in \evB_v$. Since $\permap(\p{x}) = 1$, we know $\p{y} \neq \p{x}$. As observed above, there exists $(\store, \obs) \in \Union \pspA$ such that $\store' = \store\upd{\p{x} -> v}$.
        By \cref{lemma:compatibility-constructive}, we distinguish two cases:\begin{itemize}
            \item $\p{y} = \logs$. Since $\pspA \compat \permap$, for every $\obs'$ also $(\store, \obs \app \obs') \in \Union \pspA$. Then, by definition of $\evB_v$, $(\store', \obs \app \obs') = (\store\upd{\p{x} -> v}, \obs \app \obs') \in \evB_v$.
            \item $\p{y} \neq \logs$. Since $\pspA \compat \permap$, for every $w \in \Val$ also $(\store\upd{\p{y} -> w}, \obs) \in \Union \pspA$. Then, by definition of $\evB_v$, $(\store'\upd{\p{y} -> w}, \obs) = (\store\upd{\p{y} -> w, \p{x} -> v}, \obs) \in \evB_v$.
        \end{itemize}
        \case*[$\raValid(b \raOp f)$] Since $\raValid(a \raOp f)$ we have $\permap + \permar \leq \funconst 1$. Moreover, since $\permap(\p{x}) = 1$, we get $\permar(\p{x}) = 0$.
        Fix $\evF \in \pspF$ and any $\evA \in \pspA$. There must exist $(\store, \obs) \in \evA \inters \evF$. Fix a $v \in V \subs \Val$. Because $\pspF \compat \permar$ and $\permar(\p{x}) = 0$, also $(\store\upd{\p{x} -> v}, \obs) \in \evF$. By definition of $\evB_v$, we get $(\store\upd{\p{x} -> v}, \obs) \in \evB_v$. Thus, $\evB_v \inters \evF \neq \emptyset$ for all $v \in V$, which proves $\pspB \compat \pspF$.
        \case*[$b \raOp f \raLeq \sem{\code{x:=rnd($V$)}}(S)$] By \cref{lemma:wp-state-subset-model} $S \subs \Union \pspA$. By monotonicity of denotational semantics, $\sem{\code{x:=rnd($V$)}}(S) \subs \Union \pspB$. By $a \raOp f \raLeq S$, for any $\evF \in \pspF$ and $\evA \in \pspA$ there exists $(\store, \obs) \in \evA \inters \evF \inters S$. As in the previous point, for $v \in V$ we have $(\store\upd{\p{x} -> v}, \obs) \in \evB_v \inters \evF$. Moreover, by definition of $\sem{\code{x:=rnd($V$)}}$ also $(\store, \obs) \in \sem{\code{x:=rnd($V$)}}(S)$. So $\evB_v \inters \evF \inters \sem{\code{x:=rnd($V$)}}(S) \neq \emptyset$, which proves $b \raOp f \raLeq \sem{\code{x:=rnd($V$)}}(S)$.
        \case*[$b \models \In{\p{x}}{V}\withperm{\p{x} : 1}$] $\permap(\p{x}) = 1$, so the $\perm{\p{x} : 1}$ is satisfied. To prove $b \models \In{\p{x}}{V}$ we have to show two things.
        \begin{itemize}
            \item Fix $v \in V$. By definition $\Sem[E]{\p{x}}(\evB_v) = \set{ v }$, therefore $\evB_v \subs \inv{\p{x}}(v)$ and $\evB_v \inters \inv{\p{x}}(v') =\emptyset$ for $v' \neq v$. If $v \in V$ we have $\inv{\p{x}}(v) \inters \Union \pspB = \evB_v$ so it is expressible. If $v \notin V$ then $\inv{\p{x}}(v) \inters \Union \pspB = \emptyset$, so it is expressible.
            \item
            \[
            \Sem[E]{\p{x}}(\Union \pspB) = \Union_{\val \in V} \Sem[E]{\p{x}}(\evB_\val) = \Union_{\val \in V} \set{ \val } = V
            \]
        \end{itemize}
    \end{casesplit}
\end{proof}

\begin{lemma}
    \Cref{rule:wp-log} is sound.
\end{lemma}
\begin{proof}
    Fix a valid model $a = (\psp, \permap) \models \auth{(t-1)}\withp{\permap}$. Recall that $\auth{(t-1)} = \sure{\length(\logs) = t-1}\withperm{\logs : 1}$. By \cref{lemma:model-e-equal-v} this means that $\All \state \in \Union \psp . \length(\logs(\state)) = t - 1$.

    The goal is to prove that $a \models \WP{[\code{log!($\aexpr$)}]}{\auth{t} * (t \obspointsto \aexpr)\withp{\permap}}$, that is, by definition of $\wpsymb$:
    \[
    \All S, f. (a \raOp f \raLeq S \land \sem{\code{log!($\aexpr$)}}(S) \neq \emptyset) \implies (\Exists b . b \raOp f \raLeq \sem{\code{log!($\aexpr$)}}(S) \land b \models \auth{t} * (t \obspointsto \aexpr)\withp{\permap}
    \]

    Fix $S \subseteq \State$ and $f = (\pspF, \permar) \in \PSpPm$ such that $a \raOp f \raLeq S$ and $\sem{\code{log!($\aexpr$)}}(S) \neq \emptyset$.
    Since $\raValid(a \raOp f)$ and $\permap(\logs) \ge 1$, we get $\permar(\logs) = 0$.

    Define $\evB \is \set{ (\store, \obs \app (\Sem[E]{\aexpr}(\store, \obs))) | (\store, \obs) \in \Union \psp }$, $\pspB \is \set{ \evB }$ and $b = (\pspB, \permap)$. We show that $b$ is the witness for the existential.

    We prove that $\pspB \compat \permap$. Fix \p{x} such that $\permap(\p{x}) = 0$, the only $\evB \in \pspB$, a state $(\store, \obs) \in \evB$ and a value $\val \in \Val$. Since $\permap(\logs) = 1$, we know $\p{x} \neq \logs$. By \cref{lemma:compatibility-constructive}, we need to show that $(\store\upd{\p{x} -> \val}, \obs) \in \evB$.
    By definition of $\evB$, there exists $(\store, \obs_0) \in \Union \psp$ such that $(\store, \obs) = (\store, \obs_0 \app (\Sem[E]{\aexpr}(\store, \obs_0)))$. Because $\All \p{y} \in \pvar(\aexpr) . \permap(\p{y}) > 0$, it follows that $\p{x} \notin \pvar(\aexpr)$. Therefore, $\Sem[E]{\aexpr}(\store, \obs_0) = \Sem[E]{\aexpr}(\store\upd{\p{x} -> \val}, \obs_0)$, which means
    \begin{align*}
        (\store\upd{\p{x} -> \val}, \obs_0 \app \Sem[E]{\aexpr}(\store\upd{\p{x} -> \val}, \obs_0))
        = (\store\upd{\p{x} -> \val}, \obs_0 \app \Sem[E]{\aexpr}(\store, \obs_0))
        = (\store\upd{\p{x} -> \val}, \obs)
    \end{align*}
    Moreover, since $\psp \compat \permap$, also $(\store\upd{\p{x} -> \val}, \obs_0) \in \Union \psp$, which implies $(\store\upd{\p{x} -> \val}, \obs) = (\store\upd{\p{x} -> \val}, \obs_0 \app \Sem[E]{\aexpr}(\store\upd{\p{x} -> \val}, \obs_0)) \in \evB$.

    We prove that $\raValid(b \raOp f)$. From $\raValid(a \raOp f)$ we get $\permap + \permar \leq \funconst 1$. Therefore we only need to show $\pspB \compat \pspF$. Fix $\evF \in \pspF$. By $\raValid(a \raOp f)$ we know $\psp \compat \pspF$. Therefore pick any $\evA \in \pspA$, and there must be a state $(\store, \obs) \in \evA \inters \evF$. By definition of $\evB$, $(\store, \obs \app \val) \in \evB$ for $\val = \Sem[E]{\aexpr}(\store, \obs)$. Moreover, $\permar(\logs) = 0$ and $\pspF \compat \permar$, so by \cref{lemma:compatibility-constructive} also $(\store, \obs \app \val) \in \evF$.

    Fix any state $(\store, \obs) \in S$. Because $a \raOp f \raLeq S$ we have $S \subs \Union (\psp \iprod \pspF) = \Union \psp \inters \Union \pspF$ by \cref{lemma:union-iprod}. Therefore, $(\store, \obs) \in \Union \pspF$ and as before this implies $(\store, \obs \app \val) \in \Union \pspF$ for any $\val \in \Val$. Thus, $\sem{\code{log!($\aexpr$)}}(S) \subs \Union \pspF$. By definition $\sem{\code{log!($\aexpr$)}}(S) = \evB = \Union \pspB$, therefore $\sem{\code{log!($\aexpr$)}}(S) \subs \Union (\pspB \iprod \pspF)$, which means $b \raOp f \raLeq \sem{\code{log!($\aexpr$)}}(S)$.

    Lastly, we show $b \models \auth{t} * (t \obspointsto \aexpr)\withp{\permap}$. Permissions are the same as $a$. Because $\psp \models \auth{t - 1}$ we obtain that $\All (\store, \obs) \in S \subs \Union \psp . \length(\obs) = t - 1$. Therefore, any $(\store', \obs') \in \sem{\code{log!($\aexpr$)}}(S) = \evB$ is such that $\length(\obs') = \length(\obs) + 1 = t$ and $\obs'[t] = \aexpr$. These mean, by \cref{lemma:model-e-equal-v}, that $\pspB \models \sure{\length(\logs) = t} = \auth{t}$ and $\pspB \models \sure{(\logs[t] = \aexpr)} = (t \obspointsto \aexpr)$. By \cref{lemma:model-sure-and-to-star} this implies $\pspB \models \auth{t} * (t \obspointsto \aexpr)$.
\end{proof}

\begin{lemma}
    \Cref{rule:wp-if-prim} is sound.
\end{lemma}
\begin{proof}
    We show the case for $v = 1$. The case for $v = 0$ is analogous.
    Fix a model $a \models \text{if $v$  then } \WP{[\prog_1]}{Q(1)} \text{ else } \WP{[\prog_0]}{Q(0)}$. Because $v = 1$,
    \[
    \text{if $v$  then } \WP{[\prog_1]}{Q(1)} \text{ else } \WP{[\prog_0]}{Q(0)} \lequiv \WP{[\prog_1]}{Q(1)}
    \]

    To prove that
    $a \models \WP{[\code{if $\;v\;$ then $\;\prog_1\;$ else $\;\prog_0$}]}{Q(v)}$,
    fix $S \subs \State$, $f \in \PSpPm$ such that $a \raOp f \raLeq S$ and $\sem{\code{if $\;v\;$ then $\;\prog_1\;$ else $\;\prog_0$}}(S) \neq \emptyset$. Because $v = 1$, $\sem{\code{if $\;v\;$ then $\;\prog_1\;$ else $\;\prog_0$}} = \sem{\prog_1}$.
    Thus, by the hypothesis $a \models \WP{[\prog_1]}{Q(1)}$ we get the witness $b$ such that $b \models Q(1)$ and $b \raOp f \raLeq \sem{\prog_1}(S)$, which is exactly what we need to prove the thesis.
\end{proof}

\subsection{Program Rules about $\wpsymb$: Derived Rules}
Soundness of derived program rules about $\wpsymb$.
We show here only a representative subset of the proofs.

\begin{lemma}
    \Cref{rule:wp-if-det} is sound.
\end{lemma}
\begin{proof}
    For any fixed $i \in \set{ 0 , 1 }$:
    \begin{eqexplain}
        P * \sure{\bexpr = i}
        \whichentails*
        \sure{\bexpr = i} * P * \sure{\bexpr = i}
        \byrule{sure-persistent}
        \whichentails
        \sure{\bexpr = i} * \WP{[\prog_i]}{Q * \sure{\bexpr = i}}
        \by{hypothesis}
        \whichentails
        \sure{\bexpr = i} * \big(\text{if $i$  then } \WP{[\prog_1]}{Q * \sure{\bexpr = i}} \text{ else } \WP{[\prog_0]}{Q * \sure{\bexpr = i}} \big)
        \by{$i \in \set{ 0, 1 }$}
        \whichentails
        \sure{\bexpr = i} * \WP{[\code{if $\;i\;$ then $\;\prog_1\;$ else $\;\prog_0$}]}{Q * \sure{\bexpr = i}}
        \byrule{wp-if-prim}
        \whichentails
        \WP{[\code{if $\;\bexpr\;$ then $\;\prog_1\;$ else $\;\prog_0$}]}{Q * \sure{\bexpr = i}}
        \by{wp-bind}
    \end{eqexplain}
\end{proof}
 \section{Modelling Correspondence}
\label{sec:app:modelling}

In this section, we discuss in more formal detail how the triples from \cref{sec:modelling-security} correspond to standard, trace-based notions of security.

\subsection{Adequacy of Noninterference}
\label{sec:adequacy}

Noninterference was already discussed in \cref{sec:modelling-security}. Below we prove the two implications of \cref{theorem:adequacy} separately.

\begin{theorem}\label{theorem:adequacy-soundness}
    For any program $\prog$, set of valid high inputs $H \subs \Val$ with $\bot \notin H$ and function $\declOf$, if
    \begin{align*}
        \All \obsL, \vec{v} .
        \J{}|- {(\sure{\lowOf = \obsL} * \sure{\pv{x} = \vec{v}} * \sure{\highIn \in H} * \own{\logs})\withperm{\_ : 1}}{\prog}{ \vcond{\declOf}{\own{\lowOf} * \own{\highIn}} }
    \end{align*}
    then $\TINIrel{\declOf}(\prog)$ holds.
\end{theorem}
\begin{proof}
    Fix program $\prog$, set of valid high inputs $H \subs \Val$ and function $\declOf$, and assume
    \begin{align*}
        \All \obsL, \vec{v} .
        \J{}|- {(\sure{\lowOf = \obsL} * \sure{\pv{x} = \vec{v}} * \sure{\highIn \in H} * \own{\logs})\withperm{\_ : 1}}{\prog}{ \vcond{\declOf}{\own{\lowOf} * \own{\highIn}} }
    \end{align*}

    To prove $\TINIrel{\declOf}(\prog)$ fix states $\state_1,\state_1',\state_2, \state_2'$ such that $\state_1' \in \sem{\prog}(\state_1)$, $\state_2' \in\sem{\prog}(\state_2)$, $\lowOf(\state_1) = \lowOf(\state_2)$, $\pv{x}(\state_1) = \pv{x}(\state_2)$, $\highIn(\state_1), \highIn(\state_2) \in H$, $\highIn(\state_1) \neq \highIn(\state_2)$ and $\declOf(\state_1') = \declOf(\state_2')$.

    Let $\pspA \is \set{ \set{\state_1}, \set{\state_2}}$, $\permap \is \funconst 1$, $\obsL \is \lowOf(\state_1) = \lowOf(\state_2)$ and $d \is \declOf(\state_1') = \declOf(\state_2')$. We have
    \begin{align*}
        &\pspA \compat \permap
        && (\pspA, \permap) \models \sure{\lowOf = \obsL}
        && (\pspA, \permap) \models \sure{\pv{x} = \vec{v}}
        && (\pspA, \permap) \models \sure{\highIn \in H}
        && (\pspA, \permap) \models \own{\logs}
    \end{align*}
    Therefore by \cref{lemma:model-sure-and-to-star} $(\pspA, \permap) \models (\sure{\lowOf = \obsL} * \sure{\pv{x} = \vec{v}} * \sure{\highIn \in H} * \own{\logs})\withperm{\_ : 1}$. We apply the definition of $\wpsymb$ to get that
    \begin{align*}
        &\All S, f . (a \raOp f \raLeq S \land \sem{\prog}(S) \neq \emptyset)
        \implies (\Exists b \in \PSpPm . b \raOp f \raLeq \sem{\prog}(S) \land b \models \vcond{\declOf}{\own{\lowOf} * \own{\highIn}}))
    \end{align*}
    Fix $S \is \set{ \state_1, \state_2 }$ and $f \is \raUnit$.
    We have that $\state_1' \in \sem{\prog}(\state_1) \subseteq \sem{\prog}(S)$, so $\sem{\prog}(S) \neq \emptyset$.
    We also know that $\Union \pspA = S$, so by \cref{lemma:pleq-psp-set-union} we have $\pspA \pLeq S$, which means $(\pspA, \permap) \raOp f = (\pspA, \permap) \raLeq S$.
    Therefore we get the existence of $b = (\pspB, \permaq)$ such that $b \raOp f \raLeq \sem{\prog}(S)$ and
    \begin{align*}
        &b \models (\vcond{\declOf}{\own{\lowOf} * \own{\highIn}}) = \Exists R . \cond{r}{R}{\sure{\declOf = r} * \own{\lowOf} * \own{\highIn}}
    \end{align*}
    Because $(\pspB, \permaq) = b = b \raOp f \raLeq \sem{\prog}(S)$, we have that $\sem{\prog}(S) \subs \Union \pspB$, so $\state_1', \state_2' \in \Union \pspB$.
    Let $R$ be the set witnessing the existential, and let $\set{ \pspB_r }_{r \in R}$ be the split witnessing the $\condmod$, that is $\pspB \pGeq \Union_{r \in R} \pspB_r$ and $\pspB_r \models \sure{\declOf = r} * \own{\lowOf} * \own{\highIn}$ for all $r \in R$. By \cref{lemma:model-sure-and-to-star,lemma:model-e-equal-v} we get that $\declOf(\pspB_r) = \set{ r }$. Since $\state_1' \in \sem{\prog}(S) \subs \Union \pspB \subs \Union (\Union_{r \in R} \pspB_r) = \Union_{r \in R} (\Union \pspB_r)$ we have that there exists an $r_1 \in R$ such that $\state_1' \in \pspB_{r_1}$. But this means that $d = \declOf(\state_1') = r_1$, so $d \in R$ and $\state_1' \in \Union \pspB_{d}$. Analogously for $\state_2'$. From now on, we focus on $d$ only.

    By definition of $*$ there exists $\pspB_d^{L}, \pspB_d^{H}$ such that
    \begin{align*}
        \pspB_d^{L} \iprod \pspB_d^{H}&\pLeq \pspB_d & \pspB_d^{L} &\models \own{\lowOf}              &\pspB_d^{H} &\models \own{\highIn}
    \end{align*}
    By the first point we have $\state_1', \state_2' \in \Union \pspB_{d} \subs \Union (\pspB_d^{L} \iprod \pspB_d^{H}) = (\Union \pspB_d^{L}) \inters (\Union \pspB_d^{H})$ (the last equality is \cref{lemma:union-iprod}). Therefore, there exists $B^{L} \in \pspB_d^{L}$ such that $\state_1' \in B^L$ and $B^{H} \in \pspB_d^{H}$ such that $\state_2' \in B^H$. Let $B \is  B^L \inters B^H\in \pspB_d^{L} \iprod \pspB_d^{H}$.
    From $\pspB_d^{L} \iprod \pspB_d^{H} \pLeq \pspB_d$ we know that there exists a $B' \in \pspB_d$ such that $B' \subs B$. Then $\pspB_d \subs \Union_{r \in R} \pspB_r$ implies $B' \in \Union_{r \in R} \pspB_r$. Lastly, from $\Union_{r \in R} \pspB_r \pLeq \pspB \pLeq \sem{\prog}(S)$ we have $B' \inters \sem{\prog}(S) \neq \emptyset$. Therefore, also $B \inters \sem{\prog}(S) \neq \emptyset$.
    Let $\state_2'' \in B \inters \sem{\prog}(S) = B^L \inters B^H \inters \sem{\prog}(S)$.

    We prove this $\state_2''$ is the state required by  $\TINIrel{\declOf}(\prog)$, namely
    \[
    \state_2'' \in \sem{\prog}(\state_2)
    \qquad
    \lowOf(\state_1') = \lowOf(\state_2'')
    \qquad
    \declOf(\state_2') = \declOf(\state_2'')
    \]
    We observe that $\state_2'' \in \Union \pspB_d^{L} \iprod \pspB_d^{H} \subs \Union \pspB_d$, therefore $\declOf(\state_2'') = d = \declOf(\state_2')$.
    We observe that $\state_2'' \in B^L \in \pspB_d^{L}$, and also $\state_1' \in B^L$. Because $\pspB_d^{L} \models \own{\lowOf}$, by \cref{lemma:model-own-characterization} we have $\lowOf(\state_1') = \lowOf(\state_2'')$.

    To prove that $\state_2'' \in \sem{\prog}(\state_2)$, we first observe that $\state_2'' \in B^H \in \pspB_d^{H}$, and also $\state_2' \in B^H$. Because $\pspB_d^{H} \models \own{\highIn}$, by \cref{lemma:model-in-exists-value} we have $\highIn(\state_2') = \highIn(\state_2'')$. By definition $\state_2'' \in \sem{\prog}(S) = \sem{\prog}(\set{ \state_1, \state_2 })$.
    If by contradiction $\state_2'' \notin \sem{\prog}(\state_2)$ then it must be the case that $\state_2'' \in \sem{\prog}(\state_1)$. However, since executing a program cannot change the value of $\highIn$, this leads to the contradiction $\highIn(\state_2) = \highIn(\state_2'') = \highIn(\state_1) \neq \highIn(\state_2)$.
\end{proof}

\begin{theorem}\label{theorem:adequacy-completeness}
    For any program $\prog$ such that $\TINIrel{\declOf}(\prog)$ holds,
    \begin{align*}
        \All \obsL, \vec{v} .
        \J{}|- {(\sure{\lowOf = \obsL} * \sure{\pv{x} = \vec{v}} * \sure{\highIn \in H} * \own{\logs})\withperm{\_ : 1}}{\prog}{ \vcond{\declOf}{\own{\lowOf} * \own{\highIn}} }
    \end{align*}
\end{theorem}
\begin{proof}
    Fix program $\prog$, set of valid high inputs $H \subs \Val$ and function $\declOf$, and assume $\TINIrel{\declOf}(\prog)$ holds. Moreover fix $\obsL$ and $\vec{v}$ and a model
    \[
    a = (\pspA, \permap) \models (\sure{\lowOf = \obsL} * \sure{\pv{x} = \vec{v}} * \sure{\highIn \in H} * \own{\logs})\withperm{\_ : 1} .
    \]
    From this, by \cref{lemma:model-sure-and-to-star,lemma:model-e-equal-v}, it follows that
    \begin{align*}
        &\permap = \funconst 1
        &&\lowOf( \Union \pspA ) = \set{ \obsL }
        &&{\pv{x}}(\Union \pspA) = \set{ \vec{v} }
        &&\highIn( \Union \pspA ) \subs H
        &&\pspA \models \own{\logs}
    \end{align*}

    The goal is to prove that
    \[
    (\pspA, \permap) \models \WP{\prog}{ \vcond{\declOf}{\own{\lowOf} * \own{\highIn}} }
    \]
    that is, by definition of $\wpsymb$:
    \[
    \All S, f. (a \raOp f \raLeq S \land \sem{\prog}(S) \neq \emptyset) \implies (\Exists b . b \raOp f \raLeq \sem{\prog}(S) \land b \models { \vcond{\declOf}{\own{\lowOf} * \own{\highIn}} } )
    \]

	Fix $S \subs \State$ and $f = (\pspF, \permar) \in \PSpPm$ such that $a \raOp f \raLeq S$ and $\sem{\prog}(S) \neq \emptyset$.
    Because $\raValid(a \raOp f)$ we have that $\permap + \permar = \funconst 1 + \permar \leq \funconst 1$, which means $\permar = \funconst 0$.
    Because $a \raOp f \raLeq S$, by \cref{lemma:wp-state-subset-model} we have $S \subs \Union \pspA$ and $S \subs \Union \pspF$. Since $\lowOf( \Union \pspA ) = \set{ \obsL }$ and $\Sem[E]{\pv{x}}(\Union \pspA) = \set{ \vec{v} }$, this means that $\All \state \in S . \lowOf(\state) = \obsL$ and $\pv{x}(\state) = \vec{v}$.

    Let $R \is \declOf(\sem{\prog}(S))$, and define $\restr{(\sem{\prog}(S))}{\declOf=r} \is \set{ \state \in \sem{\prog}(S) | \declOf(\state) = r }$ for any $r \in R$. It follows that $\Union_{r \in R} \restr{(\sem{\prog}(S))}{\declOf=r} = \sem{\prog}(S)$.
    Define $\pspB_{r}^L$ as the set $\restr{(\sem{\prog}(S))}{\declOf=r}$ partitioned by $\lowOf$:
    \[
    \pspB_{r}^L \is \set{ \set{ \state \in \restr{(\sem{\prog}(S))}{\declOf=r} | \lowOf(\state) = \obsL' } | \obsL' \in \lowOf(\restr{(\sem{\prog}(S))}{\declOf=r}) } .
    \]
    Analogously, define $\pspB_{r}^H$ as the set $\restr{(\sem{\prog}(S))}{\declOf=r}$ partitioned by $\highIn$:
    \[
    \pspB_{r}^H \is \set{ \set{ \state \in \restr{(\sem{\prog}(S))}{\declOf=r} | \highIn(\state) = h' } | h' \in \highIn(\restr{(\sem{\prog}(S))}{\declOf=r}) } .
    \]
    It follows that $\Union \pspB_r^L = \Union \pspB_r^H = \restr{(\sem{\prog}(S))}{\declOf=r}$.

    Let $\pspB \is \Union_{r \in R} (\pspB_r^L \iprod \pspB_r^H)$. We show that the witness for the existential is $b = (\pspB, \permap)$.
    This requires proving the two conditions
    \[
    b \models \vcond{\declOf}{\own{\lowOf} * \own{\highIn}} = \Exists R' . \cond{r}{R'} (\sure{\declOf = r} * \own{\lowOf} * \own{\highIn})
    \]
    and $b \raOp f \raLeq \sem{\prog}(S)$.
    We first prove the former, in the several points below.
    \begin{itemize}
        \item For every fixed $r \in R$, we prove $\pspB_r^L \models \sure{\declOf = r}$. By \cref{lemma:model-e-equal-v} this follows from $\declOf(\Union \pspB_r^L) = \set{ r }$, which in turn follows from $\Union \pspB_r^L = \restr{(\sem{\prog}(S))}{\declOf=r}$.
        \item For every fixed $r \in R$, we prove $\pspB_r^L \models \own{\lowOf}$. By \cref{lemma:model-own-characterization}, it suffices to prove that $\lowOf(\evB)$ is a singleton for every $\evB \in \pspB_r^L$. But this is true by definition of $\pspB_r^L$.
        \item For every fixed $r \in R$, we prove $\pspB_r^H \models \own{\highIn}$. As above, this is true by definition of $\pspB_r^H$.
        \item $\pspB_r^L \compat \pspB_r^H$. Fix $\evB_L \in \pspB_r^L$ and $\evB_H \in \pspB_r^H$. Since they are both non-empty, take $\state_1' \in \evB_L$ and $\state_2' \in \evB_H$.
        By definition of $\pspB_r^L$ we have $\state_1' \in \sem{\prog}(S)$.
        If $\highIn(\state_1') = \highIn(\state_2')$ then, by definition of $\pspB_r^H$, we have $\state_1' \in \evB_H$ so the two sets intersect.
        Because $\state_1' \in \sem{\prog}(S)$ there must exist a $\state_1 \in S$ such that $\state_1' \in  \sem{\prog}(\state_1)$. Analogously there exists $\state_2 \in S$ such that $\state_2' \in \sem{\prog}(\state_2)$. Since both $\state_1, \state_2 \in S$ we have that $\lowOf(\state_1) = \obsL = \lowOf(\state_2)$ and $\highIn(\state_1), \highIn(\state_2) \in H$. We can then apply $\TINIrel{\declOf}(\prog)$ to get $\state_2'' \in \sem{\prog}(\state_2)$ such that $\lowOf(\state_2'') = \lowOf(\state_1')$ and $\declOf(\state_2'') = \declOf(\state_2') = r$. From the latter, $\state_2'' \in  \restr{(\sem{\prog}(S))}{\declOf=r}$. Then, from the former and definition of $\pspB_r^L$, we get $\state_2'' \in \evB_L$.
        Since the program does not change $\highIn$ we get $\highIn(\state_2'') = \highIn(\state_2) = \highIn(\state_2')$.
        Therefore, by definition of $\pspB_r^H$ we have $\state_2'' \in \evB_H$. So the two sets intersect.
        \item For every fixed $r \in R$, we prove $\pspB_r^L \iprod \pspB_r^H \models \sure{\declOf = r} * \own{\lowOf} * \own{\highIn}$. By \cref{lemma:model-sure-and-to-star} and definition of * this follows from the previous points.
        \item We prove that $\set{ \pspB_r^L \iprod \pspB_r^H }_{r \in R}$ are disjoint. This follows because for every $r$ we have (by the point above and \cref{lemma:model-sure-and-to-star}) $\pspB_r^L \iprod \pspB_r^H \models \sure{\declOf = r}$.
        \item We prove that $b = (\Union_{r \in R} (\pspB_r^L \iprod \pspB_r^H), \permap) \models \Exists R' . \cond{r}{R'} (\sure{\declOf = r} * \own{\lowOf} * \own{\highIn})$. We observe that $\Union_{r \in R} (\pspB_r^L \iprod \pspB_r^H) \compat \permap$ because $\permap$ is different from $0$ on every $\p{x} \in \VarLog$, therefore it is compatible with every possibility space. We use $R = \declOf(\sem{\prog}(S))$ as the witness for the existential. Then $\pspB_r^L \iprod \pspB_r^H \models \sure{\declOf = r} * \own{\lowOf} * \own{\highIn}$ for all $r \in R$ was proved above.
    \end{itemize}

    Recall that $f = (\pspF, \funconst 0)$. Towards proving that $b \raOp f \raLeq \sem{\prog}(S)$, we first show that $\pspB \compat \pspF$.
    Fix $\evF \in \pspF$ and $\evB \in \pspB$. Take $\state' \in \evB$, which must exist because $\evB$ is non-empty. Then by definition of $\pspB = \Union_{r \in R} (\pspB_r^L \iprod \pspB_r^H)$ this means there exists some $r$ and $B_L \in \pspB_r^L$ such that $\state' \in B_L$. By definition of $\pspB_r^L$ this means that there exists $\state \in S$ such that $\state' \in \sem{\prog}(\state)$. Because $S \subs \Union \pspA$, there exists $\evA \in \pspA$ such that $\state \in \evA$. $\raValid(a \raOp f)$ means that $\pspA \compat \pspF$, therefore there must exist some $\state_1 \in \evA \inters \evF$. Since both $\state, \state_1 \in \evA$ and $\pspA \models \own{\logs}$, by \cref{lemma:model-own-characterization} we have $\logs(\state) = \logs(\state_1)$. Since $\state_1 \in \evF$ and $\permar =  \funconst 0$, by \cref{lemma:compatibility-constructive} we have also $\state \in \evF$. Moreover, since the observation log is append-only, $\logs(\state) \preceq \logs(\state')$, which again by \cref{lemma:compatibility-constructive} implies $\state' \in \evF$. Thus $\evB \inters \evF \neq \emptyset$.

    Observe that $\state' \in \sem{\prog}(S)$. Therefore, since any $\evC \in \pspB \iprod \pspF$ is $\evC = \evB \inters \evF$ for some $\evF \in \pspF$ and $\evB \in \pspB$, we have $\evC \inters \sem{\prog}(S) \neq \emptyset$.

    To prove that $b \raOp f \raLeq \sem{\prog}(S)$ we are only left with proving that $\raValid(b \raOp f)$ and $\pspB \iprod \pspF \pLeq \sem{\prog}(S)$. For the former, we observe that $\permap + \permar = \funconst 1$, so it is compatible with every possibility space.
    For the latter, we just showed that $\evC \in \pspB \iprod \pspF$ satisfies $\evC \inters \sem{\prog}(S) \neq \emptyset$. By \cref{lemma:pleq-psp-set} we only need to show that $\sem{\prog}(S) \subseteq \Union (\pspB \iprod \pspF)$. By \cref{lemma:union-iprod} this is true if and only if $\sem{\prog}(S) \subseteq \Union \pspB$ and $\sem{\prog}(S) \subseteq \Union \pspF$.
    \begin{align*}
        \Union \pspB &= \Union (\Union_{r \in R} (\pspB_r^L \iprod \pspB_r^H))
        = \Union_{r \in R} ((\Union \pspB_r^L) \inters (\Union \pspB_r^H)) \\
        &= \Union_{r \in R} (\restr{(\sem{\prog}(S))}{\declOf=r} \inters \restr{(\sem{\prog}(S))}{\declOf=r})
        = \sem{\prog}(S)
    \end{align*}
    Now fix $\state' \in \sem{\prog}(S)$. This means there exists $\state \in S$ such that $\state' \in \sem{\prog}(\state)$. But $S \subs \Union \pspF$, so by $\permar =  \funconst 0$ and \cref{lemma:compatibility-constructive} we have also $\state' \in \Union \pspF$. Therefore, $\sem{\prog}(S) \subseteq \Union \pspF$.
\end{proof}

\subsection{`When' Declassification}
\label{sec:app:modelling:when-decl}

For `when' declassification, the reference notion of security is the following:

\begin{definition}[`When' declassification]
    Let $\declOf \from \Obs \to D$ be a function of the log and $H \subs \Val$ such that $\bot \notin H$.

    A program~$\prog$ satisfies \emph{`when' declassification} if:
    \[
    \All \state_1,\state_1',\state_2, \state_2', m.
    \begin{conj}
        \state_1' \in \sem{\prog}(\state_1)
        \text{, } \quad
        \state_2' \in\sem{\prog}(\state_2)
        \land
        \lowOf(\state_1) = \lowOf(\state_2)
        \text{, } \quad
        \pv{x}(\state_1) = \pv{x}(\state_2)
        \land
        \highIn(\state_1), \highIn(\state_2) \in H
        \land
        \highIn(\state_1) \neq \highIn(\state_2)
        \land
        \obsfun{\length}(\state_1) = \obsfun{\length}(\state_2)
        \text{, } \quad
        m \geq \obsfun{\length}(\state_1)
        \land
        \restr{\declOf}{m}(\state_1') = \restr{\declOf}{m}(\state_2')
    \end{conj}
    \implies
    \Exists \state_2''.
    \begin{conj}
        \state_2'' \in \sem{\prog}(\state_2)
        \land
        \restr{\lowOf}{m}(\state_1') = \restr{\lowOf}{m}(\state_2'')
        \land
        \restr{\declOf}{m}(\state_1') = \restr{\declOf}{m}(\state_2'')
    \end{conj}
    \]
\end{definition}

This definition has three major differences from $\TINIrel{\declOf}$.
First, it requires the two initial states to agree on the ``current time'', which in our encoding is the number of events that are recorded in the past log ($\obsfun{\length}(\state_1) = \obsfun{\length}(\state_2)$).
Then, it takes an arbitrary timestamp $m$ that happens during execution of the program: $m \geq \obsfun{\length}(\state_1)$, so it is not before execution of $\prog$ starts.
If the two executions have declassified the same information until $m$ (the $\restr{\declOf}{m}(\state_1') = \restr{\declOf}{m}(\state_2')$ hypothesis) then there must be another execution $\state_2''$ starting from $\state_2$ that, by the time $m$, is indistinguishable from $\state_1'$.

\begin{theorem}
    Fix any program $\prog$, set of valid high inputs $H \subs \Val$ with $\bot \notin H$ and function $\declOf$.
    Then $\prog$ satisfies `when' declassification if and only if
    \begin{align*}
        &\All \obsL, \vec{v}, n_0, m \geq n_0 . \\
        &\T {(\sure{\lowOf = \obsL} * \sure{\pv{x} = \vec{v}} * \sure{\obsfun{\length} = n_0} * \sure{\highIn \in H} * \own{\logs})\withperm{\_ : 1}}{\prog}{ \vcond{\restr{\declOf}{m}}{\own{\restr{\lowOf}{m}} * \own{\highIn}} }
    \end{align*}
\end{theorem}
\begin{proof}[Proof sketch]
    The proof is the same as \cref{theorem:adequacy} for $\TINIrel{\declOf}$ applied for each fixed $m$. The formal proof is verified in Rocq.
\end{proof}

\subsection{Paralocks}
\label{sec:app:modelling:paralocks}

In the following, we use ``low input'' to refer to the pair $(\obsL, \vec{v})$ of the initial low-log and store. We say that a state $s = (\store, \obs)$ has low input equal to $(\obsL, \vec{v})$, written $s =_{\low^A} (\obsL, \vec{v})$, exactly when $\obsfun{view}^A(\obs) = \obsL$ and $\store = \vec{v}$.

We can define the notion of attacker knowledge from \citet[Def.~5.1]{DBLP:conf/popl/BrobergS10} for our setting as follows:
\begin{definition}[Attacker knowledge]
    Fix a program $\prog$, an attacker $A$, a low input $(\obsL, \vec{v})$ and a set of valid high inputs $H^A \subs \Val$.
    The knowledge of the attacker $A$ after observing a trace $\vec{u}$ is defined as the set of possible high inputs that produced that trace:
    \[
    k_A(\obsL, \vec{v}, \vec{u}) = \set{ h \in H^A | \exists s, s' \st s =_{\low^A} (\obsL, \vec{v}) \land \highIn^A(s) = h \land s' \in \sem{\prog}(s) \land \vec{u} \prefixOf \obsfun{view}^A(s') }
    \]

    The \emph{progress-insensitive} knowledge of the attacker $A$ is defined similarly, but ignoring high inputs that cannot produce a next event after $\vec{u}$:
    \[
    k^{+}_A(\obsL, \vec{v}, \vec{u}) = \set{ h \in H^A | \exists s, s' \st s =_{\low^A} (\obsL, \vec{v}) \land \highIn^A(s) = h \land s' \in \sem{\prog}(s) \land \exists u \st \vec{u}u \prefixOf \obsfun{view}^A(s') }
    \]
\end{definition}

Note that the set of valid high inputs $H^A$ must also depend on the attacker: different attackers induce different notions of what is secret and what is known in the input.

We write $\obsfun{open}^A_{\vec{u}}(s)$ to mean the set of locks open at the point of execution after all events in $\vec{u}$ have been seen in the log of $s$, but before the next event visible by $A$. Initially, we assume all locks are closed. Note that we write all open/close events in the log, even if the attacker may not be able to see them, so this can be computed by looking at the log of $s$. $\obsfun{open}^A_{m}(s)$ is defined by letting $\vec{u}$ be the first $m$ elements of the $A$-observable trace in $s$, and then computing $\obsfun{open}^A_{\vec{u}}(s)$.
Analogously to \citet[Def.~5.2]{DBLP:conf/popl/BrobergS10}, we define
\[
\text{Run}_A(\prog, \obsL, \vec{v}) \is \set{ (\vec{u}u, \obsfun{open}^A_{\vec{u}}(s')) | s =_{\low^A} (\obsL, \vec{v}) \land \highIn^A(s) \in H^A \land s' \in \sem{\prog}(s) \land \vec{u}u \prefixOf \obsfun{view}^A(s') }
\]
Note that, in line with their definition, we require the next observable event after $\vec{u}$ to be $u$, but we output the set of locks open after $\vec{u}$ has been emitted (and before the next A-visible event has).

The definition of Paralocks security~\citep[Def.~5.3]{DBLP:conf/popl/BrobergS10} in the original paper is termination-sensitive. To match the termination-insensitive nature of Violet, we adapt it to be termination-insensitive too.
The definition of termination-insensitive security~\citep[Def.~5.4]{DBLP:conf/popl/BrobergS10} can be instead used directly.

\begin{definition}
    We say $\prog$ is \emph{Paralocks secure} if, for all attackers $A = (a, \Sigma)$, low inputs $(\obsL, \vec{v})$ and runs $(\vec{u}u, \Delta) \in \text{Run}_A(\prog, \obsL, \vec{v})$,
    \[
    \Delta \subs \Sigma \implies
    k_A(\obsL, \vec{v}, \vec{u}u)  =  k^{+}_A(\obsL, \vec{v}, \vec{u})
    \]

    We say $\prog$ is \emph{termination-insensitive secure} if, for all attackers $A = (a, \Sigma)$, low inputs $(\obsL, \vec{v})$ and pairs of runs $(\vec{u} u, \Delta), (\vec{u} u', \Delta') \in \text{Run}_A(\prog, \obsL, \vec{v})$,
    \[
    \Delta \subs \Sigma \implies
    k_A(\obsL, \vec{v}, \vec{u} u)  =  k_A(\obsL, \vec{v}, \vec{u} u')
    \]
\end{definition}

To relate the triple above to the knowledge-based definitions above, we step through an intermediate \TINI-like definition, given below. We then prove separately that the triple implies this notion, and this notion implies the knowledge-based ones.

\begin{definition}[Paralocks trace-security]
    Fix any program $\prog$, set of attackers $\mathcal{A}$ and, for every attacker $A \in \mathcal{A}$, set of valid high inputs $H^A \subs \Val$.
    Then $\prog$ satisfies \emph{Paralocks trace-security} if and only if
    \begin{multline*}
        \forall A = (a, \Sigma), \state_1,\state_1',\state_2, \state_2', \state_3, \state_3', m \st
        \\
        \begin{conj}
            \state_1' \in \sem{\prog}(\state_1)
            \text{, } \quad
            \state_2' \in\sem{\prog}(\state_2)
            \text{, } \quad
            \state_3' \in\sem{\prog}(\state_3)
            \land
            \obsfun{view}^A(\state_2) = \obsfun{view}^A(\state_1)
            \text{, } \quad
            \pv{x}(\state_2) = \pv{x}(\state_1)
            \land
            \obsfun{view}^A(\state_3) = \obsfun{view}^A(\state_1)
            \text{, } \quad
            \pv{x}(\state_3) = \pv{x}(\state_1)
            \land
            \highIn^A(\state_1), \highIn^A(\state_2), \highIn^A(\state_3) \in H^A
            \land
            \obsfun{view}^A_m(\state_1') = \obsfun{view}^A_m(\state_2') = \obsfun{view}^A_m(\state_3')
            \land
            \obsfun{hasNxt}^A_m(\state_1')
            \text{, } \quad
            \obsfun{hasNxt}^A_m(\state_2')
            \text{, } \quad
            \obsfun{hasNxt}^A_m(\state_3')
            \land
            \obsfun{open}^A_{m}(\state_3') \subs \Sigma
        \end{conj}
        \implies
        \Exists \state_0, \state_0'.
        \begin{conj}
            \state_0' \in \sem{\prog}(\state_0)
            \land
            \obsfun{view}^A(\state_0) = \obsfun{view}^A(\state_2)
            \text{, } \quad
            \pv{x}(\state_0) = \pv{x}(\state_2)
            \land
            \highIn^A(\state_0) = \highIn^A(\state_2)
            \land
            \obsfun{view}^A_{m+1}(\state_0') = \obsfun{view}^A_{m+1}(\state_1')
        \end{conj}
    \end{multline*}
\end{definition}

Paralocks trace-security implies Paralocks security and termination-insensitive security.

\begin{theorem}
    For any fixed program $\prog$,
    \[
    \prog \text{ is Paralocks trace-secure} \implies \prog\text{ is Paralocks secure}
    \]
    and
    \[
    \prog\text{ is Paralocks trace-secure} \implies \prog\text{ is termination-insensitive secure}
    \]
\end{theorem}
\begin{proof}[Proof sketch]
    For the first statement, the inclusion $k_A(\obsL, \vec{v}, \vec{u}u) \subs k^{+}_A(\obsL, \vec{v}, \vec{u})$ is trivial. For the other inclusion, take any $h \in k^{+}_A(\obsL, \vec{v}, \vec{u})$.
    Then there must be some $w$ such that $h$ produces the sequence of observables $\vec{u} w$. Take $h'$ to be the high input of the run $(\vec{u} u, \Delta)$, so that $h' \in k_A(\obsL, \vec{v}, \vec{u}u)$. We apply Paralocks trace-security at $m = \length(\vec{u})$ on the runs starting from $h'$ and $h$. As third run we use exactly the one that produces $(\vec{u}u, \Delta)$, so that the hypothesis $\obsfun{open}^A_{m}(\state_3') \subs \Sigma$ follows from the assumption of Paralocks security that $\Delta \subs \Sigma$. Thus, we obtain another state $s_0'$ that has the high input $h$ and produces the same $\obsfun{view}^A$ as starting from $h'$ for $\length(\vec{u}) + 1$, that is exactly $\vec{u}u$.

    For the second statement, the same argument as above proves that $k_A(\obsL, \vec{v}, \vec{u} u) = k^{+}_A(\obsL, \vec{v}, \vec{u})$. To prove that $k_A(\obsL, \vec{v}, \vec{u} u') = k^{+}_A(\obsL, \vec{v}, \vec{u})$ we apply Paralocks trace-security, but using as third run the run $(\vec{u} u, \Delta)$.
\end{proof}

For brevity, in the theorem below we write \eqref{triple:paralocks} instead of the full triple from the main text.

\begin{theorem}
    For any fixed program $\prog$, if \eqref{triple:paralocks} is valid and $\prog$ does not change $\highIn^A$ then $\prog$ is Paralocks trace-secure.
\end{theorem}
\begin{proof}[Proof sketch]
    The proof is similar to the one for \TINIrel{\declOf}. The formal proof is verified in Rocq.
\end{proof}
 \section{Proof Patterns}
\label{sec:app:proof-patterns}

In this section, we show some common proof patterns on simple examples: how to handle permissions on non-log variables, how to use framing and sequencing for code without branches.
Other than in this section, these patterns are implicitly used in examples throughout the paper.

For convenience, we define the following two abbreviations:
\begin{align*}
    \frag[q]{t} &\is \sure{\obsfun{\length}(\logs) = t}\withperm{\logs : q}
    &
    \frag{t} &\is \Exists \epsilon \of \PosReal . \pure{0 < \epsilon < 1} * \frag[\epsilon]{t}
\end{align*}
Note that $\auth{t} = \frag[1]{t}$.

In the following example, and in all examples in this section, we show on the simplest possible example some common patterns of rule applications.

\begin{example}\label{ex:app:h-assigned-l}
    Consider the program \p{h:=l}. As is, this program has no observable effect. To make the assignment observable, we pair it with an explicit \code{log!} statement. Moreover, we read the initial high value from the log, as discussed in \cref{sec:key-ideas:spec}.
    \begin{center}
        \code{h:=$\highIn$; h := l; log!(high, h)}
    \end{center}

    We focus first on the assignment \p{h:=l} to show how to handle permissions on variables. Suppose the precondition is $\auth{t} * \sure{\p{l} = v_l}\withperm{\p{l} : \onehalf} * \sure{\p{h} = \highIn}\withperm{\p{h} : 1}$ with $v_l \neq \bot$. To apply \cref{rule:wp-assign} we need permission $1$ on \p{h} and any non-zero permission on \p{l}. For the former, we are forced to discard the fact $\sure{\p{h} = \highIn}$ to recover all permission on \p{h}: this ensures no frame can reference \p{h}. For \p{l}, we can just split uniformly the $\onehalf$ permission between $\sure{\p{l} = v_l}\withperm{\p{l} : \onequarter}$ and $\sure{\p{l} \neq \bot}\withperm{\p{l} : \onequarter}$: since \p{l} is not changed, a frame can record information on it. Note that we are duplicating (and weakening) $\sure{\p{l} = v_l}$: this is possible because sure assertions are persistent (\cref{rule:sure-persistent}).
    \begin{eqexplain}
        &
        \auth{t} * \sure{\p{l} = v_l}\withperm{\p{l} : \onehalf} * \sure{\p{h} = \highIn}\withperm{\p{h} : 1}
        \whichentails
        \auth{t} * \sure{\p{l} = v_l}\withperm{\p{l} : \onequarter} * \sure{\p{l} \neq \bot}\withperm{\p{l} : \onequarter} * \perm{\p{h} : 1}
        \whichentails
        \WP{[\p{h:=l}]}{\auth{t} * \sure{\p{l} = v_l}\withperm{\p{l} : \onequarter} * \sure{\p{h} = \p{l}}\withperm{\p{l} : \onequarter, \p{h} : 1}}
        \by{\cref{rule:wp-assign,rule:wp-frame}}
    \end{eqexplain}
    This is the general pattern for permission on variables: any assertion remembering facts on a modified variable is discarded, while permissions on other variables are just split uniformly with the frame. Because this basic pattern is quite mechanical, we will omit permissions on program variables from now on, and only deal with permissions on \logs.

    The derivation is carried out below, where we underline the actual precondition of the rule; everything else is a frame that is copied over with \cref{rule:wp-frame}. Note that $\highIn$ only reads the first entry of the log, therefore it needs $0$ permission on the logs.
\begin{equation*}
        \begin{proofoutline}
            \assert{ \auth{t} * \sure{\p{l} = v_l} * \underline{\sure{\p{h} = v_h}}  } \\
            \quad{ \p{h:=}$\highIn$ } \\
            \assert{ \auth{t} * \sure{\p{l} = v_l} * \underline{\sure{\p{h} = \highIn}} } \\
            \quad{ \p{h:=l} } \\
            \assert{  \underline{\auth{t}} * \sure{\p{l} = v_l} * \sure{\p{h} = \p{l}}  } \\
            \quad\text{ \code{log!(high, h)} } \\
            \assert{  \auth{t + 1} * (t+1) \obspointsto (\high, \p{h}) * \sure{\p{l} = v_l} * \sure{\p{h} = \p{l}}  }
        \end{proofoutline}
    \end{equation*}

    This derivation can then be lifted to the \eqref{triple:tini} precondition as in \cref{ex:h-assigned-l}.
\end{example}

The following example shows how to (and \emph{not} to!) extract independence from certain assertions.
\begin{example}\label{ex:app:h-assigned-l:tini}
    From \cref{ex:app:h-assigned-l} we get for \code{h:=$\highIn$; h := l; log!(high, h)} the postcondition
    \[
    \tcond[t+1]{\bigl(  \auth{t + 1} * (t+1) \obspointsto (\high, \p{h}) * \sure{\p{l} = v_l} * \sure{\p{h} = \p{l}} * \lowOf \obspointsto[t] \obsL * \own{\restr{\logs}{t}}  \bigr)}
    \]

    To recover independence of $\own{\lowOf}$ and $\own{\highIn}$, we observe that the final low log is $\lowOf \obspointsto \obsL$. This is a sure assertion (precise thanks to \cref{rule:precise-sure} when given the fraction of permission on the log that $\frag{t + 1}$ does not keep), therefore we can unframe it with \cref{rule:oplus-unframe}. We can then discard all other information and observe that $\highIn$ is a function of the log, so $\own{\logs} \entails \own{\highIn}$:
    \begin{eqexplain}
        &
        \tcond[t+1]{\bigl(  \auth{t + 1} * (t+1) \obspointsto (\high, \p{h}) * \sure{\p{l} = v_l} * \sure{\p{h} = \p{l}} * \lowOf \obspointsto[t] \obsL * \own{\restr{\logs}{t}}  \bigr)}
        \whichentails
        \lowOf \obspointsto \obsL * \tcond[t+1]{\bigl(  \frag{t + 1} * \own{\logs}  \bigr)}
        \byrule{oplus-unframe}
        \whichentails
        \lowOf \obspointsto \obsL * \own{\logs}
        \byrule{oplus-own-merge}
        \whichentails
        \lowOf \obspointsto \obsL * \own{\highIn}
    \end{eqexplain}

    Note that this is not yet the noninterference triple: $\lowOf \obspointsto \expr \not\entails \own{\lowOf}$ in general, because $\expr$ may depend on the rest of the log. In this case, we observe that $\obsL$ is a deterministic value, so we pick $S = \set{ \obsL }$ and perform the following derivation:
    \begin{eqexplain}
        \lowOf \obspointsto \obsL
        &=
        \sure{ \lowOf = \obsL }
        \whichentails
        \cond{\obsL'}{\set{ \obsL }}{\sure{ \lowOf = \obsL' }}
        \byrule{unit-l}
        \whichentails
        \In{\lowOf}{\set{ \obsL }}
        \byrule{unit-r}
        \whichentails
        \own{\lowOf}
    \end{eqexplain}
    To apply \cref{rule:unit-l}, we have to know the value $\obsL$, which in this case is a deterministic value fixed at the metalevel, outside of the triple.

    To see how this can fail, consider the insecure program \code{h:=$\highIn$; l := h; log!(low, l)}. The derivation for this program is almost the same:
    \begin{equation*}
        \begin{proofoutline}
            \assert{ (\lowOf \obspointsto \obsL * \sure{\p{l} = v_l} * \sure{\p{h} = v_h} * \own{\logs})\withperm{\logs : 1} } \\
            $\entails$ \assert{ \tcond{ \bigl(\auth{t} * \sure{\p{l} = v_l} * \sure{\p{h} = v_h} * \lowOf \obspointsto[t] \obsL * \own{\restr{\logs}{t}} \bigr) } } \\
            \begin{proofjump}[rule:oplus-triple]
                \begin{proofjump}[rule:frame]
                    \assert{ \auth{t} * \sure{\p{l} = v_l} * \sure{\p{h} = v_h}  } \\
                    \quad\text{ \code{h:=$\highIn$; l := h; log!(low, l)} } \\
                    \assert{ \auth{t + 1} * (t+1) \obspointsto (\low, \p{l}) * \sure{\p{l} = \p{h}} * \sure{\p{h} = \highIn} } \\
                \end{proofjump} \\
            \end{proofjump} \\
            \phantom{$\entails$} \assert{ \tcond[t+1]{\bigl(  \auth{t + 1} * (t+1) \obspointsto (\low, \p{l}) * \sure{\p{l} = \p{h}} * \sure{\p{h} = \highIn} * \lowOf \obspointsto[t] \obsL * \own{\restr{\logs}{t}}  \bigr)} }
        \end{proofoutline}
    \end{equation*}

    Importantly, $\lowOf \obspointsto (\obsL \app \highIn) = \sure{ \lowOf = (\obsL \app \highIn)}$ is still a sure assertion (with the permission no longer in $\frag{t + 1}$) and therefore precise: it can be unframed.
    \begin{eqexplain}
        &
        \tcond[t+1]{\bigl(  \auth{t + 1} * (t+1) \obspointsto (\low, \p{l}) * \sure{\p{l} = \p{h}} * \sure{\p{h} = \highIn} * \lowOf \obspointsto[t] \obsL * \own{\restr{\logs}{t}}  \bigr)}
        \whichentails
        \tcond[t+1]{\bigl(  \frag{t + 1} * \lowOf \obspointsto (\obsL \app \highIn) * \own{\logs}  \bigr)}
        \whichentails
        \lowOf \obspointsto (\obsL \app \highIn) * \tcond[t+1]{\bigl(  \frag{t + 1} * \own{\logs}  \bigr)}
        \byrule{oplus-unframe}
        \whichentails
        \lowOf \obspointsto (\obsL \app \highIn) * \own{\logs}
        \byrule{oplus-own-merge}
        \whichentails
        \lowOf \obspointsto (\obsL \app \highIn) * \own{\highIn}
    \end{eqexplain}

    But this program is clearly insecure. The key point is that $(\obsL \app \highIn)$ is not a deterministic value, so the application of \cref{rule:unit-l} to obtain $\In{\lowOf}{\set{ \obsL \app \highIn }}$ is invalid. In general, $\lowOf \obspointsto (\obsL \app \highIn) \not\entails \own{\lowOf}$.
\end{example}
 \section{More Examples}
\label{sec:app:more-case-studies}

Here we show in detail some triples and derivations in \thelogic{}.
The ``Password checker'' example shows the basic proof patterns in the presence of declassification.
The ``Occlusion'' example is usually considered insecure in quantitative settings, but is secure in the purely possibilistic interpretation. The fact that we can prove it in \thelogic{} both shows that \thelogic{} truly corresponds to a possibilistic interpretation and showcases some complex proof patterns that can be carried out in \thelogic{} without breaking the abstractions of the logic.
The ``Weak swap'' examples shows that, for certain programs, the weaker postcondition of \cref{rule:oplus-wp-swap-weak} is sufficient to prove TINI.

\subsection{Password Checker}
Consider the program that takes a low \p{guess} and a high \p{pwd}, compares them and outputs (on the low channel) whether they are equal or not. This program is secure because of declassification.

\begin{sourcecode}[gobble=0]
def pwdcheck:
    pwd := $\highIn$;
    guess := $\lowOf$[1];
    ok := (pwd = guess);
    log!(decl, pwd = guess);
    log!(low, ok)
\end{sourcecode}

As in \cref{ex:app:h-assigned-l} we can prove the triple
\begin{align*}
    \Jv{}|-
    {\sure{\lowOf = \obsL} * \sure{\pv{x} = \vec{v}} * \sure{\highIn \in H} * \own{\logs}}
    {\p{pwdcheck}}
    { \tcond[t+2]{ \begin{grp}
            \auth{t+2} * (t+1) \obspointsto (\decl, \lowOf[1] = \highIn) * {} \\
            (t+2) \obspointsto (\low, \lowOf[1] = \highIn) * \lowOf \obspointsto[t] \obsL * \own{\restr{\logs}{t}}
    \end{grp}  }}
\end{align*}

\Condition[ed] on $\declOf$, this program satisfies $\TINIrel{\declOf}$ because the low log becomes deterministic:
\begin{eqexplain}
    &
    \tcond[t+2]{ \begin{grp}
            \auth{t+2} * (t+1) \obspointsto (\decl, \lowOf[1] = \highIn) * {} \\
            (t+2) \obspointsto (\low, \lowOf[1] = \highIn) * \lowOf \obspointsto[t] \obsL * \own{\restr{\logs}{t}}
    \end{grp} }
    \whichentails
    \lowOf \obspointsto (\obsL \app (\lowOf[1] = \highIn)) * \sure{\declOf = (\lowOf[1] = \highIn)} * \tcond[t+2]{ \bigl(
            \auth{t+2} * \own{\restr{\logs}{t+2}}
    \bigr) }
    \byrule{oplus-unframe}
    \whichentails
    \lowOf \obspointsto (\obsL \app (\lowOf[1] = \highIn)) * \sure{\declOf = (\lowOf[1] = \highIn)} * \own{{\logs}}
    \byrule{oplus-own-merge}
    \whichentails
    \vcond[\obsD]{\declOf}{\bigl(
        \sure{\obsD = (\lowOf[1] = \highIn)} * \lowOf \obspointsto (\obsL \app \obsD) * \own{{\logs}}
    \bigr)}
    \byrule{unit-l}
    \whichentails
    \vcond[\obsD]{\declOf}{\bigl(
        \sure{\obsD = (\lowOf[1] = \highIn)} * \own{{\logs}} * \cond{\obsL'}{\set{ \obsL \app \obsD }}{\lowOf \obspointsto \obsL'}
    \bigr)}
    \byrule{unit-l}
    \whichentails
    \vcond[\obsD]{\declOf}{\bigl(
        \own{{\highIn}} * \In{\lowOf}{\set{ \obsL \app \obsD }}
    \bigr)}
    \byrule{unit-r}
\end{eqexplain}

\subsection{Occlusion}
\label{ex:app:occlusion}

Consider the occlusion program below, with valid values $H$ for the high input.
\begin{sourcecode}[gobble=0]
def occlusion:
    h := $\highIn$;
    b := rnd({0, 1});
    if b:
        l := h
    else:
        l := rnd($V$)
\end{sourcecode}

It is known that this program is insecure in a probabilistic sense, because values possible for \p{h} have higher probability. It is also known that in the possibilistic interpretation this program is secure, provided $H \subs V$: the nondeterministic sampling must cover the entire space of secrets.
In \thelogic{}, this side condition emerges naturally.

For simplicity, we focus on the precondition $\In{\highIn}{H}$ and we prove that \p{l} is independent of \p{h}; this can be extended to the full \TINIrel{\declOf} precondition as in \cref{ex:app:h-assigned-l}.

By using $\wpsymb$ rules, we can prove the triple
\begin{align*}
    \Jv{}|-
    {\In{\highIn}{H}}
    {\p{occlusion}}
    { \In{\highIn}{H} * \sure{\p{h} = \highIn} *
        \cond{b}{\set{0, 1}}{\left(
            \sure{\p{b} = b} * \begin{cases*}
                \pure{b} &$\implies \sure{\p{l} = \p{h}}$ \\
                \pure{\lnot b} &$\implies \In{\p{l}}{V}$
            \end{cases*}
            \right)
    } }
\end{align*}

We first \condition on \p{h} with \cref{rule:unit-r}. This fixes \p{h} to some deterministic value $h$. Then, by \cref{rule:unit-l,rule:unit-r}
\[
\sure{\p{l} = \p{h}} \lequiv \sure{\p{l} = h} \lequiv \cond{l}{\set{ h}} \sure{\p{l} = h} \lequiv \In{\p{l}}{\set{ h }} .
\]
The next step is to observe that $b$ is a fixed value inside the \condition[ing], so we can unify the set of values for \p{l}. Define $L_b = \text{if } b \text{ then } \set{ h } \text{ else } V$. Then, in both cases $\pure{b}$ and $\pure{\lnot b}$ we have
\[
\begin{array}{rlr}
    \pure{b} \implies \sure{\p{l} = \p{h}} &\;\entails\; \pure{b} \implies \In{\p{l}}{\set{ h }} &\entails\; \pure{b} \implies \In{\p{l}}{L_b} \\
    \pure{\lnot b} \implies \In{\p{l}}{V} &\;\entails\; &\pure{\lnot b} \implies \In{\p{l}}{L_b}
\end{array}
\]

Thus, the postcondition simplifies to
\[
\cond{h}{H}{\bigl(
    \sure{\p{h} = \highIn} * \sure{\p{h} = h} *
    \cond{b}{\set{0, 1}}{\left(   \sure{\p{b} = b} * \In{\p{l}}{L_b}   \right) }
    \bigr)}
\]

We can now \condition on \p{l} with \cref{rule:unit-r} and merge this with the \condition[ing] on $b$ with \cref{rule:oplus-fuse}: let $D_h \is \set{ (b, l) | b \in \set{0, 1}, l \in L_b }$ in
\[
\cond{h}{H}{\bigl(
    \sure{\p{h} = \highIn} * \sure{\p{h} = h} *
    \cond{b, l}{D_h}{\left(   \sure{\p{b} = b} * \sure{\p{l} = l} \right) }
    \bigr)}
\]

We observe that
\begin{equation*}
    D_h
    = \set{ (b, l) | b \in \set{0, 1}, l \in L_b }
    = \set{ (0, v) | v \in V } \cup \set{ (1, h) }
\end{equation*}
In order to apply \cref{rule:oplus-unfuse}, we would like to split this set on the second coordinate rather than the first. However, we can only do so if $h \in V$, which we can obtain from \cref{rule:oplus-pure} and $H \subs V$: we can introduce $\pure{h \in V}$ inside the \condition[ing]. Note that if we can find a value $\hat{h}\in H$ that is not in $V$, that prevents the rest of the derivation, and is in fact a source of leakage: if the attacker sees $\hat{h}$, they know that it is the secret and not noise from $V$.

We continue the rewriting and observe that
$
D_h = \set{ (0, h), (1, h) } \cup \Union_{v \neq h} \set{ (0, v) }
$,
and we define $C_h \is \set{ 0, 1 }$ and $C_v \is \set{ 0 }$ for $v \neq h$: thus, $D_h = \Union_{v \in V} \set{v} \times C_v$.
Then, by \cref{rule:oplus-unfuse} we obtain
\[
\cond{h}{H}{\bigl(
    \sure{\p{h} = \highIn} * \sure{\p{h} = h} * \sure{\p{h} \in H} *
    \cond{l}{V}{\cond{b}{C_l}{\left(   \sure{\p{b} = b} * \sure{\p{l} = l} \right)}}
    \bigr)}
\]
Now the derivation continues straightforwardly:
\begin{eqexplain}
    &
    \cond{h}{H}{\bigl(
        \sure{\p{h} = \highIn} * \sure{\p{h} = h} * \sure{\p{h} \in H} *
        \cond{l}{V}{\cond{b}{C_l}{\left(   \sure{\p{b} = b} * \sure{\p{l} = l} \right)}}
        \bigr)}
    \whichentails
    \cond{h}{H}{\bigl(
        \sure{\p{h} = \highIn} * \sure{\p{h} = h} * \sure{\p{h} \in H} *
        \cond{l}{V}{\bigl( \sure{\p{l} = l} * \cond{b}{C_l}{\sure{\p{b} = b}} \bigr)}
        \bigr)}
    \byrule{oplus-unframe}
    \whichentails
    \cond{h}{H}{\bigl(
        \sure{\p{h} = \highIn} * \sure{\p{h} = h} * \sure{\p{h} \in H} *
        \cond{l}{V}{\bigl( \sure{\p{l} = l} \bigr)}
        \bigr)}
    \whichentails
    \cond{h}{H}{\bigl(
        \sure{\p{h} = \highIn} * \sure{\p{h} = h} * \sure{\p{h} \in H} *
        \In{\p{l}}{V}
        \bigr)}
    \byrule{unit-r}
    \whichentails
    \sure{\p{h} = \highIn} * \In{\p{l}}{V} *
    \cond{h}{H}{\bigl( \sure{\p{h} = h} \bigr)}
    \byrule{oplus-unframe}
    \whichentails
    \sure{\p{h} = \highIn} * \In{\p{l}}{V} * \In{\p{h}}{H}
    \byrule{unit-r}
\end{eqexplain}

\subsection{TINI with Weak Swap}
\label{ex:app:weak-swap-ti}

Consider the code below, which leaks information on the secret \p{h} through termination only.
\begin{sourcecode}[gobble=0]
def terminationleak:
    h := $\highIn$;
    while (h = 0) do skip;
    log!(low, 0);
\end{sourcecode}

This program does not satisfy $\mayterm{\p{terminationleak}}$, so we can only apply \cref{rule:oplus-wp-swap-weak}. In this example, we show that this is still enough to prove $\TINI$.

Let $H$ be the set of possible initial values, and consider the precondition $\In{\highIn}{H} * \auth{t}$ (this lifts to the full $\TINI$ precondition as in \cref{ex:app:h-assigned-l}).

Since the guard of the \p{while} is unchanged by the body, we can make it deterministic through a \condition[ing]. We prove the same triple for both cases of $\p{h} = 0$.

If $\p{h} = 0$, let us define $P_{\true}(n) \is \sure{\p{h} = 0}$, $P_{\false}(n) \is \False$, $P(n) \is P_\true(n)$ and $E \is 0$.
Using \cref{rule:wp-while} we obtain

\begin{derivation}
    \infer*{
        \All n, i . P_i \entails \sure{(\p{h} = 0) = i}
        \\
        \All n . \sure{\p{h} = 0} \entails \WP{[\p{skip}]}{\sure{\p{h} = 0}}
        \\
        \All n . \sure{\p{h} = 0} \entails \cond{i}{\set{ \true }}{P_i} \\
        \\
        \All n . \False \entails \sure{0 = n}
    }{\sure{\p{h} = 0} \entails \WP{ \p{while (h = 0) do skip} }{ \False }
    }
\end{derivation}
and the postcondition $\False$ also entails $\sure{\p{h} = 0}$.

If $\sure{\p{h} \neq 0}$, let us define $P_{\true}(n) \is \False$, $P_{\false}(n) \is \text{if } n = 0 \text{ then } \sure{\p{h} \neq 0} \text{ else } \False$, $P(n) \is P_{\false}(n)$ and $E \is 0$.
Using \cref{rule:wp-while} we obtain

\begin{derivation}
    \infer*{
        \All n, i . P_i \entails \sure{(\p{h} = 0) = i}
        \\
        \All n . \False \entails \WP{[\p{skip}]}{ P(n+1) }
        \\
        \All n . P(n) \entails \cond{i}{\set{ \false }}{P_i} \\
        \\
        \All n . (\text{if } n = 0 \text{ then } \sure{\p{h} \neq 0} \text{ else } \False) \entails \sure{0 = n}
    }{\sure{\p{h} \neq 0} \entails \WP{ \p{while h = 0 do skip} }{ \sure{\p{h} \neq 0} }
    }
\end{derivation}

We then perform the derivation below

\begin{eqexplain}
    &
    \In{\highIn}{H} * \auth{t}
    \whichentails
    \cond{h}{H}{ \bigl( \sure{\highIn = h} * \auth{t} \bigr) }
    \byrule{unit-r}
    \whichentails
    \cond{h}{H}{ \bigl( \WP{[\code{h := } \highIn]}{ \sure{\p{h} = \highIn} * \sure{\p{h} = h} * \auth{t} } \bigr) }
    \whichentails
    \cond{h}{H}{ \bigl( \WP{[\code{h := } \highIn\code{; while (h = 0) do skip}]}{ \sure{\p{h} = \highIn} * \sure{\p{h} = h} * \auth{t} } \bigr) }
    \by{derivations above}
    \whichentails
    \cond{h}{H}{ \bigl( \WP{[\p{terminationleak}]}{ \sure{\highIn = h} * \auth{t+1} * \lowOf \obspointsto 0 } \bigr) }
\end{eqexplain}

To obtain a triple, we can now use \cref{rule:oplus-wp-swap-weak}:

\begin{eqexplain}
    &
    \cond{h}{H}{ \bigl( \WP{[\p{terminationleak}]}{ \sure{\highIn = h} * \auth{t+1} * \lowOf \obspointsto 0 } \bigr) }
    \whichentails
    \WP{[\p{terminationleak}]}{ \Exists H' . \pure{\emptyset \subsetneq H' \subs H} * \cond{h}{H'} \bigl( \sure{\highIn = h} * \auth{t+1} * \lowOf \obspointsto 0 \bigr) }
\end{eqexplain}

Still, this triple is enough to prove $\TINI$: even without knowing $H'$ exactly, we can move $(\lowOf \obspointsto 0)$ outside with \ref{rule:oplus-unframe}, then fold the remaining \condition[ing] in $\own{\highIn}$.

\begin{eqexplain}
    &
    \Exists H' . \pure{\emptyset \subsetneq H' \subs H} * \cond{h}{H'} \bigl( \sure{\highIn = h} * \auth{t+1} * \lowOf \obspointsto 0 \bigr)
    \whichentails
    \Exists H' . \pure{\emptyset \subsetneq H' \subs H} * \cond{h}{H'} \bigl( \sure{\highIn = h} * \auth{t+1} \bigr) * \lowOf \obspointsto 0
    \byrule{oplus-unframe}
    \whichentails
    \Exists H' . \In{\highIn}{H'} * \lowOf \obspointsto 0
    \byrule{unit-r}
    \whichentails
    \own{\highIn} * \own{\lowOf}
\end{eqexplain}
 \section{Abstractions}
\label{sec:app:abstractions}

\begin{figure}[t]
    \rulesection*{Original rules}
    \begin{proofrules}
        \infer*[lab=vs-skip]{}{
            \VSJ{\pclabel}|-{\code{skip}}
        }

        \infer*[lab=vs-assign]{
    \seclabelof(\p{x}) \ge \pclabel \\
    \aexpr : {\seclabelof(\p{x})}
}{
    \VSJ{\pclabel}|-{\code{x := $\aexpr$}}
}

        \infer*[lab=vs-seq]{
            \VSJ{\pclabel}|-{\prog_1} \\
            \VSJ{\pclabel}|-{\prog_2}
        }{
            \VSJ{\pclabel}|-{\prog_1\p{;} \prog_2}
        }
        \label{rule:vs-seq}

        \infer*[lab=vs-cons]{
            \VSJ{\pclabel}|-{\prog} \\
            \pclabel' \leq \pclabel
        }{
            \VSJ{\pclabel'}|-{\prog}
        }
        \label{rule:vs-cons}

        \infer*[lab=vs-if]{
    \bexpr : \pclabel \\
    \VSJ{\pclabel}|-{\prog_1} \\
    \VSJ{\pclabel}|-{\prog_2}
}{
    \VSJ{\pclabel}|-{\code{if $\;\bexpr\;$ then $\;\prog_1\;$ else $\;\prog_2$}}
}
 \end{proofrules}

    \rulesection*{Added rules}
    \begin{proofrules}
        \infer*[lab=vs-rnd]{
    \seclabelof(\p{x}) \ge \pclabel \\ V \neq \emptyset
}{
    \VSJ{\pclabel}|-{\code{x := rnd($V$)}}
} 

        \infer*[lab=vs-repeat]{
            \VSEJ|- \aexpr : \pclabel \\
            \VSJ{\pclabel}|-{\prog}
        }{
            \VSJ{\pclabel}|-{\code{repeat $\;\aexpr\;$ : $\;\prog$}}
        }
        \label{rule:vs-repeat}
    \end{proofrules}

    \caption{Rules of the \TINI{} type system. The first block is adapted from \citet[Fig.~3]{DBLP:journals/jsac/SabelfeldM03}, without the rule for \p{while}.}
    \label{fig:app:vs-type-system}
\end{figure}

As discussed in the main text, we deviate from the type system in \citet{DBLP:journals/jsac/SabelfeldM03} in two ways. First, we replace \p{while} with the bounded loop construct \p{repeat}, to ensure termination. Second, we include an explicit nondeterministic assignment \code{rnd}.

The full definition of the compilation of programs in the language of \citet{DBLP:journals/jsac/SabelfeldM03}, with the described deviations, is as follows:
\begin{align*}
    \Comp{\code{skip}} &= \code{skip} \\
    \Comp{\code{x:=$\aexpr$}} &= \code{log!($\seclabelof$(x),'x', $\aexpr$); x:=$\aexpr$} \\
    \Comp{\code{x:=rnd($V$)}} &= \code{x:=rnd($V$); log!($\seclabelof$(x),'x', x)} \\
    \Comp{\prog_1\p; \prog_2} &= \Comp{\prog_1} \p; \Comp{\prog_2} \\
    \Comp{\code{if $\;\bexpr\;$ then $\;\prog_1\;$ else $\;\prog_2$}} &= \code{if $\;\bexpr\;$ then } \Comp{\code{$\prog_1$}}\code{ else }\Comp{\code{$\prog_2$}} \\
    \Comp{\code{repeat $\;\aexpr\;$ : $\;\prog_1$}} &= \code{log!($\progcnt$, $\aexpr$); repeat $\;\aexpr\;$ : } \Comp{\code{$\prog_1$}}
\end{align*}

$\progcnt$ is an otherwise unused tag to store in the log the number of times a \p{repeat} is executed. It has no security meaning and cannot be seen at any security level.

We write $\aexpr : \secl$ to mean that $\All \p{x} \in \pvar(\aexpr) . \seclabelof(\p{x}) \le \secl$, that is, all variables that appear in $\aexpr$ have at most security label $\secl$.

\begin{proof}[Proof of \cref{theorem:vs-soundness}]
    All rules are proved as lemmas in the Rocq formalization.
\end{proof}

\end{document}